\documentclass[fleqn,usenatbib]{mnras}

\usepackage{newtxtext,newtxmath}

\usepackage[T1]{fontenc}

\DeclareRobustCommand{\VAN}[3]{#2}
\let\VANthebibliography\thebibliography
\def\thebibliography{\DeclareRobustCommand{\VAN}[3]{##3}\VANthebibliography}

\usepackage{graphicx}	
\usepackage{amsmath}	
\usepackage{xspace}
\usepackage{ulem}
\usepackage{xcolor}

\newcommand{\HI}{\ion{H}{i}\xspace}
\newcommand{\Hmol}{H$_2$\xspace}
\newcommand{\fHI}{$f_{HI}$\xspace}
\newcommand{\fmol}{$f_{H2}$\xspace}

\newcommand{\hmpc}{h^{-1}{\rm Mpc}}

\newcommand{\simba}{\mbox{{\sc Simba}}\xspace}

\newcommand{\mufasa}{\mbox{{\sc Mufasa}}\xspace}

\newcommand{\fedd}{f_{\rm Edd}}

\newcommand{\disperse}{\mbox{{\sc DisPerSE}}\xspace}
\newcommand{\caesar}{\mbox{{\sc Caesar}}\xspace}

\newcommand\tb{\textcolor{black}}
\newcommand\tbn{\textcolor{black}}

\definecolor{lightblue}{rgb}{0.12, 0.56, 1.0}
\definecolor{applegreen}{rgb}{0.55, 0.71, 0.0}

\title[How galaxy properties vary with filament proximity in the \simba simulations]{How galaxy properties vary with filament proximity in the \simba simulations}

\author[]{
Teodora-Elena Bulichi$^{1}$\thanks{E-mail: teob1823@mit.edu},
Romeel Dav{\'e}$^{1,2,3}$,
Katarina Kraljic$^4$
\\
$^{1}$ Institute for Astronomy, University of Edinburgh, Royal Observatory, Blackford Hill, Edinburgh EH9 3HJ, UK.\\
$^{2}$University of the Western Cape, Bellville, Cape Town 7535, South Africa \\
$^{3}$South African Astronomical Observatories, Observatory, Cape Town 7925, South Africa\\
$^{4}$ Observatoire Astronomique de Strasbourg, Universit\'e de Strasbourg, CNRS, UMR 7550, F-67000 Strasbourg, France
}

\date{Accepted XXX. Received YYY; in original form ZZZ}

\pubyear{2023}

\begin{document}
\label{firstpage}
\pagerange{\pageref{firstpage}--\pageref{lastpage}}
\maketitle

\begin{abstract}
We explore the dependence of global galaxy properties in the \simba simulation as a function of distance from filaments identified using \disperse. We exclude halos with mass $M_h>10^{13}M_\odot$ to mitigate the impact of group and cluster environments.  Galaxies near filaments are more massive and have more satellites, which we control for by examining deviations from best-fit scaling relations. At $z=0$, star formation (SF) is significantly suppressed within $\la 100$~kpc of filaments, more strongly for satellites, indicating substantial pre-processing in filaments.  By $z=2$, the trend is weak and if anything indicates an increase in SF activity close to filaments.  The suppression at $z\la 1$ is accompanied by lowered \HI fractions, and increased metallicities, quenched fractions, and dispersion-dominated systems.  $H_2$ fractions are not strongly suppressed when controlling for stellar mass, suggesting that star formation efficiency drives the drop in SF.  By comparing amongst different \simba feedback variant runs, we show that the majority of SF suppression owes to filamentary shock-heating, but there is a non-trivial additional effect from AGN feedback.  When looking around massive ($M_h>10^{13}M_\odot$) halos, those galaxies near filaments behave somewhat differently, indicating that filaments provide an additional environmental effect relative to halos. Finally, we compare \simba results to EAGLE and IllustrisTNG at $z=0$, showing that all models predict SF suppression within $\la100$~kpc of filaments, nonetheless, detailed differences may be observationally testable.
\end{abstract}


\section{Introduction}

The Universe on large scales is comprised of a network of galaxies, gas and dark matter forming the so-called cosmic web \citep[e.g.][]{bond_1996,aragon-calvo_2010}. This large-scale structure (LSS) consisting of void regions, sheet-like walls, filaments, and nodes is predicted by the Zel'dovich's model for the gravitational collapse of ellipsoidal fluctuations in the matter density field (\citealt{zeldovich1, zeldovich2}). 
The features of the cosmic web have been brought to light via systematic galaxy redshift surveys (e.g. \citealt{de_lapparent_1986,Geller_Huchra1989,Colless2001,Tegmark2004}), and are also supported by simulations which predict the hierarchical formation of voids, walls and filaments assuming the well-established cold dark matter (CDM) paradigm (e.g. \citealt{Springel2005}). 

Within the cosmic web, galaxies continuously grow and evolve, their properties being strongly correlated with their local environments. Denser environments show over-abundances of massive halos due to the enhanced dark matter densities and the proto-halo's earlier collapse (e.g. \citealt{bond_1996}), which favours the formation of massive galaxies. \tbn{Such environments also appear to have a non-negligible effects on galaxies, even when controlling for galaxy masses, giving rise to the long observed morphology--density and colour--density relations (\citealt{Dressler1980, PostmanGeller1984, Dressler1986,Kauffmann2004, Baldry2006}; for reviews see \citealt{BosselliGavazzi2006,BosselliGavazzi2014}).}

Traditionally, environmental effects have been studied by contrasting galaxies within groups and clusters versus those in the ``field''.  In such studies, the environment is a proxy for halo mass, with dense environments representing massive halos with virial mass $\ga 10^{13}M_\odot$.  Yet the field population itself may not be homogeneous in terms of its environmental dependence.  For instance, field galaxies within a filamentary environment could have enhanced galaxy growth due to the greater availability of gas relative to void regions, or else could be retarded if that gas were shock-heated on large-scale structure.  The filamentary web may be the site of ``pre-processing'', in which galaxy properties are altered prior to entering into group and cluster environments \citep[e.g.][]{fujita2004, wetzel2013}.  Disentangling these effects is important for fully characterising the role of the environment in galaxy evolution.

However, identifying the imprint of the cosmic web on galaxy properties beyond the dominant effect of local density and mass has been shown to be a daunting task. 
Early observational works struggled to find clear evidence of 
such signature. Among them, \cite{alpaslan2015} found that the galaxies' properties are primarily influenced by stellar mass, rather than the environment. \tb{Additionally}, \cite{eardley2015} suggested that the observed cosmic web environmental effects on galaxy properties can be explained solely by their corresponding local densities. These contradicting results may be partly explained by the inability to properly distinguish between the effects of present local densities and past large-scale environments, given their strong correlation. To sort out these issues, it is of crucial importance to distinguish between mass- and environmental-driven effects, as well as clearly separate group- and cluster-like environments from large-scale cosmic web features.



From galaxy surveys there is substantial evidence that galaxies close to filaments are more massive and show lower levels of star formation. This has been shown using the Sloan Digital Sky Survey  \citep[SDSS;][]{ababazajian2009} by \cite{chen2017,kuutma2017,pouder2017}, using the VIMOS Public Extragalactic Redshifts Survey Multi-$\lambda$ Survey \citep[VIPERS-MLS;][]{moutard2016a, scodeggio2018} by \cite{malavasi2017}, using COSMOS-2015~\citep{Laigle2015} by \cite{laigle2018}, using the Galaxy and Mass Assembly survey \citep[GAMA;][]{Driver2009} by \cite{alpaslan2015,kraljic2018}, and using the WISExSuperCOSMOS survey \citep[WISExSCOS;][]{bilicki2016} by \cite{bonjean2020} \citep[but see][for observational results finding enchanced levels of (specific) star formation in the proximity of filaments]{davish2014,vulcani2019}. These mass and SFR trends have also been supported
several studies focused on cosmic voids showing that galaxies residing within them tend to be less massive, bluer, and more star-forming \citep[e.g.][]{grogin2000,rojas2004,kreckel2011,hoyle2012,beygu2016} compared to higher density environments
\citep[but see][for claims on no significant impact of void environment on galaxy properties]{kreckel2015,ricciardelli2014,wegner2019}.

Galaxy formation simulations within a cosmological context should naturally yield such environmental trends as a consequence of the interplay between galaxy accretion and large-scale structure.  However, again, the results are mixed.  \cite{kraljic2018} and \cite{malavasi2021} investigated galaxy properties near filaments using the \textsc{HorizonAGN} (\citealt{dubois2014}) and \textsc{IllustrisTNG} (\citealt{TNG}) simulations, respectively, and generally reported suppressed star formation in agreement with some observational results.  On the other hand, when looking at high-z massive dense filaments and dwarf galaxies, \cite{zheng2022} found a slight increase in the SFR, using the Auriga simulations (\citealt{auriga}). Additionally, \cite{kotecha2022} found that galaxies close to filaments tend to be more star-forming when looking at simulated clusters from the Three Hundred Project (\citealt{300}). This proves once again that, apart from the different prescriptions of simulations, the sample selection and environment classification play a vital role in such analyses too. Reconciling all these results likely requires considering survey selection effects and the specific techniques used to characterise filamentary structure, but in principle, the properties of galaxies within the filamentary large-scale structure should provide a novel test of galaxy formation models.

Other properties have also been investigated in terms of the cosmic web environment. \cite{kuutma2017,pouder2017} showed using SDSS that at fixed environment density, the elliptical fraction is higher close to filaments. \cite{pouder2017} proposed that the differences in galaxy star formation properties result from the higher abundances of elliptical galaxies close to filaments.  A similar result was also observed by \cite{Castigani2021}, when looking out to 12 virial radii from the Virgo cluster.
\cite{salerno2020}, using the Six Degree Field Galaxy Survey (6dFGS, \citealt{6dF}), found that galaxies arriving at clusters by following filaments are more quenched than galaxies that accrete onto clusters isotropically (see also \citealt{gouin2020} and \citealt{malavasi2021}).  While these results suggest that the morphology-density and colour-density relations are established during the pre-processing phase, they are still focused on the vicinity of massive halos rather than the full cosmic web.

Gas and metal content have also been explored in terms of the filamentary web.  For the atomic hydrogen content in galaxies (\HI), \cite{kleiner2017} and \cite{Crone2018} reported different results. \cite{kleiner2017} showed, using the 6dFGS that galaxies more massive than $M_*>10^{11}M_\odot$ show higher \HI to stellar mass ratio (\HI fraction) near filaments, while no trend is observed for lower-mass galaxies. \cite{Crone2018} used the ALFALFA \HI survey (\citealt{alpha}) to show that the \HI fraction increases at increasing distance from filaments, at fixed local density and stellar mass. These last results are supported by \cite{Castigani2021}, who added that the molecular hydrogen (\Hmol) does not show a clear trend with respect to the distance from filaments. For metallicity, this has only been explored in both observations and simulations, with \cite{winkel_2021} reporting that centrals in SDSS are more metal enriched close to cosmic web structures and \cite{Donan2022} showed via IllustrisTNG simulations and SDSS that both the gas-phase and stellar metallicities are higher for galaxies closer to filaments and nodes.

One way forward to make sense of these controversies is to use state-of-the-art galaxy formation simulations to disentangle these effects.  In this context, state-of-the-art refers to simulations that reproduce the global galaxy trends in star formation rate, gas content, and metallicity versus mass; restricting to such models then provides a plausible baseline for teasing out subtle environmental effects.  To this end, this study explores how the scalar galaxy properties vary with respect to the distance to the closest filament identified using Discrete Persistent Structure Extractor (\disperse; \citealt{Sousbie2011,SousbiePK2011}) within the \simba (\citealt{Dave2019}) simulation suite.  \simba reproduces many global trends related to SFR, quenched fractions, \HI content, and many other global properties~\citep[e.g.][]{Dave2019,Dave2020}.  In this first paper, we quantify trends in 3-D space (i.e. not accounting for redshift space distortions) using all resolved \simba galaxies to better understand the intrinsic impact of filaments on galaxy properties.  We examine stellar mass ($M_*$), specific SFR (sSFR), \HI and \Hmol fractions, metallicity, and quenched fractions versus distance to the nearest filament.  We focus on galaxies outside of massive halos ($M_h\leq 10^{13}M_\odot$) in order to restrict ourselves to a classical field galaxy sample.  We study both intrinsic quantities versus distance to filament, as well as departures from the ``main sequence'' of these quantities vs. $M_*$ in order to control for the fact that galaxies near filaments are more massive. In addition, we use \simba's feedback variants to understand which trends come from large-scale structure versus particular feedback processes (as modeled in \simba).  Finally, we apply the same procedure to the EAGLE~\citep{eagle} and IllustrisTNG~\citep{Weinberger2017, Marinacci2018, Naiman2018, Nelson2018, TNG, Springel2018, Nelson2019, Pillepich2019} simulations to determine whether the trends seen with \simba are robust to variations in galaxy formation model.  We leave for future work a redshift-space comparison to observations, applying observational selection and uncertainties to robustly quantify constraints on galaxy formation models.

The rest of the paper is structured as follows: \S\ref{sect.sim_and_analysis} describes the tools and methods implemented in this study; \S\ref{Sect.prop_gen} presents the trends of the galaxy properties of interest with respect to their proximity to filaments, while \S\ref{sect:dev} investigates further the deviations of these properties from their scaling relations with $M_*$, with respect to distance from filaments; \S\ref{feedback} and \S\ref{sect.halos} compare the impact of the \simba's feedback variants and massive halos respectively with the impact of the cosmic web; \S\ref{sect.eagle,tng} compares our main findings with the results of the EAGLE and IllustrisTNG simulations and \S\ref{sect.concl} provides a summary of this study. 

We adopt the cosmological constants of the Planck Collaboration (\citealt{Planck2016}), implemented in \simba: $\Omega_m~=~0.3$, $\Omega_{\Lambda}~=~0.7$, $\Omega_b~=~0.048$, $H_0~=~60 \, \mathrm{km s^{-1} Mpc^{-1}} h^{-1}$, $\sigma_8~=~0.82$ and $n_s~=~0.97$.

\section{Simulations and Analysis} \label{sect.sim_and_analysis}

In this section, we discuss the methods employed throughout our study in order to obtain the relevant galaxy properties and cosmic web features.

\subsection{\simba}
\label{Sect:Simba}

We use the large-scale cosmological hydrodynamical \simba simulations \citep{Dave2019} for this work. \simba builds upon its predecessor \mufasa \citep{Dave2016} which uses the Meshless Finite Mass version of the \textsc{Gizmo} code \citep{Hopkins2015}, together with the \textsc{GADGET-3} tree-particle-mesh gravity solver \citep{Springel2005}. We refer the reader to \cite{Dave2019} for a full description and here summarise the relevant features of this work.

\simba models non-equilibrium cooling from primordial elements along with metal line cooling using \textsc{Grackle-3.1} (\citealt{Smith2017}), employing a spatially uniform photo-ionising background attenuated with a simple prescription for self-shielding in dense regions.  The chemical enrichment module makes use of yield tables for Type II supernovae (SNII, \citealt{Nomoto2006}), Type Ia supernovae (SNIa, \citealt{Iwamoto1999}) and asymptotic giant branch (AGB) stars \citep{OppenheimerDave2006}.  Using the metallicity and local column density, the \Hmol fraction in each gas element is computed via the subgrid recipe from \citet{KrumholzGnedin2011}.  Star formation then proceeds assuming a \citet{Schmidt1959} relation law with 2\% of the \Hmol mass being converted into stars in a local dynamical time, with a minimum density of $n_H>0.13$ cm$^{-3}$ for star formation to occur.  Galactic winds, putatively driven by SNII, are modeled in a kinetic manner, with kick probability and velocity assigned to roughly mimic scalings with galaxy stellar mass as predicted by the Feedback in Realistic Environments (FIRE) simulations~\citep{Muratov2015,Angles-Alcazar2017}.  After the kick, a wind element does not feel hydrodynamic forces or cooling until it reaches a density 1\% of the threshold density for star formation, or 2\% of a Hubble time from launch.  30\% of the winds are heated to the temperature provided by SNII, and winds are metal loaded by assigning a metallicity $dZ$ to each wind particle via $dZ = f_{\mathrm{SNII}}y_{\mathrm{SNII}}(Z)/MAX(\eta,1)$, where $f_{\mathrm{SNII}} = 0.18$ is the stellar mass fraction lost to supernova, $y_{\mathrm{SNII}}(Z)$ is the metal-dependent Type II SN yield for each species and $\eta$ is the mass loading factor. \simba locks individual metals into dust, removing them from the gas phase, following \citet{Li2019}. Taking all these aspects into consideration, \simba predicts mass-metallicity relation (MZR) evolution~\citep{Dave2019}, and star formation rate evolution~\citep{Katsianis2021} in agreement with observations, typically as well or better than other comparable simulations.

\simba simulates black hole growth via torque-limited accretion (\citealt{HopkinsQ2011,Angles-Alcazar2013,Angles-Alcazar2015}) for cool gas and Bondi accretion for hot gas. Black hole feedback is modeled as a mixture of kinetic feedback and X-ray energy feedback. The kinetic mode is designed to reproduce the observed two-mode feedback, separated via the Eddington fraction $\fedd$. In the high accretion mode, radiative AGN winds are modeled by assigning outflow velocities to the gas particles surrounding the black hole, dependent on the corresponding black hole mass. In the jet mode, initiated once $f_\mathrm{Edd} < 0.2$ and maximised when $f_\mathrm{Edd} < 0.02$, the assigned outflow velocities adopt considerably larger values than in the high accretion modes, increasing with decreasing $f_\mathrm{Edd}$. The X-ray feedback is introduced in a full-speed jet scenario, and when the ratio $M_{\mathrm{gas}}/M_{*} < 0.2$. This involves injecting energy into the surrounding gas usually via a spherical outwards push. In \simba the accretion energy determines galaxy quenching, with the jet mode feedback primarily responsible for this aspect and the X-ray feedback contributing significantly to suppressing residual star formation.  As a result of this AGN growth and feedback model, \simba reproduces the observed stellar mass function evolution from $z=6\to 0$, the star-forming main sequence, and quenched fractions in agreement with observations~\citep{Dave2019}.

Galaxies and halos are identified in post-processing using the \caesar package, as described in \citet{Dave2019}.  During the run, particles are grouped into halos using a 3D friends-of-friends (FoF) algorithm using a linking length of 0.2 times the mean interparticle spacing. Within each halo, \caesar identifies galaxies using a 6D FoF with a smaller linking length, applied to cool gas and stars only.  Stellar mass and SFR are computed as the total values among all particles grouped into a single galaxy (or halo).  The metallicity is computed in several ways, including stellar mass-weighted (from star particles) and SFR-weighted (from gas elements).  Given that considerable amounts of \HI can be present in extended regions outside the star-forming regions of galaxies, \caesar assigns each particle to the galaxy to which it is most gravitationally bound, then sums the total \HI from those particles. \Hmol is computed similarly, though the vast majority of \Hmol lies within \caesar galaxies. There is good agreement between the observations and the \simba simulated galaxy \HI and \Hmol fractions, as well as their scaling relations with stellar mass~\citep{Dave2020}.

In this work, we focus on the (100 comoving Mpc $h^{-1})^3$ main \simba run, evolved from $z = 249\to 0$ with 1024$^3$ gas elements and 1024$^3$ dark matter particles. The minimum (adaptive) gravitational softening length for this simulation is $\epsilon_{\mathrm{min}} = 0.5 h^{-1}$kpc. The mass resolution for the initial gas element and dark matter particles are $m_{\mathrm{gas}} = 1.82\times 10^7 M_\odot$ and $m_{\mathrm{DM}} = 9.60\times 10^7 M_\odot$, respectively.  This gives a minimum resolved stellar mass for galaxies as $M_{\mathrm{*,min}} = 5.8\times 10^8  M_\odot$, i.e. 32 gas\tb{/stellar} element masses (\tb{star particles have on average same mass as gas particles, $m_{\mathrm{gas}}$}). We utilize the \caesar catalog to infer the galaxy properties of interest, at redshifts $z = 0, 1, 2$ (i.e. snapshots 151, 105, 78).  We also use the feedback variant runs of \simba, which have the exact same resolution except in a 50 Mpc $h^{-1}$ box with $2\times 512^3$ particles, and whose runs turn off individual feedback modules as we will detail later.

\subsection{Tracing the cosmic web with \disperse}
\label{sect.disperse}

\begin{figure*}
    \includegraphics[width=0.8\textwidth, height=7cm]{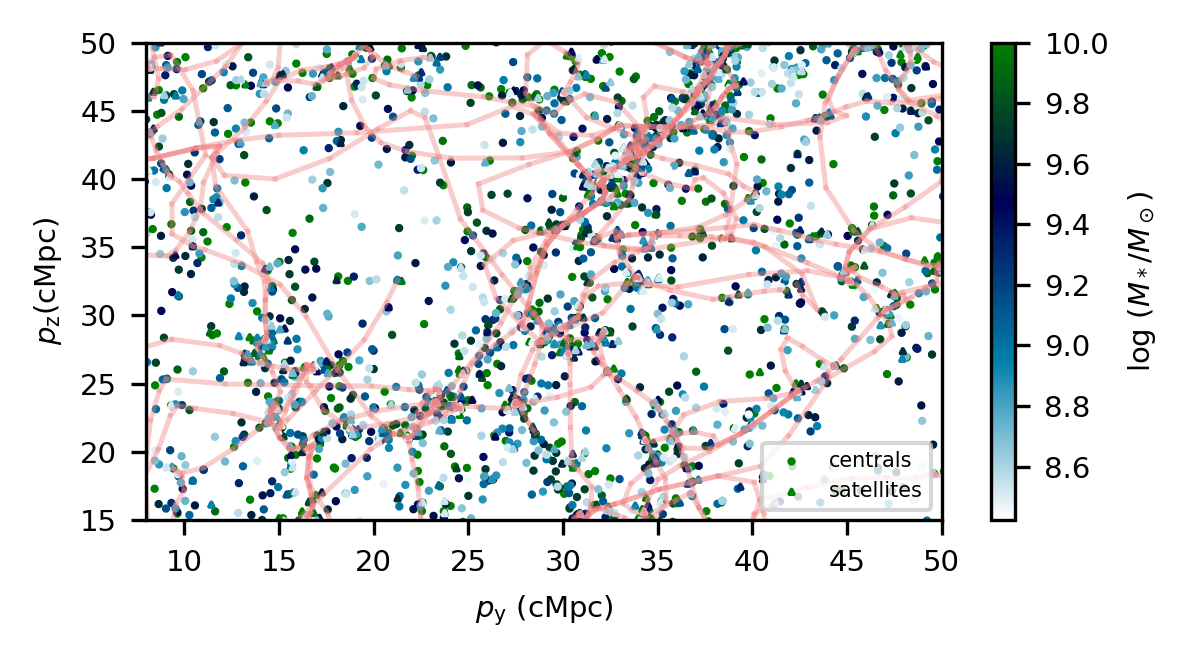}
    \caption{A 2D view of a $50 \, \hmpc$ \simba simulation slice at redshift $z$ = 0, with a slide width of $10  \, \hmpc$ . The galaxies are represented by circles (centrals) and triangles (satellites), colour coded by their total stellar mass. The filaments extracted from \disperse are plotted in pink. The figure provides a qualitative representation of the galaxies' distribution within the cosmic web, showing that the overdensities (filaments and nodes) show predominantly massive centrals and accompanying low-mass satellites.}
    \label{disperse}
\end{figure*}

To identify filaments of the cosmic web, we use the publicly available code \disperse (\citealt{Sousbie2011,SousbiePK2011}), using the Discrete Morse theory and the theory of persistence. \disperse measures the gradient of the density field via Delaunay Tessellation (e.g. \citealt{Schaap2000}) to identify the critical points, defined as the points where the gradient of the density field is null. Filaments are then constructed as segments connecting a maximum to a saddle point, representing the ridges of the Delaunay density field.

We applied \disperse to the distribution of galaxies in \simba, adopting a 3$\sigma$ persistence threshold in order to remove the filaments affected by the Poisson noise of the density distribution. As explained in e.g. \cite{Codis2018} and \cite{Kraljic2020}, we also noted that a higher threshold would result in more robust structures with a significant drop in the number of filaments generated, hence the 3$\sigma$ value adopted throughout this work represents an optimal choice. Additionally, we also applied a smoothing to the positions of the filaments' edges, by averaging their positions with those of the edges of contiguous segments. This gives rise to a smoother filamentary skeleton by reducing unphysical filamentary shapes.  These procedures generally mimic previous works that applied \disperse to simulations.

For better visualisation, Fig.~\ref{disperse} shows the filaments extracted with \disperse and a 2D view of a \simba simulation at redshift $z~=~0$, using a slice from the $50 \, \hmpc$ box with a width of $10 \, \hmpc$. The galaxies are overplotted (centrals as circles and satellites as triangles), colour-coded by their total stellar masses. This plot provides a qualitative image that the \disperse skeleton generally traces out the large-scale structure that one picks out by eye.  One can see further that the most massive galaxies tend to lie closer to filaments, together with their low-mass companions/satellites; this will be quantified in \S\ref{sect:satellites_and_centrals} and 
\S\ref{sect:masses}. 

\subsection{The galaxy sample}
\label{sect:satellites_and_centrals}
\begin{figure*}
    \centering
    \includegraphics[width=0.95\textwidth, height=8cm]{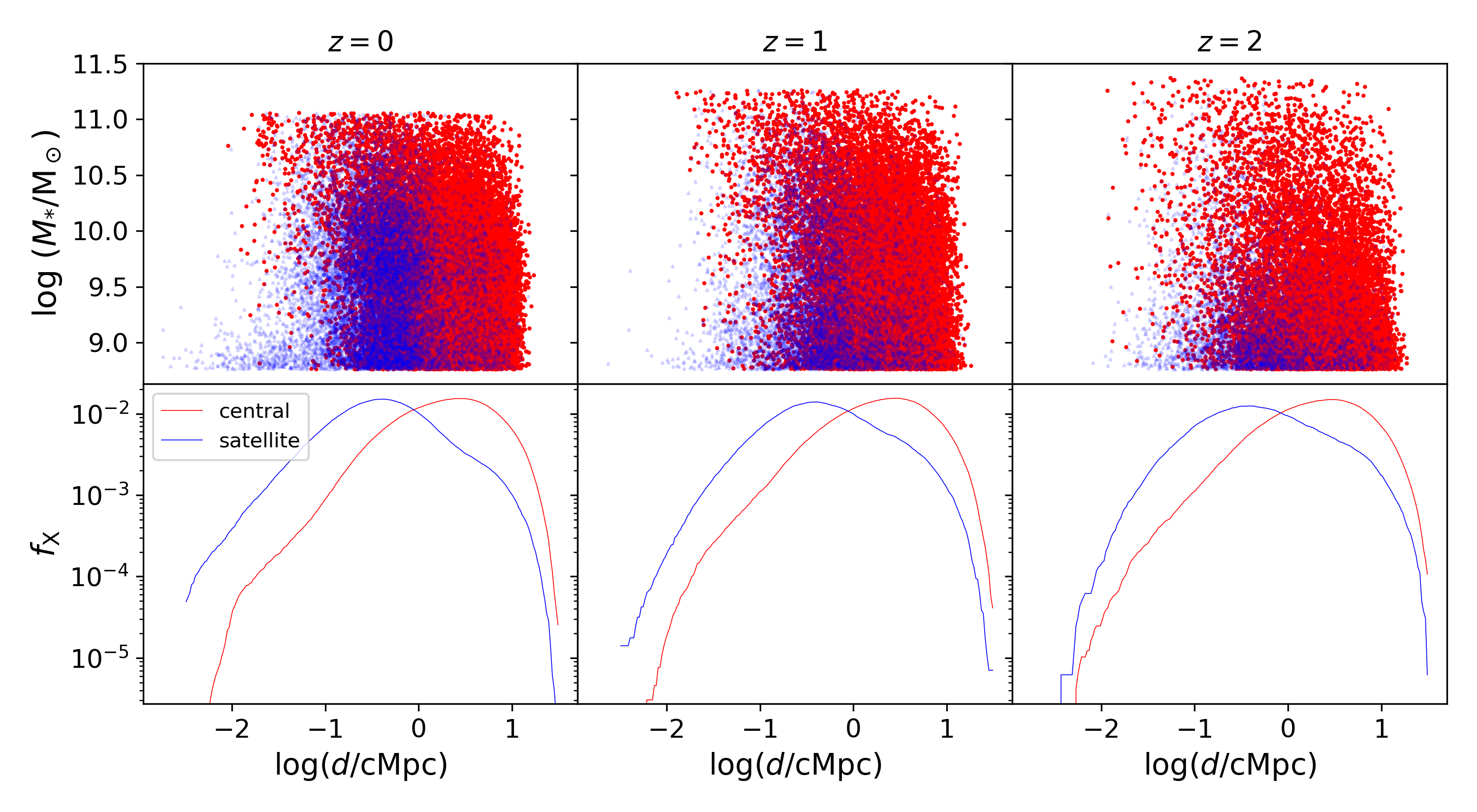}
    \caption{Upper panels show the stellar mass of satellites (blue dots) and centrals (red dots) versus the distance to their corresponding closest filament in comoving Mpc, $d$ at redshifts $z=0,1,2$.  Lower panels likewise present the probability distribution function  of satellites (blue line) and centrals (red line) vs. $d$. Within 1~cMpc there are fewer low-mass centrals and more satellites. The upper envelope visible especially at $z=0$ results from our halo mass threshold of $M_h<10^{13}M_\odot$. This figure strenghtens the qualitative findigs from Fig.~\ref{disperse}, but also shows that the largest fraction of satellites and centrals are located at $\sim$ 1 and 10 cMpc of a filament respectively.}
    \label{cen_sat_loc}
\end{figure*}

For the purpose of this work, we quantify the galaxies' positions in the cosmic web via the distance to the closest filament ($d$). \textsc{DisPerSE} reports the 3D positions of the filaments' edges, which we used to compute the minimum distance between each galaxy and the corresponding closest filament midpoint. This is slightly different than the perpendicular closest distance, but negligibly so as pointed out in e.g. \cite{madalina}, since each \disperse filament is actually comprised of a large number of small segments.

As mentioned earlier, we are specifically interested in trends versus filamentary environment within the field galaxy population. To this end, we remove all galaxies in halos with a virial mass 
above $10^{13} M_\odot$, corresponding to removing all galaxies in structures of poor group size and larger, and only report all statistics based on the remaining sample of galaxies. \tb{After doing so, we end up with 27947 centrals and 13997 satellites at $z = 0$; 18746 centrals and 8108 satellites at $z =1$, 13919 centrals and 4609 satellites at $z=2$}. This does not preclude some effect on galaxies from being in the vicinity of large halos, so we are still including any effects associated with pre-processing outside of large halos.  Nonetheless, such large halos are fairly rare, and as one can see in Fig.~\ref{disperse}, much of the filamentary structure is located in regions far from the most massive nodes.

We will further consider the impact of the environment on central and satellite galaxies separately.  Centrals are taken to be the most massive galaxy within its halo, and with few exceptions typically lie within the inner 10-20\% of their halo.  Satellites can be impacted by distinct physical processes such as ram pressure and tidal stripping, and although such processes have traditionally been associated with group and cluster environments, it is possible that the denser and hotter environment around filaments could also have an impact.  

Figure~\ref{cen_sat_loc} shows the stellar mass distribution of satellites (blue dots) and central galaxies (red dots), with respect to the distance to the closest filament, for redshifts $z=0,1,2$ (upper left to right panels). Corresponding lower panels show the probability distribution of satellites (blue curve) and centrals (red). The upper mass threshold, visible most noticeably in the $z=0$ upper left plot, comes from the aforementioned halo mass cut at $M_h<10^{13}M_\odot$. 

Overall, central galaxies skew to be more massive towards the filament spine. At short distances, only small satellites remain.  The immediate proximity of filaments ($d \lesssim 0.1$~cMpc) shows high-mass centrals accompanied by their respective low-mass satellites in the low-redshift Universe, while this distribution appears to be more spread out at redshift $z = 2$, indicating a dynamical evolution of the satellite population. At all redshifts, the satellites distribution peaks at $<1$~cMpc from the closest filament, while for centrals the corresponding peak is at a few cMpc.  This reflects the trend of the halo occupation distribution with $M_*$, since $M_*$ is well correlated with $M_h$~\citep[e.g.][]{Cui2021} and the halo occupancy rises quickly towards higher $M_h$. These galaxies represent the sample that we will use for the majority of the analysis that we discuss next.

\section{Galaxy properties in the cosmic web}

\label{Sect.prop_gen}
In this section, we correlate key global galaxy properties with their proximity to filaments as measured by $d$. We consider each property in turn: $M_*$, sSFR, $Z_*$, $f_{HI}$, and $f_{H2}$, where the latter two are the gas mass fractions in each phase with respect to stellar mass.  We also consider quenched fractions $f_Q$ and elliptical fractions $f_e$.  We present results for $z=0,1,2$, separated into centrals and satellites.

\subsection{Stellar mass}
\label{sect:masses}
\begin{figure*}
    \includegraphics[width=0.92\textwidth]{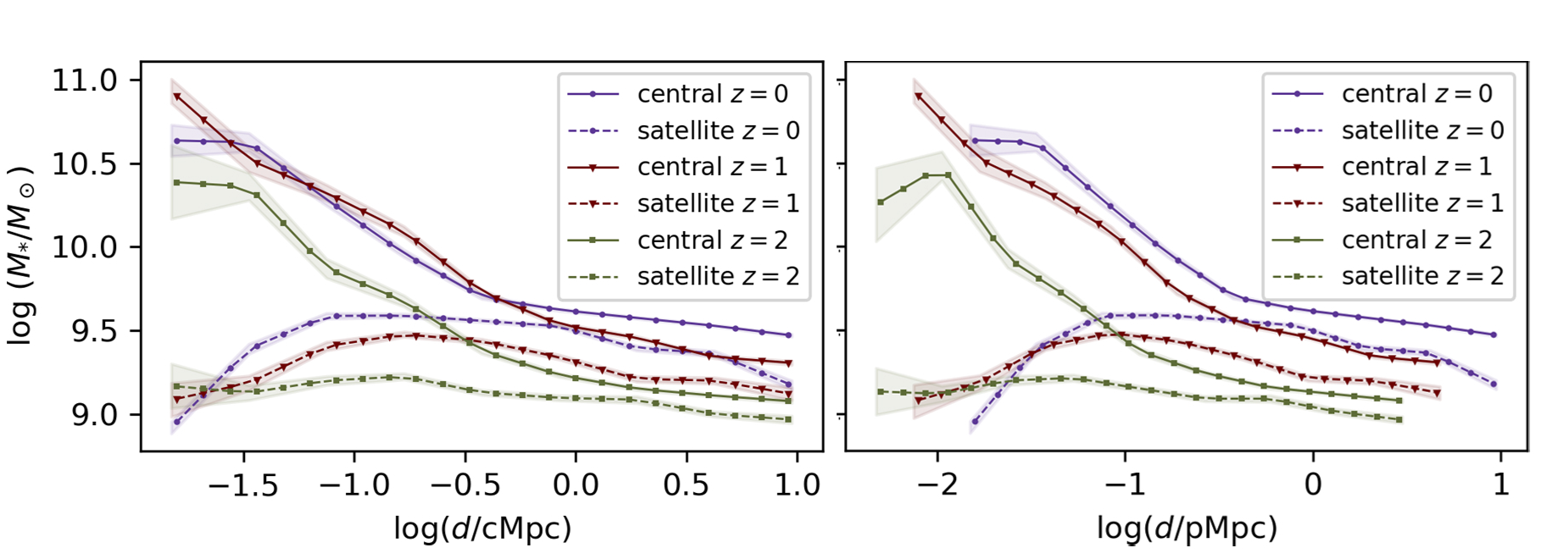}
    \caption{Distance from filaments dependence of stellar masses for centrals (continuous lines) and satellites (dashed lines) for three redshifts: $z$ = 0 (purple), $z$ = 1 (dark red), $z$ = 2 (green). The lines were obtained by binning the results in terms of distance and interpolating the corresponding medians (i.e. running medians). The shaded regions represent the corresponding standard errors in each bin. In agreement to our qualitative findings from \S\ref{sect:satellites_and_centrals}, most massive centrals lie within short distances from the filaments and host a large number of low-mass satellites. \tb{We show the distance $d$ in both cMpc (left panel) and pMpc (right panel), finding that the overall trends maintain regardless of the units of $d$.}}
    \label{M_cs_dist}
\end{figure*}

The galaxy stellar masses dependence on $d$ is shown in Fig.~\ref{M_cs_dist}, showing the medians of the binned values for redshifts $z = 0$ (purple), $z =1$ (maroon), and $z = 2$ (green). Solid and dashed lines show centrals and satellites, respectively.  The shaded regions represent the standard deviation on the mean in each bin, illustrating higher uncertainties in the filaments' proximity owing primarily to the lower number of galaxies in this region. \tb{We present the distance $d$ in both comoving Mpc (left panel) and physical Mpc (right panel). We also compared the two choices for $d$ for the other properties (presented in the following sections) and found that the trends are independent of this choice. However, the former representation offers a more visually comprehensive description of the trends and will be retained throughout this section.} 

At all three redshifts considered, the masses of the central galaxies decrease with increasing distance from the closest filament. Despite our finding from the previous section suggesting that at $z = 2$ galaxies adopt a broad range of masses in the filaments proximity, Fig.~\ref{M_cs_dist} shows that the median values of these masses are in fact higher for galaxies closer to the filaments.  This finding can be explained via the environmental effects on the halo mass function, which predicts more massive halos to lie closer to filaments, which then leads to higher mass galaxies in these regions (e.g. \citealt{john}). This highlights that we must be careful when interpreting trends with distance from filaments to ensure that they are not simply reflecting trends with stellar mass, since one of the crucial aspects of this study is to disentangle the effects of mass and environment on galaxy properties. 

The satellites curiously show a reverse trend for short distances (i.e. within log($d$/cMpc) $\lesssim$ -1) at redshifts $z = 0$ and $z = 1$, as their masses increase with distance. For larger distances, they also adopt a subtle overall decreasing trend, considerably weaker than for centrals. No clear trend is observed for satellites at redshift $z = 2$. 

In agreement with our qualitative findings from \S\ref{sect:satellites_and_centrals}, it appears that most massive central galaxies lie within short distances from the filaments and host a large number of low-mass satellites. This finding also supports the predictions of the environmental effects on the halo mass function, as the massive systems in the filaments to proximity are more likely to accrete more low-mass satellites.  The disappearance of more massive satellites very close to filaments could further owe to dynamical effects such as accelerated dynamical friction and tidal stripping in dense environs; we defer a detailed analysis of this to future works.  For now, we note this as an interesting prediction from \simba.

\subsection{(Specific) star formation rate}
\label{sect:ssfr}
\begin{figure}
    \includegraphics[width=0.5\textwidth]{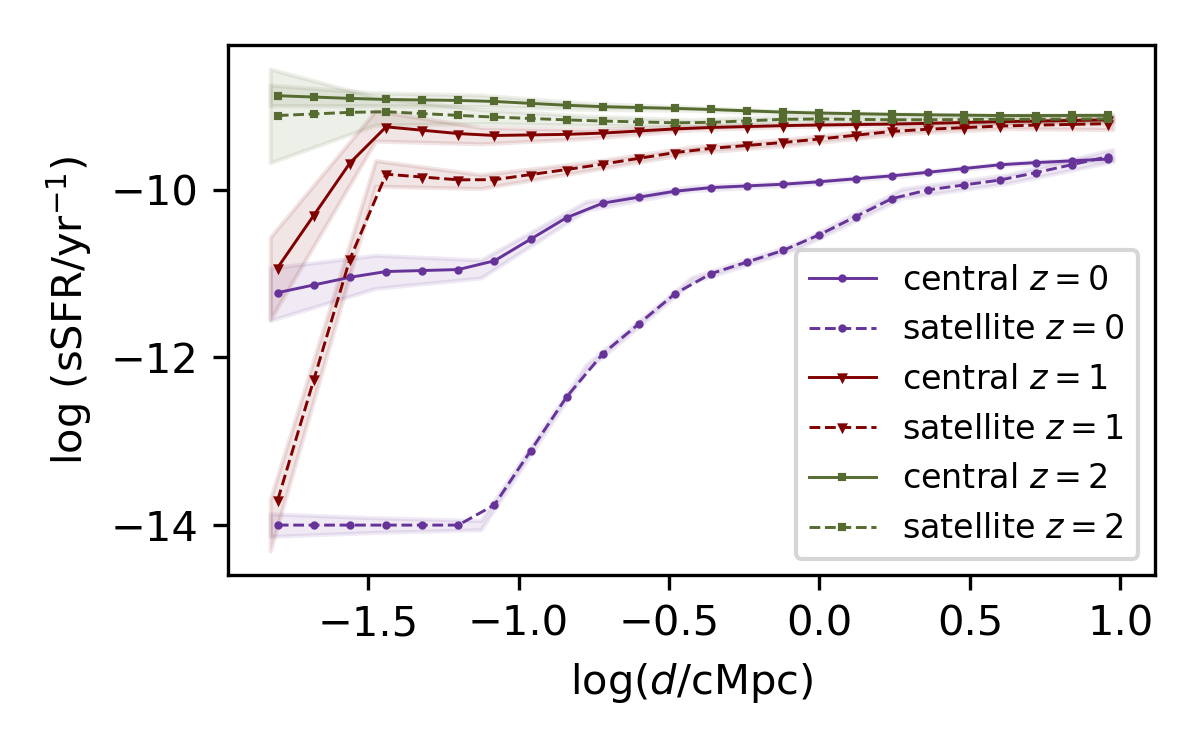}
    \vskip-0.2in
    \caption{Distance from filaments dependence of specific star formation rates for centrals (continuous lines) and satellites (dashed lines) for three redshifts: $z$ = 0 (purple), $z$ = 1 (dark red), $z$ = 2 (green). The lines were obtained by binning the results in terms of distance and interpolating the corresponding medians (i.e. running medians). The shaded regions represent the corresponding standard errors in each bin.  Galaxies close to filaments show suppressed levels of star formation, with a majority of satellites in the filaments' proximity at redshift $z = 0$ being fully quenched.}
    \label{fig.ssfr_d_all}
\end{figure}

Galaxies near filaments could be enhanced in star formation relative to those far away since there is more gas in the vicinity, or they could be suppressed because the filaments are heated which can suppress accretion.  Hence SFR provides a key barometer for the interplay between galaxy growth and filamentary large-scale structure.  To mitigate the fact that galaxies closer to filaments have larger $M_*$ as found in the previous section, we consider the specific SFR, although the broad results are similar if we consider the SFR itself.

Figure~\ref{fig.ssfr_d_all} presents the dependence of the specific star formation rate (sSFR) with respect to distance from filaments at $z=0,1,2$.  The line styles and colours mimic Fig.~\ref{M_cs_dist}. For all non-star forming galaxies (SFR=0), we set log(sSFR)$=-14$.  Since we are considering medians, the analysis is not sensitive to the exact value as long as it is very low.

As seen in Fig.~\ref{fig.ssfr_d_all}, central galaxies clearly show a reduction in sSFR at close proximity to filaments at $z=1$ and $0$ relative to galaxies far away from filaments.  At $z=1$, the dip is only seen at very small distances ($\la 30$ kpc), but by $z=0$ the sSFR is suppressed farther out to $\sim 100$~kpc.  At $z=2$, if anything there is a rise in the sSFR towards filaments, perhaps indicating a reversal in the star formation--density relation; we will explore this further in future work.  Far away from filaments, there is an overall reduction in sSFR that reflects the global decline in cosmic star formation over time~\citep[e.g.][]{Daddi2007,Dave2008} primarily due to falling accretion rates~\citep{Dekel2009}.  Hence the growth of centrals via star formation is clearly retarded even in filamentary regions, indicating evidence for pre-processing of galaxies before falling into galaxy groups or clusters.

Examining the satellites, at $z=0$ the satellites' median sSFR vanishes within 100~kpc of filament, indicating that more than half the satellites are fully quenched. Then the sSFR shows a rapid increase with increasing distance, converging subsequently to the same value as centrals by $\sim 10$~Mpc. At redshift $z=1$, the sSFR for satellites follows that of centrals but shows somewhat more suppression close to filaments, while becoming similar to centrals at $\ga 2$~Mpc.  Meanwhile, at $z=2$, satellites show only a very mild suppression with respect to centrals.  These trends indicate that satellites are more impacted by filamentary environments than centrals, albeit in a qualitatively similar way, and the additional impact grows rapidly from $\sim 1\to 0$.

To sum up, our results show that galaxies close to filaments are suppressed in stellar growth rates out to $z\ga1$, with a majority of satellites in the filaments' proximity at redshift $z = 0$ being fully quenched.  Within \simba, it is seen that galaxy circum-galactic and even intergalactic media are strongly impacted by AGN jet feedback~\citep{Appleby2021,Sorini2022}, which thereby grows the quenched galaxy population~\citep{Dave2019}.  Our results here suggest that AGN feedback also may be a contributor to suppressing SFRs, particularly in satellites close to filaments, although it could also owe to increased shock heating owing to the growth of large-scale structure. \tb{We will investigate further in \S\ref{feedback} the effects of AGN feedback compared to the effects of filaments solely. We also note that the interplay between filaments and AGN feedback is a non-trivial topic, filaments potentially influencing the AGN feedback. There are works suggesting that dense environments determine a deficit of the cold gas reservoir available to trigger the AGN activity~\citep[e.g.][]{Jaffe2015,Beyoro-Amado2021}, while others suggest that the ram pressure stripping acts as a triggering mechanism for AGNs~\citep[e.g.][]{Poggianti2017,Kouloridis2024}. This topic is above the scope of this study and we aim to investigate it in future work.} In any case, \tb{our} results clearly demonstrate that both centrals and satellites undergo pre-processing outside of galaxy groups.


\subsection {Gas content}
\label{sect.fH12}
\begin{figure}
    \includegraphics[width=0.5\textwidth]{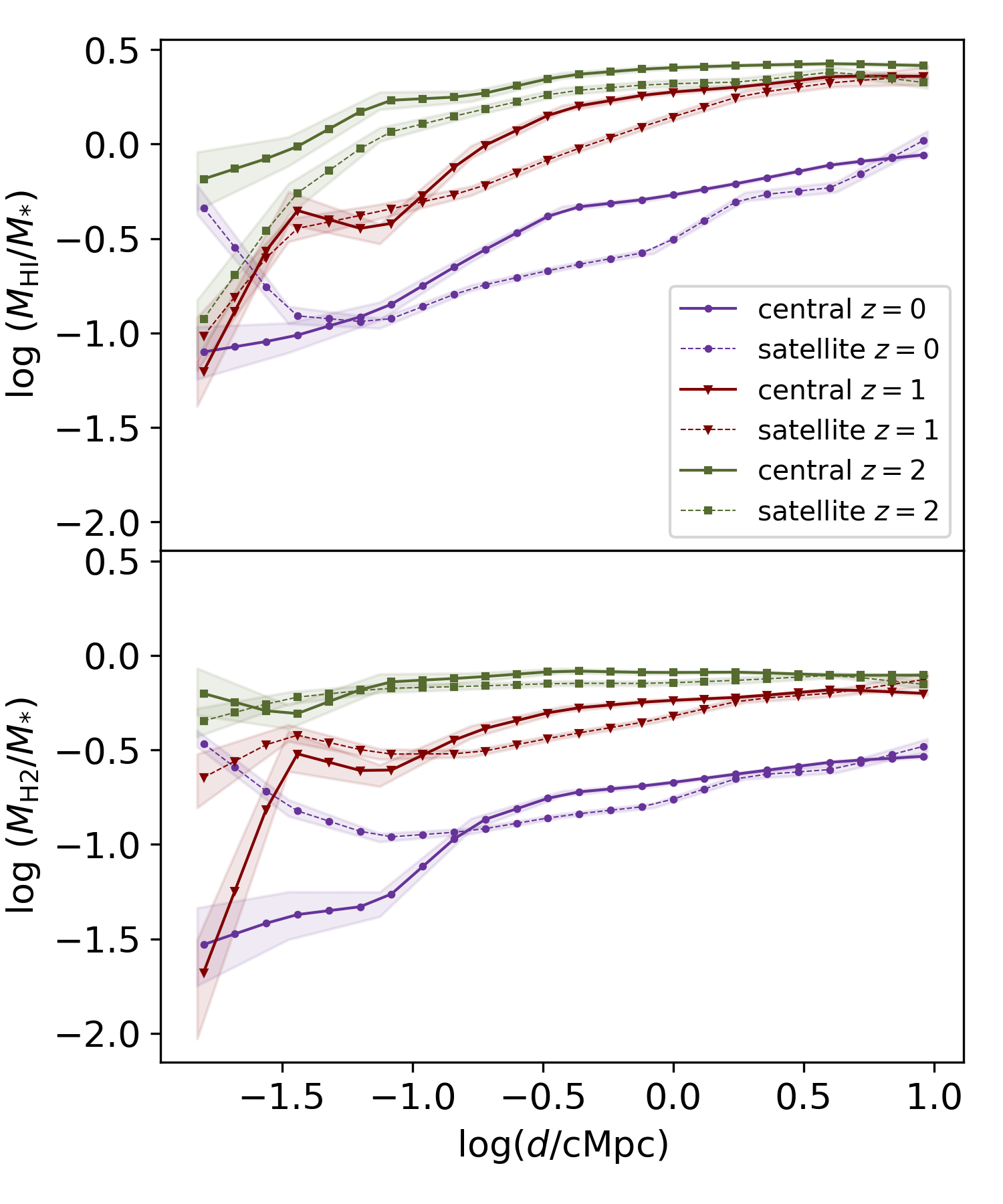}
    \vskip-0.2in
    \caption{Distance from filaments dependence of \HI (\textit{left}) and \Hmol (\textit{right}) content for centrals (continuous lines) and satellites (dashed lines) for three redshifts: $z$ = 0 (purple), $z$ = 1 (dark red), $z$ = 2 (green). The lines were obtained by binning the results in terms of distance and interpolating the corresponding medians (i.e. running medians). The shaded regions represent the corresponding standard errors in each bin. The \HI fraction increases with increasing distance from filaments at all three redshifts considered.  \fmol shows similar trends at redshifts $z = 0$ and $z = 1$, while (similar to sSFR, see Fig.~\ref{fig.ssfr_d_all}) showing no clear trend for \fmol versus distance from filament at redshift $z = 2$.}
    \label{fig.fH12_d}
\end{figure}

Given that the sSFR is suppressed towards filaments, one would expect that the gas contents that fuel star formation would also be lowered.  For molecular hydrogen, \simba directly ties the $H_2$ content to star formation, but for \HI, this provides a reservoir on larger scales that could be more influenced by environmental effects.  Hence it is interesting to examine the molecular and atomic contents of galaxies as a function of distance to filaments.  Again, to mitigate the overall trend that the gas contents increase with mass (at least among star-forming systems), we consider the atomic and molecular Hydrogen fractions $f_{\mathrm{HI}} \equiv M_\mathrm{{HI}}/M_*$ and $f_{\mathrm{H2}} \equiv M_\mathrm{{H2}}/M_*$.

Fig.~\ref{fig.fH12_d} shows \fHI (top) and \fmol (bottom) as a function of filamentary distance, using the same colour and line type scheme as the previous plots.  For central galaxies (solid lines), the trends broadly mimic that for the sSFR: galaxies have suppressed gas contents close to filaments, with that suppression increasing in strength and extent towards lower redshifts.  In detail, \Hmol traces sSFR more faithfully, while \HI even shows suppression near filaments even at $z=2$, and a larger range of suppression.

The satellites however behave substantially differently from sSFR.  At $z=2$, there is little difference in \fmol between centrals and satellites, but for \fHI the satellites show an increased suppression near filaments.  However, these trends tend to reverse near filaments at lower redshifts: the satellite gas fractions are generally comparable to (at $z=1$) or even higher ($z=0$) near filaments versus the centrals. This is an odd turn, which may have to do with the way that the gas fractions are computed by including all gas within halos that is most bound to a given galaxy.  In galaxy groups, it has been observed that \HI can be present throughout the group environment~\citep[e.g.][]{Lucero2015}, and in our \HI assignment scheme that gas may have been associated with group satellites.  If the same effect is happening in the densest regions of filaments, this \HI may in detail not be associated with individual galaxies, but rather the overall environment. It is less easy to understand the upturn in \fmol at a low distance. This could partially be explained by the HI association for satellites, but in order to disentangle these effects and make proper predictions for comparison to data we would need to create mock data cubes of these systems~\citep[as in][]{Glowacki2021}, which is beyond the scope here. 

Overall, the galaxies do not seem to show as much suppression in gas content as in sSFR (more visible for satellites). Since sSFR can be decomposed into \fmol times the star formation efficiency (SFE $\equiv M_{H2}/M_*$), this suggests that there are variations in the efficiency of converting gas into stars, with the SFE generally being lower closer to filaments. 


In summary, our results show that the \HI fraction increases with increasing distance from filaments at all three redshifts considered, with satellites at redshift $z = 0$ showing an initial decrease in gas fractions near filaments.  \fmol shows similar trends at redshifts $z = 0$ and $z = 1$, while (like sSFR) showing no clear trend for \fmol versus distance from filament at redshift $z = 2$.  The satellites show a curious increase in gas fractions at $z\leq1$ very close to filaments, which may owe to analysis methodology, and highlights the difficulty of unambiguously assigning particularly \HI to galaxies in dense environments.


\subsection {Metallicity}
\label{sect:metallicity}
\begin{figure}
    \includegraphics[width=0.5\textwidth]{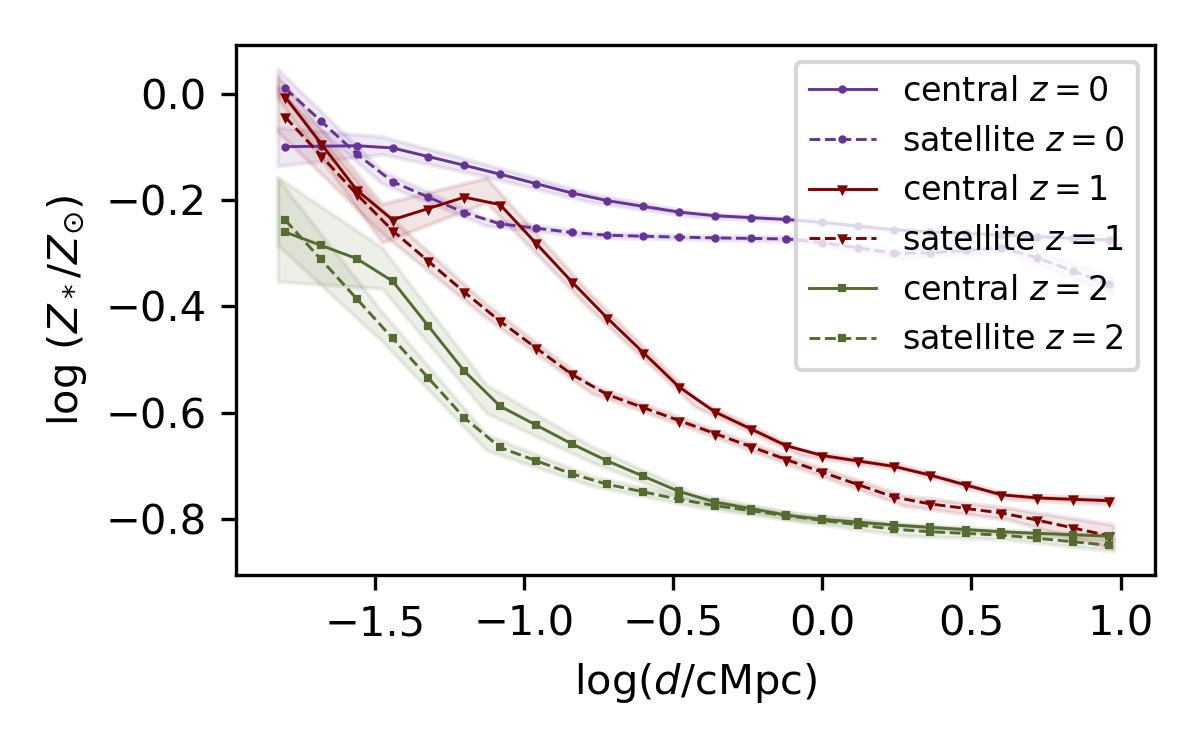}
    \caption{Distance from filaments dependence of stellar metallicities for centrals (continuous lines) and satellites (dashed lines) for three redshifts: $z$ = 0 (purple), $z$ = 1 (dark red), $z$ = 2 (green). The lines were obtained by binning the results in terms of distance and interpolating the corresponding medians (i.e. running medians). The shaded regions represent the corresponding standard errors in each bin.  Galaxies close to filaments are more metal-enriched, the centrals and satellites showing similar behaviours.}
    \label{fig.Zs_d_all}
\end{figure}

Another key global galaxy property is its metallicity.  Galaxies are known to have a strong correlation of $M_*$ with $Z$ known as the mass-metallicity relation~\citep[e.g.][]{MZR,Tremonti2004,MZR_review}.  This is believed to be set by a competition between pristine inflow diluting metallicity, star formation enhancing metallicity, outflows removing metals, and the re-accretion of outflows providing an additional source of metals~\citep[e.g.][]{FinlatorDave2008,Dave2012}.  Particularly the latter effect could be impacted by environment, because the gas in denser regions may be more enriched from previous star-formation activity and may retard outflows leading to more re-accretion. For satellites in denser regions, one expects there to be less pristine inflow and more enriched recycling, leading to higher metallicities; indeed this is found in simulations~\citep[e.g.][]{Dave2011b}. These trends are for the gas-phase metallicity, but models generally predict the stellar metallicity traces this reasonably well.  Here, because we want to compare across both gas-rich and gas-poor galaxies, we will employ the stellar metallicity ($Z_*$) versus filamentary distance since this can be computed for all galaxies.


Fig.~\ref{fig.Zs_d_all} shows the median $Z_*(d)$, using the same colour and line scheme as in previous plots. In all cases, the metallicities are overall higher in the filaments proximity. Given the MZR, or even the Fundamental Metallicity Relation (FMR, \citealt{mannucci_et_al_2010}), our results can be explained via the higher mass and low SFR galaxies present in the filaments' proximity.

The centrals show an abrupt drop in metallicity when moving away from filaments, and the differences between centrals and satellites, in this case, are less obvious compared to the previous quantities (\S\ref{sect:ssfr}).  In general, the satellites have slightly lower metallicity, but this is likely most explained by their lower $M_*$.  However, when comparing to Fig.~\ref{M_cs_dist} and considering the MZR, this similarity between centrals and satellites is actually somewhat surprising.  This is because the satellites' median $M_*$'s are quite flat with distance, yet their metallicity continues to increase strongly towards filaments just like the centrals. This indicates that there is a strong effect from the suppression of SFR (i.e. the FMR), and perhaps also an effect from the environment.

Additionally, the trends observed appear to be weaker at low redshift. A possible explanation for this might be the steeper mass-metallicity correlation observed in \simba at high redshift, $z \sim 2$ (see \citealt{Dave2019}).  However, the change is quite dramatic from $z=1\to 0$, with galaxies far from filaments being much more enriched at late times.  This suggests another effect, such as wind recycling of enriched materials even far from filaments, being more important since $z\sim 1$.  We plan to investigate the detailed evolution of metallicity as a function of environment in future work.

In short, our results show that galaxies close to filaments are more metal-enriched, the centrals and satellites showing similar behaviours.  The trend with the satellites, combined with the lack of a trend in $M_*$ in contrast with the dramatic evolution in satellites' SFR, suggests that the FMR is an important driver in setting the environmental trends. The trend substantially weakens by $z=0$, owing to some complex interplay between the environment and the various physical processes governing galaxy metallicities. 


\subsection{Quenching and morphology}
\label{q&e}
\begin{figure}
    \includegraphics[width=0.5\textwidth]{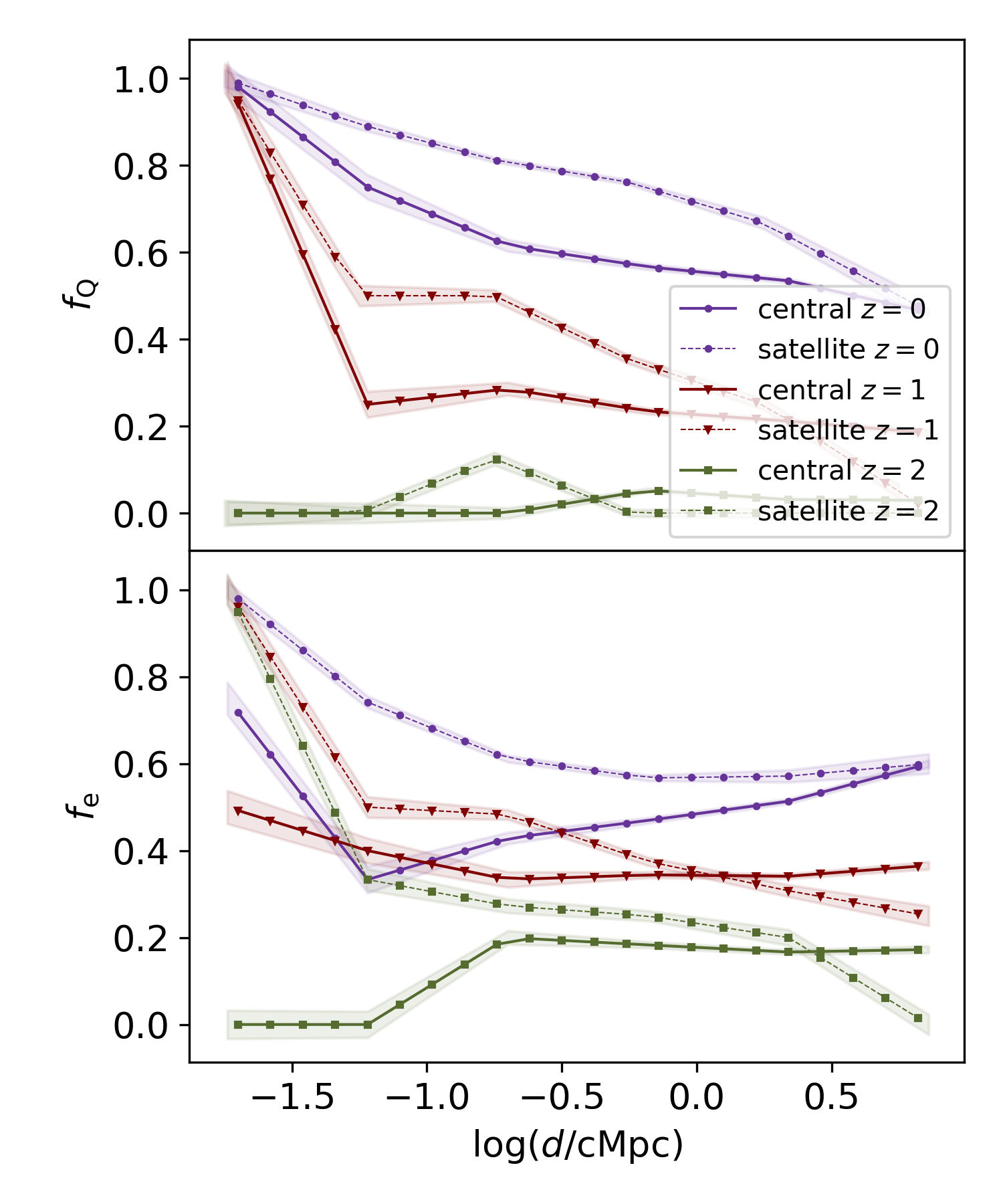}
    \caption{Distance from filaments dependence of quenched fractions (\textit{left}) and elliptical fractions (\textit{right}) for centrals (continuous lines) and satellites (dashed lines) for three redshifts: $z = 0$ (purple), $z = 1$ (dark red), $z = 2$ (green). The lines were obtained by binning the results in terms of distance and interpolating the corresponding medians (i.e. running medians). Galaxies close to filaments are more elliptical and quenched at low redshift, the trends being more prominent for satellites}
    \label{fig.fQe}
\end{figure}

We have seen that galaxies, and satellites in particular, have lower sSFRs near filaments at $z\la 1$.  Another way to quantify this is using the quenched fraction $f_\mathrm{Q}$, which we compute via the \cite{williams} UVJ diagram. Quenching is also well-known to be correlated with morphology. Hence we also examine the elliptical fraction $f_\mathrm{e}$, which we define using the fraction of kinetic energy in rotation \citep{sales2012} since \citet{Kraljic2020} found that this was the most well-correlated measure reflecting visual morphology in \simba. We chose a threshold of 0.3, meaning that galaxies with a smaller fraction of kinetic energy in rotation are considered elliptical, but varying this from 0.25 or 0.35 does not affect the results significantly.

Fig.~\ref{fig.fQe} shows the galaxies' quenched fraction ($f_\mathrm{Q}$, top panel) and elliptical fraction ($f_\mathrm{e}$, bottom) as a function of distance from filament, using the same scheme as in previous plots.  We note that the distance range adopted for this analysis is smaller than in the previous cases. This is because at large enough distances from filaments (log($d$/cMpc) $\gtrsim$ 0.5) the corresponding behaviours are reasonably well converged. 

As expected from \S\ref{sect:ssfr}, both centrals and satellites are more quenched close to filaments. A decreasing trend is not visible at redshift $z=2$, but becomes prominent at $z\leq 1$.  The quenched fraction increases with time, which follows the trend seen in the overall galaxy population in \simba, as well as in observations.

The trend is similar for both centrals and satellites, but the satellites have a higher overall $f_\mathrm{Q}$.  Within $\la 100$~ckpc of filaments at $z=0$, $\sim~90\%$ of satellites are quenched, as opposed to $\sim 70\%$ of centrals.  This shows again the important role that the environment plays in quenched satellite galaxies, which is already prominent around filamentary structures.  At $z=2$ however, little difference is seen between the central and satellite quenched fractions.  Hence the environmental effects are restricted to lower redshifts, where large-scale structure causes more shock heating and AGN feedback provides significant energy into the IGM in \simba.  These findings echo our results from \S\ref{sect:ssfr}, quantified in a different way that provides another testable prediction from \simba.


In general, $f_\mathrm{e}$ increases with time, and the satellites' $f_\mathrm{e}$ are higher than that of the centrals, with the difference between them also growing with time.  However, the elliptical fraction as defined here shows overall less evolution than the quenched fractions, and particularly at $z\leq 1$ the elliptical fractions for centrals very close to filaments show no evolution.  At large distances, centrals and satellites show similar elliptical fractions.  In general, it is more difficult to compare our elliptical fractions defined via $\kappa_{\rm rot}$ to observations since this is not so easy to measure, but this at least qualitatively demonstrates that in general \simba galaxies become less rotationally supported with time and proximity to filament.


To sum up, our results show that galaxies close to filaments are more elliptical and quenched, the trends being more prominent for satellites. For both centrals and satellites, the quenched fraction are similar at redshift $z = 2$, although the elliptical fractions are still a bit higher for satellites at $z = 2$. The higher elliptical fraction suggests galaxies that are likely to be more massive, passive, and metal-enriched, with lower gas content, tend to have less rotational support, concordant with the results noticed in previous sections (\S\ref{sect:masses}--\S\ref{sect:metallicity}).  The fact that $f_\mathrm{e}$ and $f_\mathrm{Q}$ do not mimic each other exactly even qualitatively in their trends with $d$ and redshift suggests that quenching and morphological transformation are not happening in exactly the same galaxies at the same time within \simba.  However, the crudeness of the morphological measure begs further investigation to disentangle the relation between quenching and morphology, perhaps requiring higher resolution simulations capable of resolving the scale height of typical disks.


\section{Deviations from the scaling relations}
\label{sect:dev}

We have seen that proximity to \disperse-identified filaments has the effect of lowering star formation and gas content, raising the metallicity, and increasing the elliptical fraction of galaxies at $z\la 1$, effects that are enhanced amongst satellite galaxies.  However, these effects with $d$ are qualitatively degenerate with stellar mass -- that is, galaxies with higher mass which tend to be found closer to filaments also share these general trends.  Thus it is important to examine whether the effects are truly due to the location in the cosmic web.

To do this, as mentioned earlier, in this section we investigate many of the same quantities and their trends with $d$, but now at fixed stellar mass $M_*$. For this purpose, we compute the deviations of the quantities of interest from their scaling relations with stellar mass, specifically the star formation main sequence (SFMS), mass metallicity relation (MZR), and the \HI and \Hmol fractions relations with $M_*$.  By comparing these to the corresponding ones in the previous section, we can see how much of the effect owes to its stellar mass dependence and how much owes to the impact of filamentary large-scale structure.

\subsection{Deviation from star-forming main sequence}
\label{sect:SFMS}
\begin{figure}
    \includegraphics[width=0.5\textwidth]{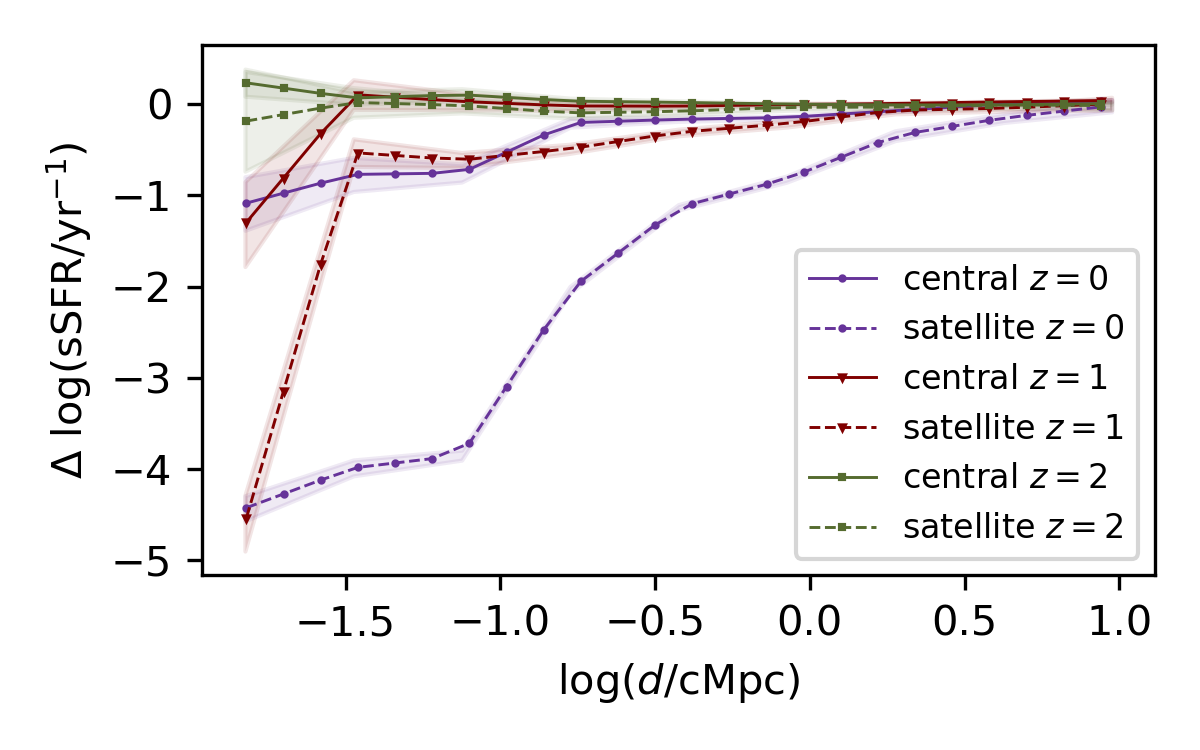}
    \caption{Distance from filaments dependence of the specific star formation rates' deviations from the star formation main sequence for centrals (continuous lines) and satellites (dashed lines) for three redshifts: $z$ = 0 (purple), $z$ = 1 (dark red), $z$ = 2 (green). The lines were obtained by binning the results in terms of distance and interpolating the corresponding medians (i.e. running medians). The shaded regions represent the corresponding standard errors in each bin. The centrals show an increasing suppression of star formation activity with time very close to filament centres, with no strong trend at $z=2$ while by $z=0$ the typical galaxy at the centre of a filament is quenched. The satellites likewise show a strong trend with redshift, with suppression of star formation activity extending to quite large distances by $z=0$ (see also Fig.~\ref{fig.ssfr_d_all})}.
    \label{fig.deltassfr}
\end{figure}

The (specific) star formation rate is expected to have a clear dependence on mass via the SFMS (see e.g. \citealt{SFR}), which, as mentioned in \S\ref{Sect:Simba}, is reasonably reproduced in \simba. We obtain the SFMS by fitting the medians of sSFR in bins of $M_*$ for all star-forming galaxies defined as log(sSFR/yr$^{-1}) > -10.8 + 0.3z$, same as in \cite{Dave2019}. We have kept our definition of star-forming galaxies simple despite there being more sophisticated ways to define the SFMS~\cite[see e.g.][]{HahnStarkenburg2019} in order to be more straightforwardly comparable with observations in the future.  For instance, changing the $z = 0$ threshold from $-10.5 \to -11$ has little effect on the results, since we are only concerned with the relative deviation of galaxies near filaments versus those of the overall galaxy population.

Fig.~\ref{fig.deltassfr} shows the deviation from the SFMS ($\Delta\log$~sSFR) as a function of $d$ at $z=0,1,2$, for centrals (solid) and satellites (dashed).  This can be compared to Fig.~\ref{fig.ssfr_d_all}, which shows the corresponding plot for sSFR itself.

Overall, the trends look qualitatively, and for the most part quantitatively, similar to those for sSFR: the centrals show an increasing suppression of star formation activity with time very close to filament centres, with no strong trend at $z=2$ while by $z=0$ the typical galaxy at the centre of a filament is quenched. The satellites likewise show a strong trend with redshift, with suppression of star formation activity extending to quite large distances by $z=0$.  This indicates that the effects in sSFR seen previously do not owe primarily to any mass dependence in galaxies as a function of $d$.

In detail, the trends in $\Delta\log$~sSFR are slightly weaker than those seen in sSFR.  For instance, at $z=0$, the central galaxies close to filaments lie $\sim1$~dex below those at $z=2$ in $\Delta\log$~sSFR, while in Fig.~\ref{fig.ssfr_d_all} the difference is closer to $\sim 2$~dex.  But much of that difference is explained by the fact that sSFR's at $z=2$ are generally higher than at $z=0$.  We conclude that the trends in $M_*$ as a function of $d$ are not an important factor in establishing the suppression of star formation activity near filaments and that such effects genuinely owe to environmental effects from the cosmic web.


\subsection{Distance from \HI and \Hmol mean relations}
\label{sect:H1/2MS}
 \begin{figure}
    \includegraphics[width=0.5\textwidth]{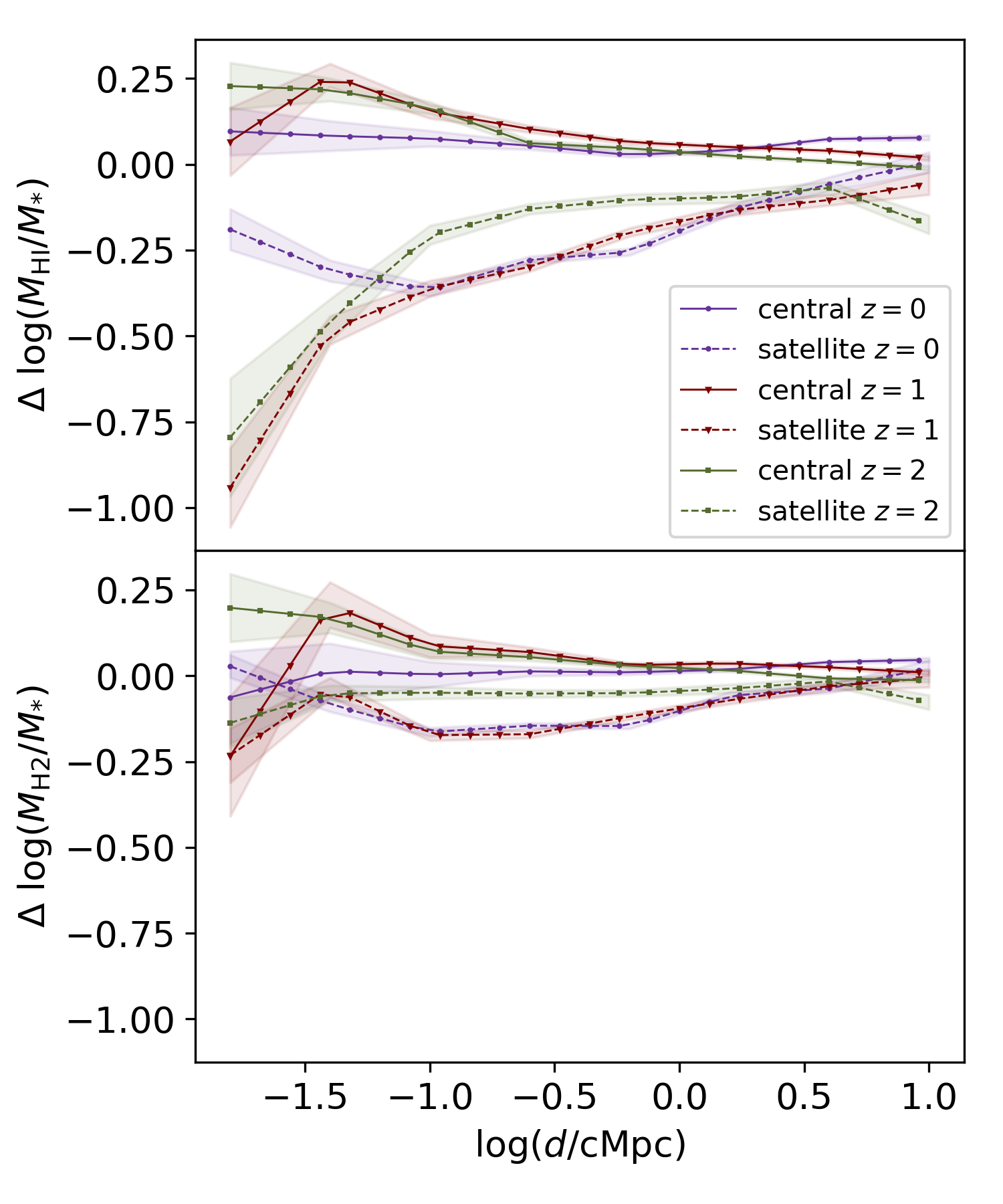}
    \caption{Distance from filaments dependence of the \HI (\textit{left}) and \Hmol (\textit{right}) fractions' deviations from the corresponding scaling relations with mass, for centrals (continuous lines) and satellites (dashed lines) for three redshifts: $z = 0$ (purple), $z = 1$ (dark red), $z = 2$ (green). The lines were obtained by binning the results in terms of distance and interpolating the corresponding medians (i.e. running medians). The shaded regions represent the corresponding standard errors in each bin. Satellites are depleted in \HI at a given $M_*$ near to filaments at all three redshifts considered, $z = 0, 1$ and 2, while \fmol at a given $M_*$ is not significantly depleted.}
    \label{fig.delta_fH12}
\end{figure}

The \HI and \Hmol fractions are known to have a clear dependence on galaxy stellar mass (e.g. \citealt{Obreschkow2009,catinella2010,maddox2015}). \simba reproduces these trends fairly well, as shown in \citet{Dave2020}. We obtain the underlying scaling relations by fitting a running median of each gas fraction as a function of $M_*$. This then allows us to compute the corresponding deviations $\Delta$log($M_\mathrm{HI}$/$M_*$) and $\Delta$log($M_\mathrm{H2}$/$M_*$).

Figure~\ref{fig.delta_fH12} shows these quantities $\Delta$\fHI and $\Delta$\fmol as a function of distance from filament $d$, at $z=0,1,2$ for centrals and satellites.  This plot can be compared to Fig.~\ref{fig.fH12_d}, which uses the same line style scheme.

Unlike $\Delta$sSFR discussed in the previous section, there are noticeable differences between $\Delta$\fHI and $\Delta$\fmol versus $d$ and the corresponding trends in the gas fractions themselves.  For central galaxies, \fHI and \fmol both show clear declines towards filaments, but the declines are much weaker or absent when considering $\Delta$\fHI and $\Delta$\fmol.  Indeed, at $z=2$, the gas fractions at a given $M_*$ are actually enhanced close to filaments, and this remains true for \HI even at $z=0$.  At $z=1$ there is an odd feature, present in both gas fractions as well as the sSFR, which is difficult to explain and may owe to small number statistics.  That aside, it appears that at high redshifts, the filamentary large-scale structure brings in more cool gas to supply galaxies, rather than suppressing it via shock heating.  Overall, the central galaxies do not show any significant suppression of gas contents near filaments. \tb{We also note here that the trends for centrals are flatter than for satellites and also compared to the trends found in the previous section when looking at the deviations from the SFMS (Fig.~\ref{fig.deltassfr}). However, even though the deviations here are relatively small, they are statistically significant, as illustrated by the shaded regions.}

For satellites, $\Delta$\fHI shows significant suppression towards filaments, following similar trends for \fHI, including the upturn in \HI content close to filaments at $z=0$.  This indicates that for \HI, the mass dependence of satellites is not critical in establishing trends of gas fractions with $d$.

In contrast, satellites show decidedly less amounts of suppression in $\Delta$\fmol close to filaments than \fmol.  This is particularly surprising since sSFR and $\Delta$sSFR both show significant suppression, and the \Hmol content is directly responsible for feeding star formation.  This shows that the molecular gas fractions have a stronger $M_*$ dependence which yields more significant trends with $d$ in \fmol than in $\Delta$\fmol. Combined with the variations in star formation efficiency discussed in \S\ref{sect.fH12}, this leads to a lack of a strong trend in $\Delta$\fmol$(d)$.


To sum up, we can state that satellites are depleted in \HI, \tb{relative to centrals,} at a given $M_*$ near to filaments at all three redshifts considered, $z = 0, 1$, and 2. In contrast, \fmol at a given $M_*$ is not significantly depleted. Meanwhile, centrals show only weak trends, with a hint of an enhanced gas content close to filaments particularly at higher redshifts.  With respect to quenching, it can be seen that the gas depletion for satellites starts at higher redshifts than the star formation suppression (\S\ref{sect:SFMS}), in agreement with the recent results of \cite{hasan2023}. This may be expected as gas removal does not immediately result in reduced star formation, hence we expect the gas depletion to start at earlier epochs. 

\subsection{Deviations from mass metallicity relation}
\label{sect:MZR}
\begin{figure}
    \includegraphics[width=0.5\textwidth]{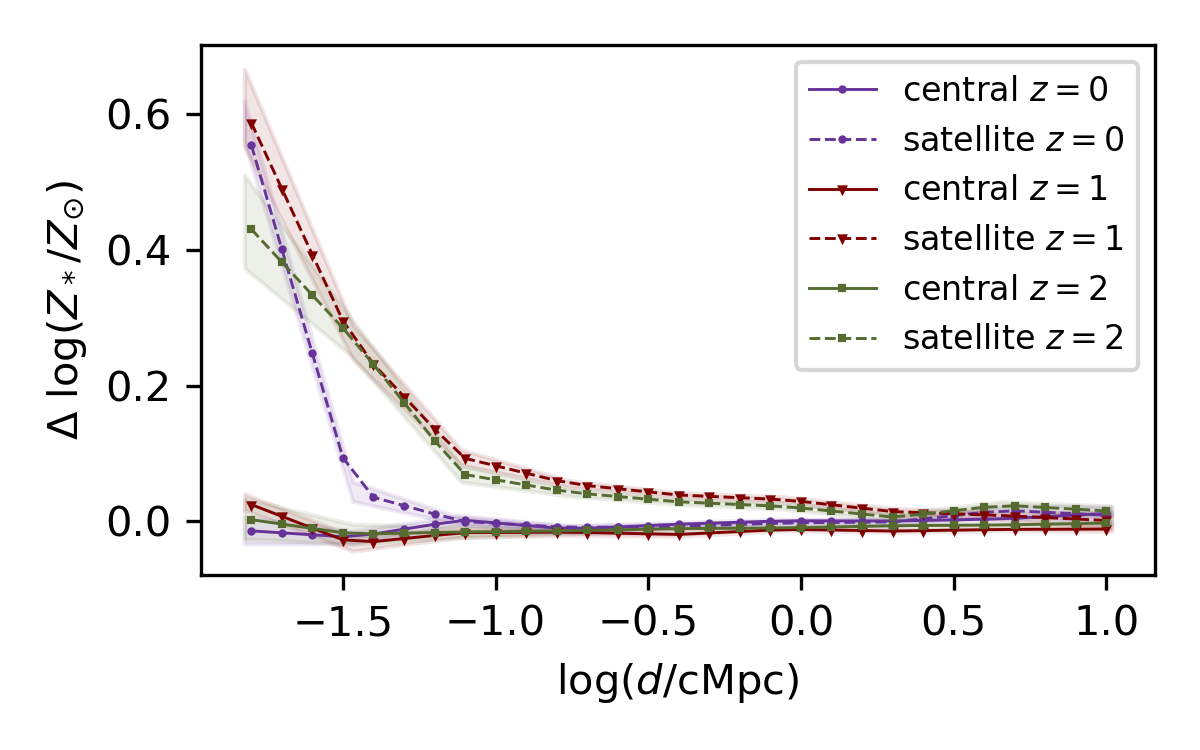}
    \caption{Distance from satellites dependence of the stellar metallicities' deviations from the mass-metallicity relation for centrals (continuous lines) and satellites (dashed lines) for three redshifts: $z = 0$ (purple), $z = 1$ (dark red), $z = 2$ (green). The lines were obtained by binning the results in terms of distance and interpolating the corresponding medians (i.e. running medians). The shaded regions represent the corresponding standard errors in each bin. Satellites close to filaments show significantly higher metallicities than the MZR expectations.}
    \label{fig.delta_Zs}
\end{figure}

We likewise investigate the deviations from the mass metallicity relation (MZR).  As mentioned in \S\ref{Sect:Simba}, the MZR in \simba is seen to be in good agreement with observational results at all three redshifts considered in this study. We compute the MZR by fitting a running median for the mass-metallicity distribution of all the galaxies considered (both centrals and satellites), and then we compute the deviation $\Delta$MZR from this fit for each galaxy, interpolated to its $M_*$. As previously discussed in \S\ref{sect:metallicity}, we examine the mass-weighted stellar metallicity here.

Figure~\ref{fig.delta_Zs} shows $\Delta Z_*$ with respect to the distance from filaments $d$.  This can be compared to the plot of $Z_*(d)$ shown in Fig.~\ref{fig.Z_comp}, which employs the same colour scheme.

The trends in $\Delta Z_*(d)$ are noticeably different than for $Z_*(d)$.  Most obviously, centrals show essentially no trend of $\Delta Z_*$ with $d$ at any redshift, whereas there was a strong trend of metallicity increasing closer to filaments.  The implication is that the trend seen in the MZR owes entirely to the fact that more massive centrals live closer to filaments, and intrinsically there is no effect of centrals' metallicity caused by large-scale structure.  Given that the sSFR is significantly impacted, the implication is that the fundamental metallicity relation~\citep{mannucci_et_al_2010} is in \simba predicted to be dependent on location within the cosmic web.  We aim to explore a comparison of this prediction to observations in future work.

The satellites, in contrast, continue to show a significant dependency of metallicity with $d$ at all redshifts.  Nonetheless, one sees significant differences comparing the trends in the dashed lines in Fig.~\ref{fig.Zs_d_all} versus Fig.~\ref{fig.delta_Zs}.  For instance, at $z=0$, $Z_*(d)$ shows a very flat dependence, while $\Delta Z_*(d)$ shows essentially no deviation until reaching very close to the filament centre, and then a strong upturn.  The trends at $z=1,2$ are more similar between these two quantities in terms of the trend, although the overall evolution of the MZR is removed by considering $\Delta Z_*$.


To sum up, our results show that only satellites are more metal-enriched than the mass-metallicity relation predictions in filaments' proximity, at all redshifts considered. A plausible explanation relies on the FMR, given the low levels of star formation at fixed stellar mass for satellites close to filaments (\S\ref{sect:SFMS}). We note however that satellites do not show evidence of strong star formation suppression at redshift $z = 2$, which again suggests that the FMR is not invariant with the environment.  We speculate that this finding might be caused by the early chemical enrichment expected to primarily influence satellites in dense regions, as reported in e.g. \cite{highZ1} and \cite{highZ2}.   Meanwhile, centrals at a given $M_*$ show no enhancement close to filaments.


Overall, for redshift $z = 0$ and $z = 1$ we have seen that there is a significant effect on many galaxy properties close to filaments, with the effects being much stronger for satellite galaxies than for centrals.  Close to filaments, galaxies (particularly satellites) tend to be less star-forming, less gas-rich, more metal-rich and are more likely to be quenched and dispersion-dominated.  In detail, the suppression in gas content (and particularly \fmol) is not as strong as seen for sSFR, and the stellar metallicities are not impacted by location in the cosmic web once the mass dependence of the MZR is taken out.  These predictions provide a comprehensive view of the growing impact of filaments on galaxies, which could potentially be compared to observations.

\section{\simba feedback variants}
\label{feedback}
\begin{figure}
    \includegraphics[width=0.5\textwidth]{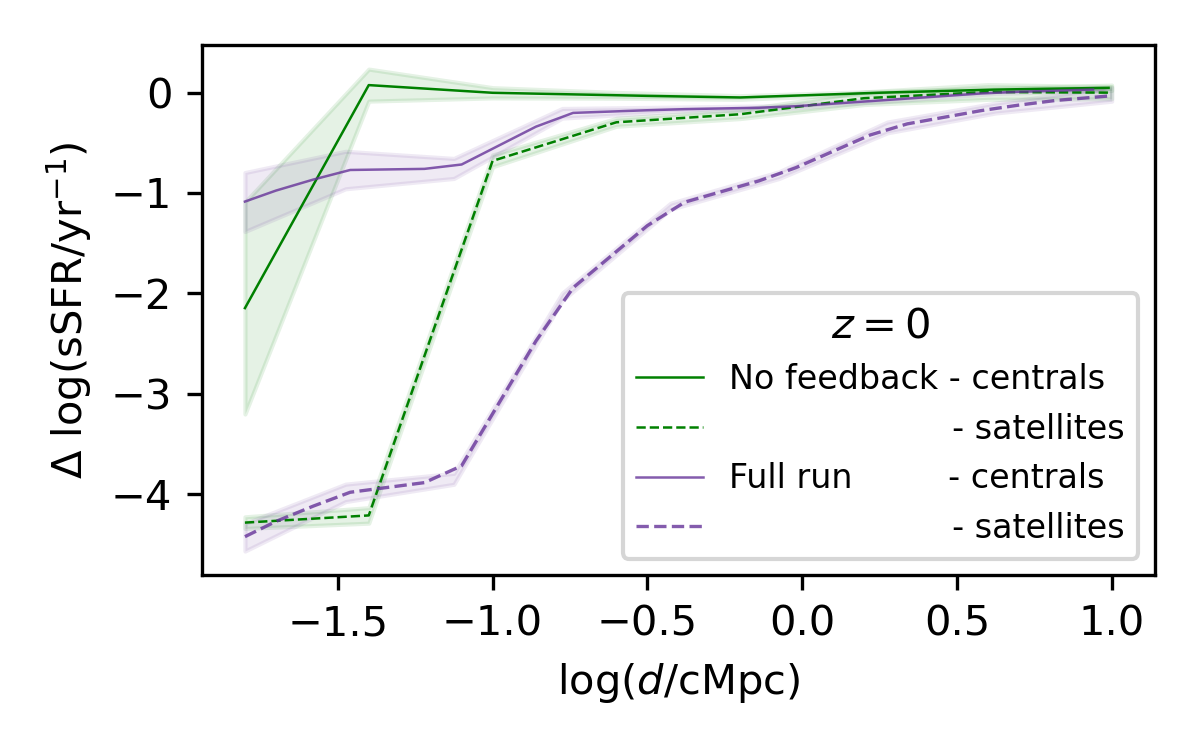}
    \caption{Distance from filaments dependence of the specific star formation rates’ deviations from the star formation main sequence for centrals (continuous lines) and satellites (dashed lines) at redshift $z$ = 0. The green lines represent no feedback scenarios, while the purple lines resulted from the full \simba runs (same as the purple lines in Fig.~\ref{fig.deltassfr}). The lines were obtained by binning the results in terms of distance and interpolating the corresponding medians (i.e. running medians). The shaded regions represent the corresponding standard errors in each bin. Galaxies have sSFR  suppressed near filaments even when feedback is excluded, suggesting that large-scale structure heating is the dominant primary of the suppression.}
    \label{fig.nofd_ssfr}
\end{figure} 

In the next two sections, we investigate the underlying cause(s) of why galaxies near filaments have systematically different properties.  We focus on the suppression of star formation since this is the clearest trend, and is correlated with the trends in other properties.  As discussed earlier, two possibilities for why the sSFR is lower near filaments are that it owes to shock heating from large-scale structure, or feedback heating from either star formation or AGN. In particular, AGN feedback is circumstantially implicated because the environmental effects become strong at $z\la 2$ and particularly at $z\la 1$, which matches up with the era in which AGN feedback increasingly quenches galaxies.  

With the \simba suite, we have an opportunity to directly test the impact of feedback mechanisms on galaxy properties in the cosmic web using the feedback variant runs.  In this section, we investigate the star formation properties of central and satellites at redshift $z = 0$, in the case when individual feedback modes of \simba are turned off as described by \S\ref{Sect:Simba}.  These runs, done in a $(50\hmpc)^3$ volume with otherwise the same resolution and input physics as the main $(100\hmpc)^3$ volume run, exclude feedback modes one at a time.

The main motivation behind this analysis is, on the one hand, to gain a better understanding of the main causes of quenching found in \S\ref{sect:ssfr} and \S\ref{sect:SFMS}, but also to separate the potential cosmic web effects from the feedback ones. As explained in \S\ref{sect:dev}, we chose to present the results here as deviations from their corresponding scaling relations with stellar mass, in order to minimise the mass effects on our findings. \tb{We computed the SFMS for each feedback variant and then calculated the corresponding deviations. The SFMS is dependent on feedback, given that all the various feedback channels implemented in \simba can suppress the star formation. For a more detailed description of these effects, we direct the reader to \cite{Dave2019}.}

Figure~\ref{fig.nofd_ssfr} presents the deviations from the star formation main sequence with respect to distance from filaments, when feedback is excluded (green lines) with the previous results from \S\ref{sect:SFMS} (Fig.~\ref{fig.deltassfr}) overplotted for reference (purple lines). As before, the satellites are represented by dashed lines and centrals by solid lines. It can be seen that when feedback is not included, star formation suppression is still evident in the filaments' proximity, this effect being stronger for satellites. 

\begin{figure}
    \includegraphics[width=0.5\textwidth]{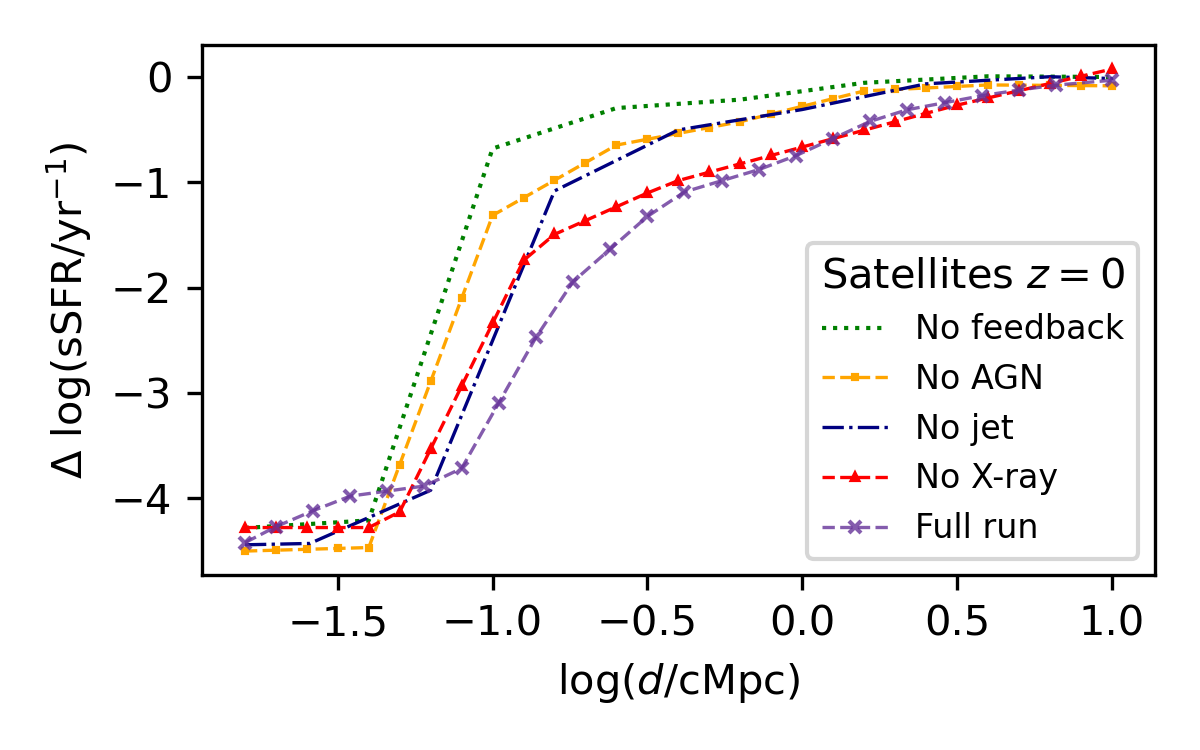}
    \caption{Distance from filaments dependence of the specific star formation rates’ deviations from the star formation main sequence for and satellites at redshift $z$ = 0, in different feedback scenarios: no feedback (green lines) -- same as in Fig.~\ref{fig.nofd_ssfr}; no AGN (yellow lines); no jet (blue lines); no X-ray feedback (red lines); full run (purple lines) -- same as in Fig.~\ref{fig.deltassfr} and Fig.~\ref{fig.nofd_ssfr}. The lines were obtained by binning the results in terms of distance and interpolating the corresponding medians (i.e. running medians). Note that no errors are plotted, due to the lines being very close to each other, but the approximate size of the errors in these determinations can be inferred from Fig.~\ref{fig.nofd_ssfr}. Among feedback processes, AGN feedback has the most significant impact on providing extra suppression of sSFR near filaments.}
    \label{masterpiece}
\end{figure}

In order to gain a better understanding of how/where feedback impacts star formation, Fig.~\ref{masterpiece} shows the deviations from the SFMS for satellites at redshift $z = 0$ in different feedback scenarios. Note that no errors are shown in this plot, due to how close the lines are, but the approximate size of the errors in these determinations can be inferred from Fig.~\ref{fig.nofd_ssfr}. It can be seen that in all cases, star formation is suppressed for a broader range of distances when some feedback is included, with AGN feedback having the stronger contribution in the filaments' proximity.  

It is worth mentioning that when feedback is excluded, the distance range for which star formation is suppressed is shorter than in the full run/partial feedback cases, as only galaxies in the immediate proximity of filaments (i.e. log($d$/cMpc) $\lesssim$ -1.4) show evidence of star formation suppression. This finding shows that feedback effects play indeed a role in quenching, however, for galaxies close to filaments we need a different explanation for the suppressed levels of star formation.

\section{The cosmic web around massive halos} \label{sect.halos}

As explained in \S\ref{Sect.prop_gen}, we have only studied galaxies within non-group halos ($M_h \leq 10^{13} M_\odot$) in order to avoid the influence of massive halos on the galaxy properties in the resulting cosmic web. However, the mass and spatial range over which massive halos influence surrounding galaxies remains uncertain.  For instance, \citet{Gabor2015} showed that galaxies can undergo ``neighbourhood quenching'' out to several virial radii, owing to elongated satellite orbits or being within the shock-heated region around a massive halo that can extend beyond the halo virial radius for quite massive systems.

\tb{Similarly, \cite{Bahe2012} and \cite{Wetzel2012} investigated how the halos' effects on quenching vary with distance, finding a non-trivial connection. Specifically, \cite{Bahe2012} found that confinement pressure by the intracluster medium can lead to an increase in the mass of hot gas, hence galaxy quenching being a competition between ram-pressure stripping and confinement pressure. \cite{Wetzel2012} also found that halos can quench galaxies even outside the virial radius, this effect fading out at $\sim$ two virial radii. Overall, the effects halos have on galaxy quenching are complex and still a debated topic in the literature \citep[e.g.][]{Schawinski2014,Zolotov2015,Behroozi2019}}.

In this section, we igim to explore how massive halos impact their surroundings in relation to the cosmic web. Specifically, we investigate further the quenching trend found near filaments in the previous sections (\S\ref{sect:ssfr}, \S\ref{sect:SFMS}), contrasted versus the influence of simply being near a massive halo regardless of the relation to a filament.  Certainly, we expect the massive halo to be the dominant environmental influence in its vicinity, however, we would like to determine if there is an additional influence from being close to a filament near the massive halo.

\subsection{Star formation near massive halos}

\begin{figure}
    \includegraphics[width=0.5\textwidth]{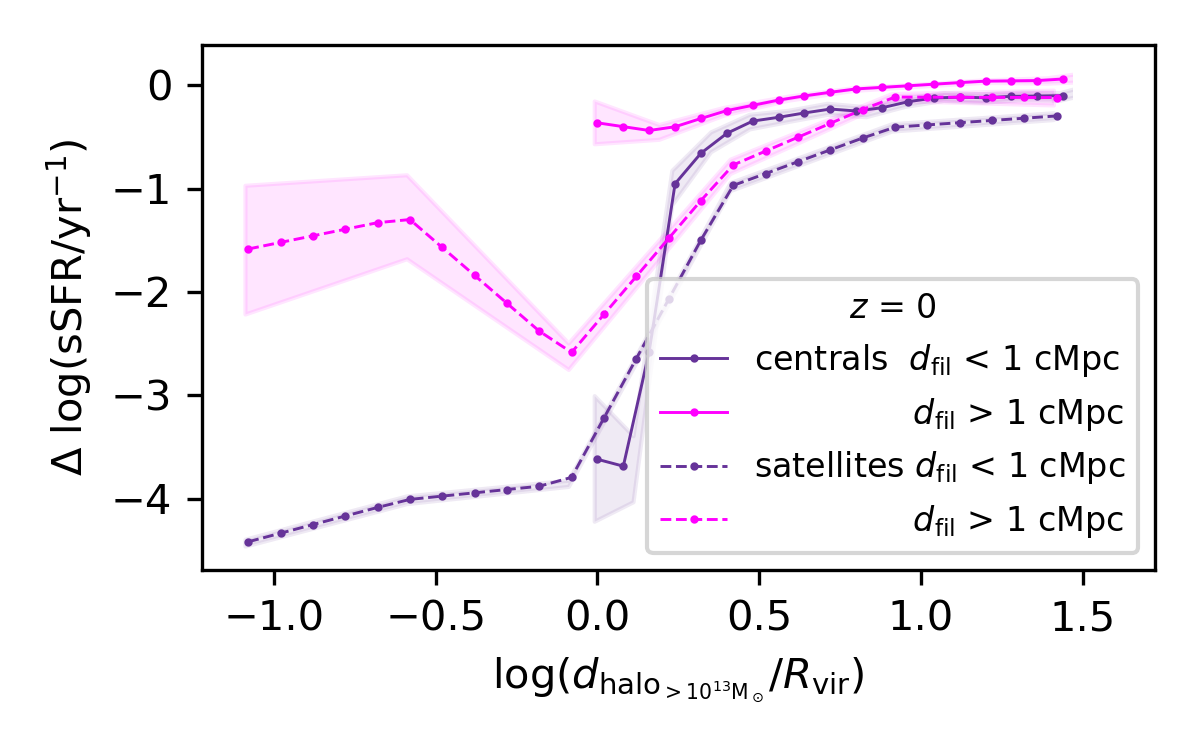}
    \caption{The deviations of the specific star formation rate from the star formation main sequence dependence on the distance to the closest massive halo ($M_h > 10^{13} M_\odot$) for centrals (continuous lines) and satellites (dashed lines), at redshift $z = 0$. The galaxies outside filaments ($d > $1 cMpc) are considered separately, represented by the magenta lines. In addition to the halo effects, filamentary environments also play a non-negligible role in sSFR suppression.}
    \label{fig.deltahalos}
\end{figure}

Figure~\ref{fig.deltahalos} shows how the deviations of the specific star formation rate from the SFMS for the galaxies in our sample varies with the distance from the closest massive halo ($M_h > 10^{13} M_\odot$), normalised by the virial radius $R_\mathrm{vir}$ \tb{of each corresponding closest massive halo}, at $z=0$. We define the virial radius as enclosing 200 times the critical density. \tb{We note that satellites can be found within distances as small as $\sim 0.1 R_\mathrm{vir}$ of a massive halo. This happens for satellites located within the outskirts of their host halo and in the neighborhood of a massive halo. This is not the case for centrals, since they are located at the centre of their host halos.}
In the purple lines, we show the results for the centrals (solid) and satellites (dashed) within 1 cMpc of a filament. 
In the magenta lines, we show those galaxies selected to be away from filaments, at a distance of at least 1 cMpc of a filament. Changing this threshold value does not qualitatively change the results, but lowering it makes the curves noisier due to the small number statistics while raising it lessens the effect.  The fraction of all galaxies within 1 cMpc of a filament is 46.24\%.

Figure~\ref{fig.deltahalos} shows, as expected, that the specific star formation rate is suppressed in the proximity of massive halos, more so for satellites than centrals. Centrals converge towards the global SFMS beyond a few $R_\mathrm{vir}$ from massive halos, showing that galaxies' star formation is impacted even outside the virial radius, though modestly so.  For satellites, the impact is more dramatic, and clearly extends out to $\sim 2R_\mathrm{vir}$. 

A curious feature seen in the centrals close to filaments is that the sSFR does not monotonically deviate from the global main sequence when going close to the halo centre.  
Instead, the deviation from the sSFR lessens at small radii. 
On the one hand, these galaxies are more quenched, hence their sSFR is lower compared to the other centrals in the sample. On top of this, they are also more massive because massive galaxies tend to cluster around the most massive halos, due to the stellar-to-halo mass relation (SFMR, see e.g., \citealt{Moster2010,Girelli2020}).
Hence the centrals close to filaments and also close to the centre of a massive halo ($\sim 1 R_{\rm{vir}}$ from the massive halo's centre) tend to lie significantly below the SFMS. 
It is also true that the small-number statistics can influence our results, but nonetheless the overall trend is consistent within the standard error.


Comparing between the galaxies 1 cMpc within/outside of filaments, we see that the broad trends are similar whether close to filaments or not.  Hence the dominant environmental effect around massive halos is set by proximity to the halo.  However, there are noticeable differences.  For instance, central galaxies outside filaments remain close to the global sSFR well within halos, while those within 1 cMpc of a filament are suppressed in SFR.  It appears that being in a filament may initially shield a galaxy from the environmental effects of shock-heated gas generated by the massive halo, but eventually ends up seeing an increased suppression in sSFR from being near a filament.  Meanwhile, satellites show a modest but distinct effect, in that being near a filament causes a cessation of star formation at a slightly larger radius.

We conclude that the location within the cosmic web has a non-negligible albeit sub-dominant effect on star formation in galaxies around massive halos.  
While in general star formation is increasingly suppressed closer to big halos, \tb{ filaments appear to play a role too, suppressing star formation further in their vicinity.} For satellites near filaments, the effects around central galaxies extend slightly further out as compared to satellites not near filaments.  These subtle environmental trends provide interesting directions for future work to both see if this is detectable in observations and to compare to other models.

\section{Comparisons with EAGLE and TNG}
\label{sect.eagle,tng}

\begin{figure}
    \includegraphics[width=0.5\textwidth]{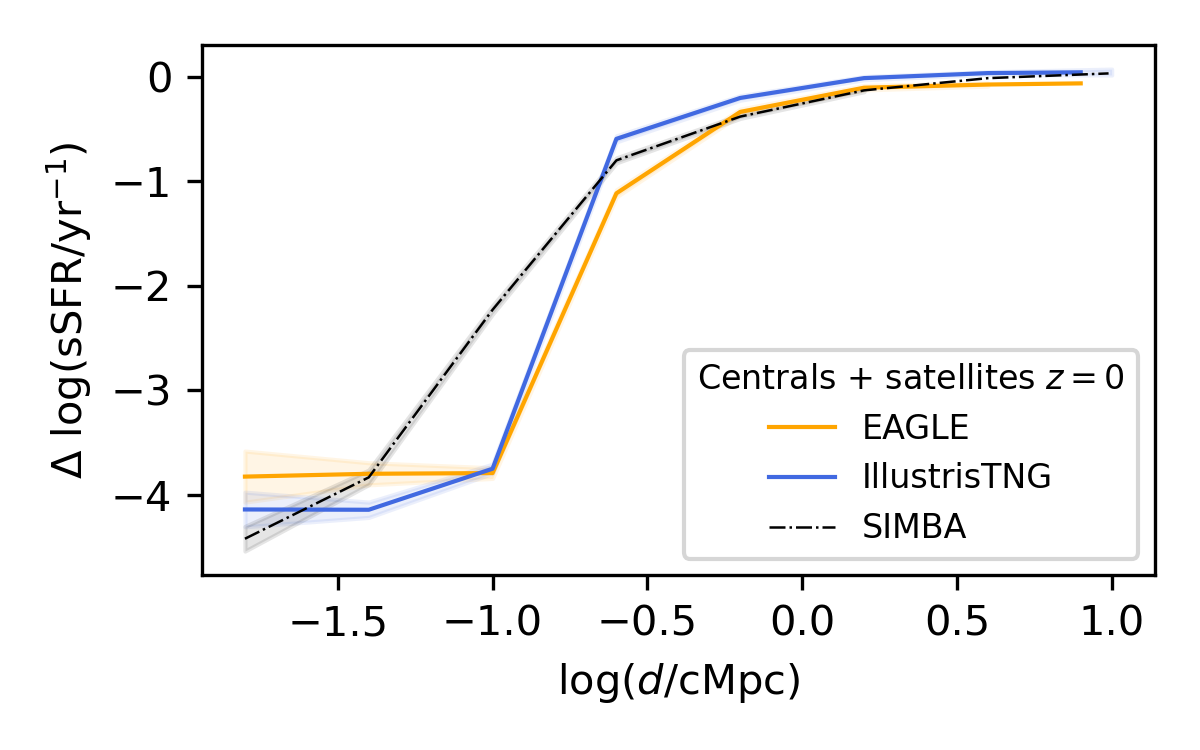}
    \caption{Distance from filaments dependence of the specific star formation rates' deviations from the star formation main sequence for all galaxies computed in EAGLE (orange line) and IllustrisTNG (blue line). The lines were obtained by binning the results in terms of distance and interpolating the corresponding medians. All the determinations are made for redshift $z$ = 0. sSFR is significantly suppressed near filaments in all models, with a broad agreement between EAGLE, IllustrisTNG, and SIMBA, modulo differences at $d\sim 100$~kpc.}
    \label{fig.ssfr_comp}
\end{figure}

\begin{figure}
    \includegraphics[width=0.5\textwidth, height=5.5cm]{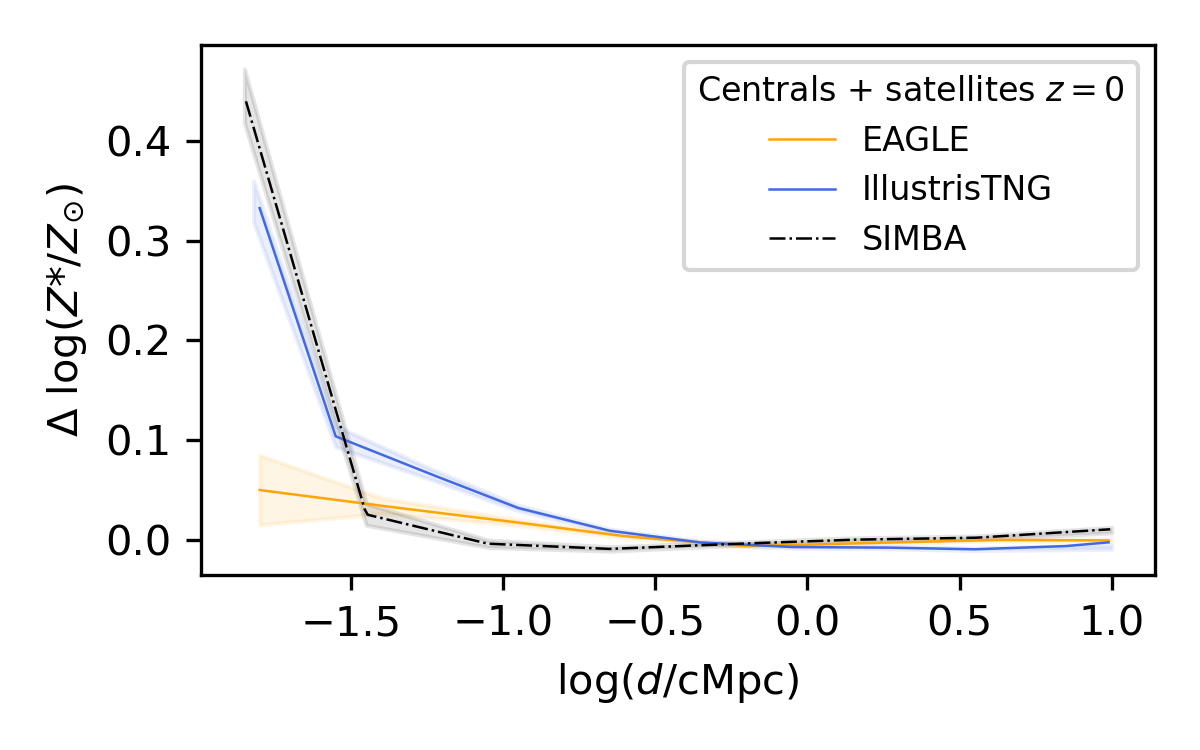}
    \caption{Distance from filaments dependence of the stellar metallicities' deviations from the mass-metallicity relation for all galaxies computed in EAGLE (orange line) and IllustrisTNG (blue line). The lines were obtained by binning the results in terms of distance and interpolating the corresponding medians. All the determinations are made for redshift $z$ = 0. Galaxies in all simulations are more metal-enriched (mainly satellites) in the filaments proximity, but IllustrisTNG and \simba show strong increases close to filaments, while EAGLE shows only a modest increase. }
    \label{fig.Z_comp}
\end{figure}

State-of-the-art cosmological simulations today tend to do a good job reproducing the stellar contents of galaxies, often because they are tuned to match the global stellar mass function.  However, different models achieve this agreement using very different feedback prescriptions, which could have differing effects on galaxy sub-populations. In this section, we compare our main findings with \textsc{EAGLE} and \textsc{IllustrisTNG} simulations, to assess how different their predictions are in terms of galaxies around filaments, towards potentially using this as a way to discriminate between models. \tb{Both \textsc{EAGLE} and \textsc{IllustrisTNG} show similar stellar mass functions and SFMS (see Fig.~2 in \citealt{Dave2020}). We also investigated the $M_*$ dependence on $d$ and found that the overall trend agrees between the three simulation suites -- i.e. the stellar masses decrease when moving away from filaments.} \tbn{However, due to the different satellites/centrals prescriptions in the three simulation suites, we cannot draw definite conclusions regarding the differences between their stellar mass distributions around filaments and limit at saying that the overall trend agrees between the three simulations.}  We focus on the deviations from the SFMS and MZR, as these are the quantities publicly available for both these models, \tb{and are expected to provide reliable results, given that they incorporate the mass effects. The subsequent plots in this section show the combined trends of centrals and satellites (i.e. we do not distinguish between them).}

For both \textsc{EAGLE} and \textsc{IllustrisTNG}, we identified filaments by applying \textsc{DisPerSE} to their galaxy catalogs exactly as described in \S\ref{sect.disperse}. Additionally, we used the same mass cuts as for the previous analysis -- the lower stellar mass limit comes from the \simba resolution and the upper limit from the halo mass cut (see \S\ref{Sect:Simba}). As before (\S\ref{sect:ssfr}) for all the galaxies showing a null SFR, we set log (sSFR/ yr$^{-1}$) = -14  and we compute the SFMS based on the galaxies with log (sSFR/ yr$^{-1}$) $>-10.8 + 0.3z$ (see \S\ref{sect:SFMS}). Using this \disperse skeleton we investigate the overall trends of all galaxies (centrals and satellites) at redshift $z = 0$ among these two models and compare to our results from \simba.

Figure~\ref{fig.ssfr_comp} shows $\Delta$sSFR versus filamentary distance $d$ at $z=0$ for \simba (grey), IllustrisTNG (blue), and EAGLE (orange).  Errors on the running medians are indicated by the shaded regions.

In general, all models predict a strong departure towards quenched galaxies when within $\la 100$~ckpc of a filament.  However, the details of trends show significant differences.  The decline in \simba is much more gradual than for EAGLE and IllustrisTNG, which show a very rapid transition from all galaxies being essentially on the SFMS to all galaxies being quenched.  In \simba, this occurs for the satellites, but less so for the centrals (see Fig.~\ref{fig.deltassfr}). This suggests that the typical sSFR or quenched fractions in central galaxies $\sim 100$~ckpc away from filaments may be a good discriminator between models.


Figure~\ref{fig.Z_comp} presents the metallicity trends, investigating as before the deviation of the stellar metallicity from the stellar MZR with distance from the nearest filament.  In this case, the differences are only strongly visible out to a few tens of kpc from the filament, and \simba and IllustrisTNG show substantial increases in $Z_*$ while EAGLE shows almost none\footnote{We note that the underlying MZR in these three models are quite different in both shape and amplitude, owing to differences in assumed yields, feedback efficiencies, and metal loading. It is beyond the scope of this work to examine in detail the origin of these variations; here we aim to mitigate the effects of such differences by considering only the deviations from the MZR self-consistently computed within each simulation.}.  It is further interesting that the deviations from the global mean relations extend much farther out in sSFR than in $Z_*$, showing that the two quantities are not simply inversely correlated via some sort of stellar fundamental metallicity relation, but rather have a more complex relationship.  These predictions provide clear testable ways to distinguish between models if stellar metallicities can be measured for such samples.


To sum up, overall we see general agreement between the trends resulting from \textsc{Eagle}, \textsc{IllustrisTNG} and \simba in sSFR and $Z_*$. Specifically, we find that galaxies close to filaments are less star-forming and more metal-enriched, though only satellites are expected to lie above the MZR predictions in this region (see \S\ref{sect:MZR}).  These results strengthen the hypothesis of more quenched galaxies in the filaments' proximity, as noticed in the previous sections (\S\ref{Sect.prop_gen}, \S\ref{sect:dev} and \S\ref{feedback}). Nonetheless, there are some distinct differences that could potentially be testable with present-day or upcoming large spectroscopic surveys.  In the future, we will investigate how these trends are diluted when confronted with observational limitations such as redshift space distortions.

\section{Summary}\label{sect.concl}

Using \simba simulations and the cosmic web extractor \textsc{DisPerSE}, we have investigated how galaxy properties depend on their location with respect to the filaments spines of the cosmic web.  We have done so at various redshifts from $z=2\to 0$, and have examined how these trends are governed by specific modes of feedback as implemented in \simba.  We have specifically excluded the cosmic web within halos of $M_h>10^{13}M_\odot$ so as to focus on environments outside of galaxy groups.  We also compared \simba's predictions with those from \textsc{EAGLE} and \textsc{IllustrisTNG} results derived using the same methodology with the same mass cuts. The main findings of this work are summarised as follows:

\begin{itemize}
    \item Central galaxies close to filaments have typically higher stellar mass and are surrounded by more satellites than those far away from filaments, similarly at all redshifts considered $z=0,1,2$ (Figs.~\ref{cen_sat_loc} and \ref{M_cs_dist}), in agreement with various previous literature results, \citep[e.g.][]{chen2017,malavasi2017,kraljic2018}. It is important to control for these variations in mass and halo occupancy in order to isolate the effects owing to the cosmic web environment.  We do so by considering centrals and satellites separately, and by either normalising to stellar mass or by computing deviations of quantities at a given $M_*$. This represents a novel aspect of this study since disentangling between centrals and satellites is very challenging in observational studies.
    
    \item At redshifts $z = 0$ and 1, the specific star formation rate or just sSFR is suppressed for satellites and centrals close to filaments, and increasing with distance. This trend has been reported before \citep[e.g.][]{kuutma2017,pouder2017}. We additionally note that this effect is considerably stronger for satellites and at later epochs, showing that satellites are more strongly impacted by the cosmic web environment over time.  For instance, at redshift $z = 0$ satellites are typically fully quenched within several hundreds kpc of a filament and do not converge to the sSFR's of centrals until one reaches $\sim 10$~cMpc from filaments (Fig.~\ref{fig.ssfr_d_all}).  This shows that pre-processing of satellites is already prevalent in filaments at $z\la 1$, prior to reaching group environments.  The effects on centrals are also noticeable, particularly within $\la 100$~ckpc of filaments.  One can thus regard $100$~ckpc as a rough scale over which filamentary environment impacts star formation in galaxies.
    
    \item The cold gas fractions, characterised in this study via $M_\mathrm{{HI}}$/$M_*$ and $M_\mathrm{{H2}}$/$M_*$, show more subtle and challenging trends with the distance from the closest filaments (Fig.~\ref{fig.fH12_d}), especially for centrals.  Broadly, cold gas is suppressed towards filament spines, increasingly so to lower redshifts as with sSFR, in qualitative agreement with \cite{Crone2018} and disagreement with \cite{kleiner2017}.  However, centrals can be more or less suppressed in cold gas than satellites depending on distance.  One aspect that may be confusing is that it can be difficult to associate particularly \HI with any particular galaxy within a denser environment, as \HI arises in relatively diffuse gas.  Hence a proper investigation of \HI contents may require creating mock observations for a particular setup and conducting side-by-side analyses with data, which is beyond the scope here but will be feasible using upcoming multi-wavelength radio surveys.

    \item The stellar metallicity is higher close to filaments for both centrals and satellites, at all three redshifts considered (Fig.~\ref{fig.Zs_d_all}), in agreement with \cite{winkel_2021,Donan2022}.  However, we additionally note that the trend with distance is much steeper at $z=2,1$, and is diluted by $z=0$.  The trends for centrals and satellites are not markedly different.
    
    \item The quenched fraction and the elliptical fractions are both anti-correlated with distance to filament, for both satellites and centrals, generally tracking the trends for sSFR as expected. The quenched fraction trend fades out at redshift $z = 2$, while the elliptical fraction trend is more consistent at all three redshifts considered $z = 0, 1$, and 2 (Fig.~\ref{fig.fQe}).  At a given distance, satellites at $z=0,1$ tend to be more quenched and elliptical.  This is in spite of the fact that they are lower mass than the centrals (we have not controlled for stellar mass in this plot).  Hence cosmic web environment impacts both colour and morphology.

    \item The trends noted above for sSFR are broadly similar when considering deviations from mean scaling relations rather than the quantities themselves (Fig.~\ref{fig.deltassfr}). This gives us confidence that the trends seen previously did not owe simply to trends with $M_*$, and are instead genuinely due to being close to a filament.  However, there are more significant differences in the case of metallicity $Z_*$; unlike for the overall metallicity $Z_*$ for which we saw a clear increase towards filaments, once we control for the $M_*$ dependence via the MZR, we now see no deviation from the mean MZR with distance for central galaxies, while satellites show strong deviations from the mean MZR only within $\la 100$~ckpc (Fig.~\ref{fig.delta_Zs}). Also, the trends in $H_2$ fraction with distance are not very strong when considering deviations from the mean $M_{H2}-M_*$ relation, indicating that the reduction in sSFRs close to filaments must owe primarily to a reduction in the star formation efficiencies of those galaxies (Fig.~\ref{fig.delta_fH12}).

    \item We investigate whether the trends in sSFR owe to feedback or cosmic web growth by comparing amongst identical \simba runs with individual feedback modes turned off.  We find that the predominant effect owes to the cosmic web, presumably via shock heating the gas to retard star formation near filaments.  However, the effects of feedback are not negligible; they add to the effects of large-scale structure and increase the range out to which satellites are quenched at $z=0$ by $\sim\times 2$ (Fig.~\ref{fig.nofd_ssfr}).  The bulk of this extra suppression comes from AGN feedback; star formation feedback has a minor impact (Fig.~\ref{masterpiece}).

    \item While we have mostly excluded massive halos from this analysis, we examine whether the impact of filaments is still noticeable around halos with $M_h>10^{13}M_\odot$ by comparing trends in sSFR vs. halo-centric distance for galaxies near filaments and far from filaments.  We find that the majority of the suppression of sSFR owes to the fact that these galaxies live around massive halos.  However, there are significant differences for the galaxies close to filaments feedback the massive halos; they show a different pattern of sSFR suppression for centrals, and slightly more extended suppression for satellites (Fig.~\ref{fig.deltahalos}).  Hence location within the cosmic web generates an effect over and above that arising simply due to being close to a massive halos.
    
    \item We compare our \simba results to those from the EAGLE and IllustrisTNG simulations at $z=0$, focusing on deviations from the mean sSFR and $Z_*$ relations with mass, and considering centrals and satellites together.  In general, all models show similar levels of sSFR suppression close to halos, although \simba's trend is more gradual while EAGLE and IllustrisTNG show a very sharp drop in median sSFR at $\sim 100$~ckpc (Fig.~\ref{fig.ssfr_comp}).  Meanwhile, $Z_*$ shows a significant increase very close to filaments for IllustrisTNG and \simba, but such a trend is not seen in EAGLE (Fig.~\ref{fig.Z_comp}).  These highlight possible avenues by which galaxy statistics relative to the cosmic web may provide discriminatory power between forefront simulations.
\end{itemize} 

The overall distribution of galaxies with respect to filaments, specifically high-mass centrals with their accompanying low-mass satellites in the filaments' proximity can be explained by the environmental effects on the halo mass function (e.g. \citealt{john}). The satellites' suppressed star formation, enriched metallicity and suppressed gas content trends at fixed stellar mass (\S\ref{sect:SFMS},~\ref{sect:MZR} and~\ref{sect:H1/2MS}) are consistent with a scenario where satellites close to filaments are quenched via \HI reservoir depletion and lowered efficiency in converting $H_2$ into stars putatively owing to an increase in shock-heated gas near filaments.  This also results in higher metallicities owing to the lack of infalling (relatively) pristine gas.  The corresponding trends are considerably weaker for centrals, indicating that the cosmic web effects are less efficient in this case, possibly because centrals live within denser gas near the bottom of their halos' potential wells.

Star formation suppression starts around $z\sim 2$, since environmental trends are not evident then, and if anything show a reversed trend in which all galaxies close to filaments have slightly higher sSFR's.  At earlier epochs, we find results very similar to those at $z=2$, so we did not explicitly show them. Given that gas depletion and star formation suppression are mostly present around filaments even when feedback is excluded (\S\ref{feedback}), we argue that the interactions between satellites and the hot gas of cosmic web cause quenching via a combination of gas stripping, on shorter timescales, and starvation, on longer timescales.  Nonetheless, feedback has an additional non-negligible impact, providing a way to constrain models of (particularly) AGN feedback.  

Overall, we find that the cosmic web plays a non-negligible role in shaping galaxy properties, though these effects are secondary to, i.e. weaker than, the mass effects. Our results generally agree with similar recent studies \citep[e.g.][]{kraljic2018,malavasi2021,Bhambhani2022} and provide new perspectives by clearly separating centrals and satellites, controlling for stellar mass, considering only galaxies within low-mass halos and investigating the impact of various feedback scenarios. Observational results are required to test these predictions and potentially identify areas for improvements in simulations.  Identifying the cosmic web ideally requires large-area spectroscopic surveys (although it may be possible to extract signals from the 2-D cosmic web from photo-z's), which exist now with SDSS and GAMA but will soon be greatly boosted with new facilities like Euclid \citep{euclid1,euclid2,euclid3}, WEAVE \citep{weave}, PFS \citep{PFS2014}, 4MOST \citep{4MOST}, and DESI \citep{desi}. Combining these with multi-wavelength surveys to characterise the various physical properties of galaxies, provides an exciting new frontier to explore how galaxy evolution models can be constrained using the cosmic web.


\section*{Acknowledgements}
\tb{The authors thank the anonymous referee for insightful comments.}
The authors would also like to thank Katja Fahrion for helpful discussions and the developers of \disperse and \caesar for making their codes public. This work was supported by the Science and Technology Facilities Council (STFC).  For the purpose of open access, the author has applied a Creative
Commons Attribution (CC BY) licence to any Author Accepted Manuscript version arising from this submission.

\section*{Data Availability}

The \simba simulation data and \caesar galaxy catalogues are publicly available at \url{https://simba.roe.ac.uk}. The derived data and \disperse outputs underlying this article will be shared on reasonable request to the corresponding author.



\bibliographystyle{mn2e}
\bibliography{references} 

\begin{thebibliography}{137}
\expandafter\ifx\csname natexlab\endcsname\relax\def\natexlab#1{#1}\fi

\bibitem[{{Abazajian} {et~al}\mbox{.}(2009){Abazajian}, {Adelman-McCarthy},
  {Ag{\"u}eros}, {Allam}, {Allende Prieto}, {An}, {Anderson}, {Anderson},
  {Annis}, {Bahcall}, {Bailer-Jones}, {Barentine}, {Bassett}, {Becker},
  {Beers}, {Bell}, {Belokurov}, {Berlind}, {Berman}, {Bernardi}, {Bickerton},
  {Bizyaev}, {Blakeslee}, {Blanton}, {Bochanski}, {Boroski}, {Brewington},
  {Brinchmann}, {Brinkmann}, {Brunner}, {Budav{\'a}ri}, {Carey}, {Carliles},
  {Carr}, {Castander}, {Cinabro}, {Connolly}, {Csabai}, {Cunha}, {Czarapata},
  {Davenport}, {de Haas}, {Dilday}, {Doi}, {Eisenstein}, {Evans}, {Evans},
  {Fan}, {Friedman}, {Frieman}, {Fukugita}, {G{\"a}nsicke}, {Gates},
  {Gillespie}, {Gilmore}, {Gonzalez}, {Gonzalez}, {Grebel}, {Gunn},
  {Gy{\"o}ry}, {Hall}, {Harding}, {Harris}, {Harvanek}, {Hawley}, {Hayes},
  {Heckman}, {Hendry}, {Hennessy}, {Hindsley}, {Hoblitt}, {Hogan}, {Hogg},
  {Holtzman}, {Hyde}, {Ichikawa}, {Ichikawa}, {Im}, {Ivezi{\'c}}, {Jester},
  {Jiang}, {Johnson}, {Jorgensen}, {Juri{\'c}}, {Kent}, {Kessler}, {Kleinman},
  {Knapp}, {Konishi}, {Kron}, {Krzesinski}, {Kuropatkin}, {Lampeitl},
  {Lebedeva}, {Lee}, {Lee}, {French Leger}, {L{\'e}pine}, {Li}, {Lima}, {Lin},
  {Long}, {Loomis}, {Loveday}, {Lupton}, {Magnier}, {Malanushenko},
  {Malanushenko}, {Mandelbaum}, {Margon}, {Marriner}, {Mart{\'\i}nez-Delgado},
  {Matsubara}, {McGehee}, {McKay}, {Meiksin}, {Morrison}, {Mullally}, {Munn},
  {Murphy}, {Nash}, {Nebot}, {Neilsen}, {Newberg}, {Newman}, {Nichol},
  {Nicinski}, {Nieto-Santisteban}, {Nitta}, {Okamura}, {Oravetz}, {Ostriker},
  {Owen}, {Padmanabhan}, {Pan}, {Park}, {Pauls}, {Peoples}, {Percival}, {Pier},
  {Pope}, {Pourbaix}, {Price}, {Purger}, {Quinn}, {Raddick}, {Re Fiorentin},
  {Richards}, {Richmond}, {Riess}, {Rix}, {Rockosi}, {Sako}, {Schlegel},
  {Schneider}, {Scholz}, {Schreiber}, {Schwope}, {Seljak}, {Sesar}, {Sheldon},
  {Shimasaku}, {Sibley}, {Simmons}, {Sivarani}, {Allyn Smith}, {Smith},
  {Smol{\v{c}}i{\'c}}, {Snedden}, {Stebbins}, {Steinmetz}, {Stoughton},
  {Strauss}, {SubbaRao}, {Suto}, {Szalay}, {Szapudi}, {Szkody}, {Tanaka},
  {Tegmark}, {Teodoro}, {Thakar}, {Tremonti}, {Tucker}, {Uomoto}, {Vanden
  Berk}, {Vandenberg}, {Vidrih}, {Vogeley}, {Voges}, {Vogt}, {Wadadekar},
  {Watters}, {Weinberg}, {West}, {White}, {Wilhite}, {Wonders}, {Yanny},
  {Yocum}, {York}, {Zehavi}, {Zibetti}, \& {Zucker}}]{ababazajian2009}
{Abazajian} K.~N. {et~al.}, 2009, \apjs, 182, 543

\bibitem[{{Alam} {et~al}\mbox{.}(2019){Alam}, {Zu}, {Peacock}, \&
  {Mandelbaum}}]{john}
{Alam} S., {Zu} Y., {Peacock} J.~A., {Mandelbaum} R., 2019, \mnras, 483, 4501

\bibitem[{{Alpaslan} {et~al}\mbox{.}(2015){Alpaslan}, {Driver}, {Robotham},
  {Obreschkow}, {Andrae}, {Cluver}, {Kelvin}, {Lange}, {Owers}, {Taylor},
  {Andrews}, {Bamford}, {Bland-Hawthorn}, {Brough}, {Brown}, {Colless},
  {Davies}, {Eardley}, {Grootes}, {Hopkins}, {Kennedy}, {Liske},
  {Lara-L{\'o}pez}, {L{\'o}pez-S{\'a}nchez}, {Loveday}, {Madore}, {Mahajan},
  {Meyer}, {Moffett}, {Norberg}, {Penny}, {Pimbblet}, {Popescu}, {Seibert}, \&
  {Tuffs}}]{alpaslan2015}
{Alpaslan} M. {et~al.}, 2015, \mnras, 451, 3249

\bibitem[{{Angl{\'e}s-Alc{\'a}zar}
  {et~al}\mbox{.}(2017){Angl{\'e}s-Alc{\'a}zar}, {Faucher-Gigu{\`e}re},
  {Kere{\v{s}}}, {Hopkins}, {Quataert}, \& {Murray}}]{Angles-Alcazar2017}
{Angl{\'e}s-Alc{\'a}zar} D., {Faucher-Gigu{\`e}re} C.-A., {Kere{\v{s}}} D.,
  {Hopkins} P.~F., {Quataert} E., {Murray} N., 2017, \mnras, 470, 4698

\bibitem[{{Angl{\'e}s-Alc{\'a}zar}, {{\"O}zel} \&
  {Dav{\'e}}(2013){Angl{\'e}s-Alc{\'a}zar}, {{\"O}zel}, \&
  {Dav{\'e}}}]{Angles-Alcazar2013}
{Angl{\'e}s-Alc{\'a}zar} D., {{\"O}zel} F., {Dav{\'e}} R., 2013, \apj, 770, 5

\bibitem[{{Angl{\'e}s-Alc{\'a}zar}
  {et~al}\mbox{.}(2015){Angl{\'e}s-Alc{\'a}zar}, {{\"O}zel}, {Dav{\'e}},
  {Katz}, {Kollmeier}, \& {Oppenheimer}}]{Angles-Alcazar2015}
{Angl{\'e}s-Alc{\'a}zar} D., {{\"O}zel} F., {Dav{\'e}} R., {Katz} N.,
  {Kollmeier} J.~A., {Oppenheimer} B.~D., 2015, \apj, 800, 127

\bibitem[{{Appleby} {et~al}\mbox{.}(2021){Appleby}, {Dav{\'e}}, {Sorini},
  {Storey-Fisher}, \& {Smith}}]{Appleby2021}
{Appleby} S., {Dav{\'e}} R., {Sorini} D., {Storey-Fisher} K., {Smith} B., 2021,
  \mnras, 507, 2383

\bibitem[{{Arag{\'o}n-Calvo}, {van de Weygaert} \&
  {Jones}(2010){Arag{\'o}n-Calvo}, {van de Weygaert}, \&
  {Jones}}]{aragon-calvo_2010}
{Arag{\'o}n-Calvo} M.~A., {van de Weygaert} R., {Jones} B. J.~T., 2010, \mnras,
  408, 2163

\bibitem[{{Bah{\'e}} {et~al}\mbox{.}(2012){Bah{\'e}}, {McCarthy}, {Crain}, \&
  {Theuns}}]{Bahe2012}
{Bah{\'e}} Y.~M., {McCarthy} I.~G., {Crain} R.~A., {Theuns} T., 2012, \mnras,
  424, 1179

\bibitem[{{Bah{\'e}} {et~al}\mbox{.}(2017){Bah{\'e}}, {Schaye}, {Crain},
  {McCarthy}, {Bower}, {Theuns}, {McGee}, \& {Trayford}}]{highZ1}
{Bah{\'e}} Y.~M., {Schaye} J., {Crain} R.~A., {McCarthy} I.~G., {Bower} R.~G.,
  {Theuns} T., {McGee} S.~L., {Trayford} J.~W., 2017, \mnras, 464, 508

\bibitem[{{Baldry} {et~al}\mbox{.}(2006){Baldry}, {Balogh}, {Bower},
  {Glazebrook}, {Nichol}, {Bamford}, \& {Budavari}}]{Baldry2006}
{Baldry} I.~K., {Balogh} M.~L., {Bower} R.~G., {Glazebrook} K., {Nichol} R.~C.,
  {Bamford} S.~P., {Budavari} T., 2006, \mnras, 373, 469

\bibitem[{{Behroozi} {et~al}\mbox{.}(2019){Behroozi}, {Wechsler}, {Hearin}, \&
  {Conroy}}]{Behroozi2019}
{Behroozi} P., {Wechsler} R.~H., {Hearin} A.~P., {Conroy} C., 2019, \mnras,
  488, 3143

\bibitem[{{Beygu} {et~al}\mbox{.}(2016){Beygu}, {Kreckel}, {van der Hulst},
  {Jarrett}, {Peletier}, {van de Weygaert}, {van Gorkom}, \&
  {Aragon-Calvo}}]{beygu2016}
{Beygu} B., {Kreckel} K., {van der Hulst} J.~M., {Jarrett} T.~H., {Peletier}
  R., {van de Weygaert} R., {van Gorkom} J.~H., {Aragon-Calvo} M.~A., 2016,
  \mnras, 458, 394

\bibitem[{{Beyoro-Amado} {et~al}\mbox{.}(2021){Beyoro-Amado}, {Povi{\'c}},
  {S{\'a}nchez-Portal}, {Belay Tessema}, {Getachew-Woreta}, \& {Glace
  Team}}]{Beyoro-Amado2021}
{Beyoro-Amado} Z., {Povi{\'c}} M., {S{\'a}nchez-Portal} M., {Belay Tessema} S.,
  {Getachew-Woreta} T., {Glace Team}, 2021, in Nuclear Activity in Galaxies
  Across Cosmic Time, {Povi{\'c}} M., {Marziani} P., {Masegosa} J., {Netzer}
  H., {Negu} S.~H., {Tessema} S.~B., eds., Vol. 356, pp. 163--168

\bibitem[{{Bhambhani} {et~al}\mbox{.}(2022){Bhambhani}, {Baldry}, {Brough},
  {Hill}, {Lara-Lopez}, {Loveday}, \& {Holwerda}}]{Bhambhani2022}
{Bhambhani} P.~C., {Baldry} I.~K., {Brough} S., {Hill} A.~D., {Lara-Lopez}
  M.~A., {Loveday} J., {Holwerda} B.~W., 2022, arXiv e-prints, arXiv:2210.16112

\bibitem[{{Bilicki} {et~al}\mbox{.}(2016){Bilicki}, {Peacock}, {Jarrett},
  {Cluver}, {Maddox}, {Brown}, {Taylor}, {Hambly}, {Solarz}, {Holwerda},
  {Baldry}, {Loveday}, {Moffett}, {Hopkins}, {Driver}, {Alpaslan}, \&
  {Bland-Hawthorn}}]{bilicki2016}
{Bilicki} M. {et~al.}, 2016, \apjs, 225, 5

\bibitem[{{Bond}, {Kofman} \& {Pogosyan}(1996){Bond}, {Kofman}, \&
  {Pogosyan}}]{bond_1996}
{Bond} J.~R., {Kofman} L., {Pogosyan} D., 1996, \nat, 380, 603

\bibitem[{{Bonjean} {et~al}\mbox{.}(2020){Bonjean}, {Aghanim}, {Douspis},
  {Malavasi}, \& {Tanimura}}]{bonjean2020}
{Bonjean} V., {Aghanim} N., {Douspis} M., {Malavasi} N., {Tanimura} H., 2020,
  \aap, 638, A75

\bibitem[{{Boselli} \& {Gavazzi}(2006)}]{BosselliGavazzi2006}
{Boselli} A., {Gavazzi} G., 2006, \pasp, 118, 517

\bibitem[{{Boselli} \& {Gavazzi}(2014)}]{BosselliGavazzi2014}
{Boselli} A., {Gavazzi} G., 2014, \aapr, 22, 74

\bibitem[{{Castignani} {et~al}\mbox{.}(2022){Castignani}, {Combes}, {Jablonka},
  {Finn}, {Rudnick}, {Vulcani}, {Desai}, {Zaritsky}, \&
  {Salom{\'e}}}]{Castigani2021}
{Castignani} G. {et~al.}, 2022, \aap, 657, A9

\bibitem[{{Catinella} {et~al}\mbox{.}(2010){Catinella}, {Schiminovich},
  {Kauffmann}, {Fabello}, {Wang}, {Hummels}, {Lemonias}, {Moran}, {Wu},
  {Giovanelli}, {Haynes}, {Heckman}, {Basu-Zych}, {Blanton}, {Brinchmann},
  {Budav{\'a}ri}, {Gon{\c{c}}alves}, {Johnson}, {Kennicutt}, {Madore},
  {Martin}, {Rich}, {Tacconi}, {Thilker}, {Wild}, \& {Wyder}}]{catinella2010}
{Catinella} B. {et~al.}, 2010, \mnras, 403, 683

\bibitem[{{Chen} {et~al}\mbox{.}(2017){Chen}, {Ho}, {Mandelbaum}, {Bahcall},
  {Brownstein}, {Freeman}, {Genovese}, {Schneider}, \& {Wasserman}}]{chen2017}
{Chen} Y.-C. {et~al.}, 2017, \mnras, 466, 1880

\bibitem[{{Cimatti} {et~al}\mbox{.}(2009){Cimatti}, {Robberto}, {Baugh},
  {Beckwith}, {Content}, {Daddi}, {De Lucia}, {Garilli}, {Guzzo}, {Kauffmann},
  {Lehnert}, {Maccagni}, {Mart{\'\i}nez-Sansigre}, {Pasian}, {Reid}, {Rosati},
  {Salvaterra}, {Stiavelli}, {Wang}, {Zapatero Osorio}, {Balcells},
  {Bersanelli}, {Bertoldi}, {Blaizot}, {Bottini}, {Bower}, {Bulgarelli},
  {Burgasser}, {Burigana}, {Butler}, {Casertano}, {Ciardi}, {Cirasuolo},
  {Clampin}, {Cole}, {Comastri}, {Cristiani}, {Cuby}, {Cuttaia}, {de Rosa},
  {Sanchez}, {di Capua}, {Dunlop}, {Fan}, {Ferrara}, {Finelli}, {Franceschini},
  {Franx}, {Franzetti}, {Frenk}, {Gardner}, {Gianotti}, {Grange}, {Gruppioni},
  {Gruppuso}, {Hammer}, {Hillenbrand}, {Jacobsen}, {Jarvis}, {Kennicutt},
  {Kimble}, {Kriek}, {Kurk}, {Kneib}, {Le Fevre}, {Macchetto}, {MacKenty},
  {Madau}, {Magliocchetti}, {Maino}, {Mandolesi}, {Masetti}, {McLure},
  {Mennella}, {Meyer}, {Mignoli}, {Mobasher}, {Molinari}, {Morgante}, {Morris},
  {Nicastro}, {Oliva}, {Padovani}, {Palazzi}, {Paresce}, {Perez Garrido},
  {Pian}, {Popa}, {Postman}, {Pozzetti}, {Rayner}, {Rebolo}, {Renzini},
  {R{\"o}ttgering}, {Schinnerer}, {Scodeggio}, {Saisse}, {Shanks}, {Shapley},
  {Sharples}, {Shea}, {Silk}, {Smail}, {Span{\'o}}, {Steinacker},
  {Stringhetti}, {Szalay}, {Tresse}, {Trifoglio}, {Urry}, {Valenziano},
  {Villa}, {Villo Perez}, {Walter}, {Ward}, {White}, {White}, {Wright}, {Wyse},
  {Zamorani}, {Zacchei}, {Zeilinger}, \& {Zerbi}}]{euclid3}
{Cimatti} A. {et~al.}, 2009, Experimental Astronomy, 23, 39

\bibitem[{{Codis} {et~al}\mbox{.}(2018){Codis}, {Jindal}, {Chisari}, {Vibert},
  {Dubois}, {Pichon}, \& {Devriendt}}]{Codis2018}
{Codis} S., {Jindal} A., {Chisari} N.~E., {Vibert} D., {Dubois} Y., {Pichon}
  C., {Devriendt} J., 2018, \mnras, 481, 4753

\bibitem[{{Colless} {et~al}\mbox{.}(2001){Colless}, {Dalton}, {Maddox},
  {Sutherland}, {Norberg}, {Cole}, {Bland-Hawthorn}, {Bridges}, {Cannon},
  {Collins}, {Couch}, {Cross}, {Deeley}, {De Propris}, {Driver}, {Efstathiou},
  {Ellis}, {Frenk}, {Glazebrook}, {Jackson}, {Lahav}, {Lewis}, {Lumsden},
  {Madgwick}, {Peacock}, {Peterson}, {Price}, {Seaborne}, \&
  {Taylor}}]{Colless2001}
{Colless} M. {et~al.}, 2001, \mnras, 328, 1039

\bibitem[{{Crone Odekon} {et~al}\mbox{.}(2018){Crone Odekon}, {Hallenbeck},
  {Haynes}, {Koopmann}, {Phi}, \& {Wolfe}}]{Crone2018}
{Crone Odekon} M., {Hallenbeck} G., {Haynes} M.~P., {Koopmann} R.~A., {Phi} A.,
  {Wolfe} P.-F., 2018, \apj, 852, 142

\bibitem[{{Cui} {et~al}\mbox{.}(2021){Cui}, {Dav{\'e}}, {Peacock},
  {Angl{\'e}s-Alc{\'a}zar}, \& {Yang}}]{Cui2021}
{Cui} W., {Dav{\'e}} R., {Peacock} J.~A., {Angl{\'e}s-Alc{\'a}zar} D., {Yang}
  X., 2021, Nature Astronomy, 5, 1069

\bibitem[{{Cui} {et~al}\mbox{.}(2018){Cui}, {Knebe}, {Yepes}, {Pearce},
  {Power}, {Dave}, {Arth}, {Borgani}, {Dolag}, {Elahi}, {Mostoghiu}, {Murante},
  {Rasia}, {Stoppacher}, {Vega-Ferrero}, {Wang}, {Yang}, {Benson}, {Cora},
  {Croton}, {Sinha}, {Stevens}, {Vega-Mart{\'\i}nez}, {Arthur}, {Baldi},
  {Ca{\~n}as}, {Cialone}, {Cunnama}, {De Petris}, {Durando}, {Ettori},
  {Gottl{\"o}ber}, {Nuza}, {Old}, {Pilipenko}, {Sorce}, \& {Welker}}]{300}
{Cui} W. {et~al.}, 2018, \mnras, 480, 2898

\bibitem[{{Daddi} {et~al}\mbox{.}(2007){Daddi}, {Dickinson}, {Morrison},
  {Chary}, {Cimatti}, {Elbaz}, {Frayer}, {Renzini}, {Pope}, {Alexander},
  {Bauer}, {Giavalisco}, {Huynh}, {Kurk}, \& {Mignoli}}]{Daddi2007}
{Daddi} E. {et~al.}, 2007, \apj, 670, 156

\bibitem[{{Dalton} {et~al}\mbox{.}(2012){Dalton}, {Trager}, {Abrams}, {Carter},
  {Bonifacio}, {Aguerri}, {MacIntosh}, {Evans}, {Lewis}, {Navarro}, {Agocs},
  {Dee}, {Rousset}, {Tosh}, {Middleton}, {Pragt}, {Terrett}, {Brock}, {Benn},
  {Verheijen}, {Cano Infantes}, {Bevil}, {Steele}, {Mottram}, {Bates},
  {Gribbin}, {Rey}, {Rodriguez}, {Delgado}, {Guinouard}, {Walton}, {Irwin},
  {Jagourel}, {Stuik}, {Gerlofsma}, {Roelfsma}, {Skillen}, {Ridings},
  {Balcells}, {Daban}, {Gouvret}, {Venema}, \& {Girard}}]{weave}
{Dalton} G. {et~al.}, 2012, in Society of Photo-Optical Instrumentation
  Engineers (SPIE) Conference Series, Vol. 8446, Ground-based and Airborne
  Instrumentation for Astronomy IV, {McLean} I.~S., {Ramsay} S.~K., {Takami}
  H., eds., p. 84460P

\bibitem[{{Darvish} {et~al}\mbox{.}(2014){Darvish}, {Sobral}, {Mobasher},
  {Scoville}, {Best}, {Sales}, \& {Smail}}]{davish2014}
{Darvish} B., {Sobral} D., {Mobasher} B., {Scoville} N.~Z., {Best} P., {Sales}
  L.~V., {Smail} I., 2014, \apj, 796, 51

\bibitem[{{Dav{\'e}}(2008)}]{Dave2008}
{Dav{\'e}} R., 2008, \mnras, 385, 147

\bibitem[{{Dav{\'e}} {et~al}\mbox{.}(2019){Dav{\'e}}, {Angl{\'e}s-Alc{\'a}zar},
  {Narayanan}, {Li}, {Rafieferantsoa}, \& {Appleby}}]{Dave2019}
{Dav{\'e}} R., {Angl{\'e}s-Alc{\'a}zar} D., {Narayanan} D., {Li} Q.,
  {Rafieferantsoa} M.~H., {Appleby} S., 2019, \mnras, 486, 2827

\bibitem[{{Dav{\'e}} {et~al}\mbox{.}(2020){Dav{\'e}}, {Crain}, {Stevens},
  {Narayanan}, {Saintonge}, {Catinella}, \& {Cortese}}]{Dave2020}
{Dav{\'e}} R., {Crain} R.~A., {Stevens} A. R.~H., {Narayanan} D., {Saintonge}
  A., {Catinella} B., {Cortese} L., 2020, \mnras, 497, 146

\bibitem[{{Dav{\'e}}, {Finlator} \& {Oppenheimer}(2011){Dav{\'e}}, {Finlator},
  \& {Oppenheimer}}]{Dave2011b}
{Dav{\'e}} R., {Finlator} K., {Oppenheimer} B.~D., 2011, \mnras, 416, 1354

\bibitem[{{Dav{\'e}}, {Finlator} \& {Oppenheimer}(2012){Dav{\'e}}, {Finlator},
  \& {Oppenheimer}}]{Dave2012}
{Dav{\'e}} R., {Finlator} K., {Oppenheimer} B.~D., 2012, \mnras, 421, 98

\bibitem[{{Dav{\'e}}, {Thompson} \& {Hopkins}(2016){Dav{\'e}}, {Thompson}, \&
  {Hopkins}}]{Dave2016}
{Dav{\'e}} R., {Thompson} R., {Hopkins} P.~F., 2016, \mnras, 462, 3265

\bibitem[{{de Jong} {et~al}\mbox{.}(2019){de Jong}, {Agertz}, {Berbel}, {Aird},
  {Alexander}, {Amarsi}, {Anders}, {Andrae}, {Ansarinejad}, {Ansorge},
  {Antilogus}, {Anwand-Heerwart}, {Arentsen}, {Arnadottir}, {Asplund}, {Auger},
  {Azais}, {Baade}, {Baker}, {Baker}, {Balbinot}, {Baldry}, {Banerji},
  {Barden}, {Barklem}, {Barth{\'e}l{\'e}my-Mazot}, {Battistini}, {Bauer},
  {Bell}, {Bellido-Tirado}, {Bellstedt}, {Belokurov}, {Bensby}, {Bergemann},
  {Bestenlehner}, {Bielby}, {Bilicki}, {Blake}, {Bland-Hawthorn}, {Boeche},
  {Boland}, {Boller}, {Bongard}, {Bongiorno}, {Bonifacio}, {Boudon}, {Brooks},
  {Brown}, {Brown}, {Br{\"u}ggen}, {Brynnel}, {Brzeski}, {Buchert},
  {Buschkamp}, {Caffau}, {Caillier}, {Carrick}, {Casagrande}, {Case}, {Casey},
  {Cesarini}, {Cescutti}, {Chapuis}, {Chiappini}, {Childress}, {Christlieb},
  {Church}, {Cioni}, {Cluver}, {Colless}, {Collett}, {Comparat}, {Cooper},
  {Couch}, {Courbin}, {Croom}, {Croton}, {Daguis{\'e}}, {Dalton}, {Davies},
  {Davis}, {de Laverny}, {Deason}, {Dionies}, {Disseau}, {Doel}, {D{\"o}scher},
  {Driver}, {Dwelly}, {Eckert}, {Edge}, {Edvardsson}, {Youssoufi}, {Elhaddad},
  {Enke}, {Erfanianfar}, {Farrell}, {Fechner}, {Feiz}, {Feltzing}, {Ferreras},
  {Feuerstein}, {Feuillet}, {Finoguenov}, {Ford}, {Fotopoulou}, {Fouesneau},
  {Frenk}, {Frey}, {Gaessler}, {Geier}, {Gentile Fusillo}, {Gerhard},
  {Giannantonio}, {Giannone}, {Gibson}, {Gillingham},
  {Gonz{\'a}lez-Fern{\'a}ndez}, {Gonzalez-Solares}, {Gottloeber}, {Gould},
  {Grebel}, {Gueguen}, {Guiglion}, {Haehnelt}, {Hahn}, {Hansen}, {Hartman},
  {Hauptner}, {Hawkins}, {Haynes}, {Haynes}, {Heiter}, {Helmi}, {Aguayo},
  {Hewett}, {Hinton}, {Hobbs}, {Hoenig}, {Hofman}, {Hook}, {Hopgood},
  {Hopkins}, {Hourihane}, {Howes}, {Howlett}, {Huet}, {Irwin}, {Iwert},
  {Jablonka}, {Jahn}, {Jahnke}, {Jarno}, {Jin}, {Jofre}, {Johl}, {Jones},
  {J{\"o}nsson}, {Jordan}, {Karovicova}, {Khalatyan}, {Kelz}, {Kennicutt},
  {King}, {Kitaura}, {Klar}, {Klauser}, {Kneib}, {Koch}, {Koposov},
  {Kordopatis}, {Korn}, {Kosmalski}, {Kotak}, {Kovalev}, {Kreckel}, {Kripak},
  {Krumpe}, {Kuijken}, {Kunder}, {Kushniruk}, {Lam}, {Lamer}, {Laurent},
  {Lawrence}, {Lehmitz}, {Lemasle}, {Lewis}, {Li}, {Lidman}, {Lind}, {Liske},
  {Lizon}, {Loveday}, {Ludwig}, {McDermid}, {Maguire}, {Mainieri}, {Mali},
  {Mandel}, {Mandel}, {Mannering}, {Martell}, {Martinez Delgado}, {Matijevic},
  {McGregor}, {McMahon}, {McMillan}, {Mena}, {Merloni}, {Meyer}, {Michel},
  {Micheva}, {Migniau}, {Minchev}, {Monari}, {Muller}, {Murphy},
  {Muthukrishna}, {Nandra}, {Navarro}, {Ness}, {Nichani}, {Nichol}, {Nicklas},
  {Niederhofer}, {Norberg}, {Obreschkow}, {Oliver}, {Owers}, {Pai},
  {Pankratow}, {Parkinson}, {Paschke}, {Paterson}, {Pecontal}, {Parry},
  {Phillips}, {Pillepich}, {Pinard}, {Pirard}, {Piskunov}, {Plank},
  {Pl{\"u}schke}, {Pons}, {Popesso}, {Power}, {Pragt}, {Pramskiy}, {Pryer},
  {Quattri}, {Queiroz}, {Quirrenbach}, {Rahurkar}, {Raichoor}, {Ramstedt},
  {Rau}, {Recio-Blanco}, {Reiss}, {Renaud}, {Revaz}, {Rhode}, {Richard},
  {Richter}, {Rix}, {Robotham}, {Roelfsema}, {Romaniello}, {Rosario},
  {Rothmaier}, {Roukema}, {Ruchti}, {Rupprecht}, {Rybizki}, {Ryde}, {Saar},
  {Sadler}, {Sahl{\'e}n}, {Salvato}, {Sassolas}, {Saunders}, {Saviauk},
  {Sbordone}, {Schmidt}, {Schnurr}, {Scholz}, {Schwope}, {Seifert}, {Shanks},
  {Sheinis}, {Sivov}, {Sk{\'u}lad{\'o}ttir}, {Smartt}, {Smedley}, {Smith},
  {Smith}, {Sorce}, {Spitler}, {Starkenburg}, {Steinmetz}, {Stilz}, {Storm},
  {Sullivan}, {Sutherland}, {Swann}, {Tamone}, {Taylor}, {Teillon}, {Tempel},
  {ter Horst}, {Thi}, {Tolstoy}, {Trager}, {Traven}, {Tremblay}, {Tresse},
  {Valentini}, {van de Weygaert}, {van den Ancker}, {Veljanoski}, {Venkatesan},
  {Wagner}, {Wagner}, {Walcher}, {Waller}, {Walton}, {Wang}, {Winkler},
  {Wisotzki}, {Worley}, {Worseck}, {Xiang}, {Xu}, {Yong}, {Zhao}, {Zheng},
  {Zscheyge}, \& {Zucker}}]{4MOST}
{de Jong} R.~S. {et~al.}, 2019, The Messenger, 175, 3

\bibitem[{{de Lapparent}, {Geller} \& {Huchra}(1986){de Lapparent}, {Geller},
  \& {Huchra}}]{de_lapparent_1986}
{de Lapparent} V., {Geller} M.~J., {Huchra} J.~P., 1986, \apjl, 302, L1

\bibitem[{{Dekel} {et~al}\mbox{.}(2009){Dekel}, {Birnboim}, {Engel},
  {Freundlich}, {Goerdt}, {Mumcuoglu}, {Neistein}, {Pichon}, {Teyssier}, \&
  {Zinger}}]{Dekel2009}
{Dekel} A. {et~al.}, 2009, \nat, 457, 451

\bibitem[{{DESI Collaboration} {et~al}\mbox{.}(2016){DESI Collaboration},
  {Aghamousa}, {Aguilar}, {Ahlen}, {Alam}, {Allen}, {Allende Prieto}, {Annis},
  {Bailey}, {Balland}, {Ballester}, {Baltay}, {Beaufore}, {Bebek}, {Beers},
  {Bell}, {Bernal}, {Besuner}, {Beutler}, {Blake}, {Bleuler}, {Blomqvist},
  {Blum}, {Bolton}, {Briceno}, {Brooks}, {Brownstein}, {Buckley-Geer},
  {Burden}, {Burtin}, {Busca}, {Cahn}, {Cai}, {Cardiel-Sas}, {Carlberg},
  {Carton}, {Casas}, {Castander}, {Cervantes-Cota}, {Claybaugh}, {Close},
  {Coker}, {Cole}, {Comparat}, {Cooper}, {Cousinou}, {Crocce}, {Cuby},
  {Cunningham}, {Davis}, {Dawson}, {de la Macorra}, {De Vicente}, {Delubac},
  {Derwent}, {Dey}, {Dhungana}, {Ding}, {Doel}, {Duan}, {Ealet}, {Edelstein},
  {Eftekharzadeh}, {Eisenstein}, {Elliott}, {Escoffier}, {Evatt}, {Fagrelius},
  {Fan}, {Fanning}, {Farahi}, {Farihi}, {Favole}, {Feng}, {Fernandez},
  {Findlay}, {Finkbeiner}, {Fitzpatrick}, {Flaugher}, {Flender}, {Font-Ribera},
  {Forero-Romero}, {Fosalba}, {Frenk}, {Fumagalli}, {Gaensicke}, {Gallo},
  {Garcia-Bellido}, {Gaztanaga}, {Pietro Gentile Fusillo}, {Gerard},
  {Gershkovich}, {Giannantonio}, {Gillet}, {Gonzalez-de-Rivera},
  {Gonzalez-Perez}, {Gott}, {Graur}, {Gutierrez}, {Guy}, {Habib}, {Heetderks},
  {Heetderks}, {Heitmann}, {Hellwing}, {Herrera}, {Ho}, {Holland}, {Honscheid},
  {Huff}, {Hutchinson}, {Huterer}, {Hwang}, {Illa Laguna}, {Ishikawa},
  {Jacobs}, {Jeffrey}, {Jelinsky}, {Jennings}, {Jiang}, {Jimenez}, {Johnson},
  {Joyce}, {Jullo}, {Juneau}, {Kama}, {Karcher}, {Karkar}, {Kehoe}, {Kennamer},
  {Kent}, {Kilbinger}, {Kim}, {Kirkby}, {Kisner}, {Kitanidis}, {Kneib},
  {Koposov}, {Kovacs}, {Koyama}, {Kremin}, {Kron}, {Kronig}, {Kueter-Young},
  {Lacey}, {Lafever}, {Lahav}, {Lambert}, {Lampton}, {Landriau}, {Lang},
  {Lauer}, {Le Goff}, {Le Guillou}, {Le Van Suu}, {Lee}, {Lee}, {Leitner},
  {Lesser}, {Levi}, {L'Huillier}, {Li}, {Liang}, {Lin}, {Linder}, {Loebman},
  {Luki{\'c}}, {Ma}, {MacCrann}, {Magneville}, {Makarem}, {Manera}, {Manser},
  {Marshall}, {Martini}, {Massey}, {Matheson}, {McCauley}, {McDonald},
  {McGreer}, {Meisner}, {Metcalfe}, {Miller}, {Miquel}, {Moustakas}, {Myers},
  {Naik}, {Newman}, {Nichol}, {Nicola}, {Nicolati da Costa}, {Nie}, {Niz},
  {Norberg}, {Nord}, {Norman}, {Nugent}, {O'Brien}, {Oh}, {Olsen}, {Padilla},
  {Padmanabhan}, {Padmanabhan}, {Palanque-Delabrouille}, {Palmese},
  {Pappalardo}, {P{\^a}ris}, {Park}, {Patej}, {Peacock}, {Peiris}, {Peng},
  {Percival}, {Perruchot}, {Pieri}, {Pogge}, {Pollack}, {Poppett}, {Prada},
  {Prakash}, {Probst}, {Rabinowitz}, {Raichoor}, {Ree}, {Refregier}, {Regal},
  {Reid}, {Reil}, {Rezaie}, {Rockosi}, {Roe}, {Ronayette}, {Roodman}, {Ross},
  {Ross}, {Rossi}, {Rozo}, {Ruhlmann-Kleider}, {Rykoff}, {Sabiu}, {Samushia},
  {Sanchez}, {Sanchez}, {Schlegel}, {Schneider}, {Schubnell}, {Secroun},
  {Seljak}, {Seo}, {Serrano}, {Shafieloo}, {Shan}, {Sharples}, {Sholl},
  {Shourt}, {Silber}, {Silva}, {Sirk}, {Slosar}, {Smith}, {Smoot}, {Som},
  {Song}, {Sprayberry}, {Staten}, {Stefanik}, {Tarle}, {Sien Tie}, {Tinker},
  {Tojeiro}, {Valdes}, {Valenzuela}, {Valluri}, {Vargas-Magana}, {Verde},
  {Walker}, {Wang}, {Wang}, {Weaver}, {Weaverdyck}, {Wechsler}, {Weinberg},
  {White}, {Yang}, {Yeche}, {Zhang}, {Zhao}, {Zheng}, {Zhou}, {Zhou}, {Zhu},
  {Zou}, \& {Zu}}]{desi}
{DESI Collaboration} {et~al.}, 2016, arXiv e-prints, arXiv:1611.00036

\bibitem[{{Donnan}, {Tojeiro} \& {Kraljic}(2022){Donnan}, {Tojeiro}, \&
  {Kraljic}}]{Donan2022}
{Donnan} C.~T., {Tojeiro} R., {Kraljic} K., 2022, Nature Astronomy, 6, 599

\bibitem[{{Dressler}(1980)}]{Dressler1980}
{Dressler} A., 1980, \apj, 236, 351

\bibitem[{{Dressler}(1986)}]{Dressler1986}
{Dressler} A., 1986, \apj, 301, 35

\bibitem[{{Driver} {et~al}\mbox{.}(2009){Driver}, {Norberg}, {Baldry},
  {Bamford}, {Hopkins}, {Liske}, {Loveday}, {Peacock}, {Hill}, {Kelvin},
  {Robotham}, {Cross}, {Parkinson}, {Prescott}, {Conselice}, {Dunne}, {Brough},
  {Jones}, {Sharp}, {van Kampen}, {Oliver}, {Roseboom}, {Bland-Hawthorn},
  {Croom}, {Ellis}, {Cameron}, {Cole}, {Frenk}, {Couch}, {Graham}, {Proctor},
  {De Propris}, {Doyle}, {Edmondson}, {Nichol}, {Thomas}, {Eales}, {Jarvis},
  {Kuijken}, {Lahav}, {Madore}, {Seibert}, {Meyer}, {Staveley-Smith},
  {Phillipps}, {Popescu}, {Sansom}, {Sutherland}, {Tuffs}, \&
  {Warren}}]{Driver2009}
{Driver} S.~P. {et~al.}, 2009, Astronomy and Geophysics, 50, 5.12

\bibitem[{{Dubois} {et~al}\mbox{.}(2014){Dubois}, {Pichon}, {Welker}, {Le
  Borgne}, {Devriendt}, {Laigle}, {Codis}, {Pogosyan}, {Arnouts}, {Benabed},
  {Bertin}, {Blaizot}, {Bouchet}, {Cardoso}, {Colombi}, {de Lapparent},
  {Desjacques}, {Gavazzi}, {Kassin}, {Kimm}, {McCracken}, {Milliard},
  {Peirani}, {Prunet}, {Rouberol}, {Silk}, {Slyz}, {Sousbie}, {Teyssier},
  {Tresse}, {Treyer}, {Vibert}, \& {Volonteri}}]{dubois2014}
{Dubois} Y. {et~al.}, 2014, \mnras, 444, 1453

\bibitem[{{Eardley} {et~al}\mbox{.}(2015){Eardley}, {Peacock},
  {McNaught-Roberts}, {Heymans}, {Norberg}, {Alpaslan}, {Baldry},
  {Bland-Hawthorn}, {Brough}, {Cluver}, {Driver}, {Farrow}, {Liske}, {Loveday},
  \& {Robotham}}]{eardley2015}
{Eardley} E. {et~al.}, 2015, \mnras, 448, 3665

\bibitem[{{Finlator} \& {Dav{\'e}}(2008)}]{FinlatorDave2008}
{Finlator} K., {Dav{\'e}} R., 2008, \mnras, 385, 2181

\bibitem[{{Fujita}(2004)}]{fujita2004}
{Fujita} Y., 2004, \pasj, 56, 29

\bibitem[{{Gabor} \& {Dav{\'e}}(2015)}]{Gabor2015}
{Gabor} J.~M., {Dav{\'e}} R., 2015, \mnras, 447, 374

\bibitem[{{Geller} \& {Huchra}(1989)}]{Geller_Huchra1989}
{Geller} M.~J., {Huchra} J.~P., 1989, Science, 246, 897

\bibitem[{{Giovanelli} {et~al}\mbox{.}(2005){Giovanelli}, {Haynes}, {Kent},
  {Perillat}, {Saintonge}, {Brosch}, {Catinella}, {Hoffman}, {Stierwalt},
  {Spekkens}, {Lerner}, {Masters}, {Momjian}, {Rosenberg}, {Springob},
  {Boselli}, {Charmandaris}, {Darling}, {Davies}, {Garcia Lambas}, {Gavazzi},
  {Giovanardi}, {Hardy}, {Hunt}, {Iovino}, {Karachentsev}, {Karachentseva},
  {Koopmann}, {Marinoni}, {Minchin}, {Muller}, {Putman}, {Pantoja}, {Salzer},
  {Scodeggio}, {Skillman}, {Solanes}, {Valotto}, {van Driel}, \& {van
  Zee}}]{alpha}
{Giovanelli} R. {et~al.}, 2005, \aj, 130, 2598

\bibitem[{{Girelli} {et~al}\mbox{.}(2020){Girelli}, {Pozzetti}, {Bolzonella},
  {Giocoli}, {Marulli}, \& {Baldi}}]{Girelli2020}
{Girelli} G., {Pozzetti} L., {Bolzonella} M., {Giocoli} C., {Marulli} F.,
  {Baldi} M., 2020, \aap, 634, A135

\bibitem[{{Glowacki}, {Elson} \& {Dav{\'e}}(2021){Glowacki}, {Elson}, \&
  {Dav{\'e}}}]{Glowacki2021}
{Glowacki} M., {Elson} E., {Dav{\'e}} R., 2021, \mnras, 507, 3267

\bibitem[{{Gouin} {et~al}\mbox{.}(2020){Gouin}, {Aghanim}, {Bonjean}, \&
  {Douspis}}]{gouin2020}
{Gouin} C., {Aghanim} N., {Bonjean} V., {Douspis} M., 2020, \aap, 635, A195

\bibitem[{{Grand} {et~al}\mbox{.}(2017){Grand}, {G{\'o}mez}, {Marinacci},
  {Pakmor}, {Springel}, {Campbell}, {Frenk}, {Jenkins}, \& {White}}]{auriga}
{Grand} R. J.~J. {et~al.}, 2017, \mnras, 467, 179

\bibitem[{{Grogin} \& {Geller}(2000)}]{grogin2000}
{Grogin} N.~A., {Geller} M.~J., 2000, \aj, 119, 32

\bibitem[{{Hahn} {et~al}\mbox{.}(2019){Hahn}, {Starkenburg}, {Choi},
  {Dav{\'e}}, {Dickey}, {Geha}, {Genel}, {Hayward}, {Maller}, {Mandyam},
  {Pandya}, {Popping}, {Rafieferantsoa}, {Somerville}, \&
  {Tinker}}]{HahnStarkenburg2019}
{Hahn} C. {et~al.}, 2019, \apj, 872, 160

\bibitem[{{Hasan} {et~al}\mbox{.}(2023){Hasan}, {Burchett}, {Abeyta},
  {Hellinger}, {Mandelker}, {Primack}, {Faber}, {Koo}, {Elek}, \&
  {Nagai}}]{hasan2023}
{Hasan} F. {et~al.}, 2023, arXiv e-prints, arXiv:2303.08088

\bibitem[{{Hopkins}(2015)}]{Hopkins2015}
{Hopkins} P.~F., 2015, \mnras, 450, 53

\bibitem[{{Hopkins} \& {Quataert}(2011)}]{HopkinsQ2011}
{Hopkins} P.~F., {Quataert} E., 2011, \mnras, 415, 1027

\bibitem[{{Hoyle}, {Vogeley} \& {Pan}(2012){Hoyle}, {Vogeley}, \&
  {Pan}}]{hoyle2012}
{Hoyle} F., {Vogeley} M.~S., {Pan} D., 2012, \mnras, 426, 3041

\bibitem[{{Iwamoto} {et~al}\mbox{.}(1999){Iwamoto}, {Brachwitz}, {Nomoto},
  {Kishimoto}, {Umeda}, {Hix}, \& {Thielemann}}]{Iwamoto1999}
{Iwamoto} K., {Brachwitz} F., {Nomoto} K., {Kishimoto} N., {Umeda} H., {Hix}
  W.~R., {Thielemann} F.-K., 1999, \apjs, 125, 439

\bibitem[{{Jaff{\'e}} {et~al}\mbox{.}(2015){Jaff{\'e}}, {Smith}, {Candlish},
  {Poggianti}, {Sheen}, \& {Verheijen}}]{Jaffe2015}
{Jaff{\'e}} Y.~L., {Smith} R., {Candlish} G.~N., {Poggianti} B.~M., {Sheen}
  Y.-K., {Verheijen} M. A.~W., 2015, \mnras, 448, 1715

\bibitem[{{Jones} {et~al}\mbox{.}(2004){Jones}, {Saunders}, {Colless}, {Read},
  {Parker}, {Watson}, {Campbell}, {Burkey}, {Mauch}, {Moore}, {Hartley},
  {Cass}, {James}, {Russell}, {Fiegert}, {Dawe}, {Huchra}, {Jarrett}, {Lahav},
  {Lucey}, {Mamon}, {Proust}, {Sadler}, \& {Wakamatsu}}]{6dF}
{Jones} D.~H. {et~al.}, 2004, \mnras, 355, 747

\bibitem[{{Katsianis} {et~al}\mbox{.}(2021){Katsianis}, {Xu}, {Yang}, {Luo},
  {Cui}, {Dav{\'e}}, {Lagos}, {Zheng}, \& {Zhao}}]{Katsianis2021}
{Katsianis} A. {et~al.}, 2021, \mnras, 500, 2036

\bibitem[{{Kauffmann} {et~al}\mbox{.}(2004){Kauffmann}, {White}, {Heckman},
  {M{\'e}nard}, {Brinchmann}, {Charlot}, {Tremonti}, \&
  {Brinkmann}}]{Kauffmann2004}
{Kauffmann} G., {White} S. D.~M., {Heckman} T.~M., {M{\'e}nard} B.,
  {Brinchmann} J., {Charlot} S., {Tremonti} C., {Brinkmann} J., 2004, \mnras,
  353, 713

\bibitem[{{Kleiner} {et~al}\mbox{.}(2017){Kleiner}, {Pimbblet}, {Jones},
  {Koribalski}, \& {Serra}}]{kleiner2017}
{Kleiner} D., {Pimbblet} K.~A., {Jones} D.~H., {Koribalski} B.~S., {Serra} P.,
  2017, \mnras, 466, 4692

\bibitem[{{Kotecha} {et~al}\mbox{.}(2022){Kotecha}, {Welker}, {Zhou},
  {Wadsley}, {Kraljic}, {Sorce}, {Rasia}, {Roberts}, {Gray}, {Yepes}, \&
  {Cui}}]{kotecha2022}
{Kotecha} S. {et~al.}, 2022, \mnras, 512, 926

\bibitem[{{Koulouridis}, {Gkini} \& {Drigga}(2024){Koulouridis}, {Gkini}, \&
  {Drigga}}]{Kouloridis2024}
{Koulouridis} E., {Gkini} A., {Drigga} E., 2024, arXiv e-prints,
  arXiv:2401.05747

\bibitem[{{Kraljic} {et~al}\mbox{.}(2018){Kraljic}, {Arnouts}, {Pichon},
  {Laigle}, {de la Torre}, {Vibert}, {Cadiou}, {Dubois}, {Treyer}, {Schimd},
  {Codis}, {de Lapparent}, {Devriendt}, {Hwang}, {Le Borgne}, {Malavasi},
  {Milliard}, {Musso}, {Pogosyan}, {Alpaslan}, {Bland-Hawthorn}, \&
  {Wright}}]{kraljic2018}
{Kraljic} K. {et~al.}, 2018, \mnras, 474, 547

\bibitem[{{Kraljic}, {Dav{\'e}} \& {Pichon}(2020){Kraljic}, {Dav{\'e}}, \&
  {Pichon}}]{Kraljic2020}
{Kraljic} K., {Dav{\'e}} R., {Pichon} C., 2020, \mnras, 493, 362

\bibitem[{{Kreckel} {et~al}\mbox{.}(2015){Kreckel}, {Croxall}, {Groves}, {van
  de Weygaert}, \& {Pogge}}]{kreckel2015}
{Kreckel} K., {Croxall} K., {Groves} B., {van de Weygaert} R., {Pogge} R.~W.,
  2015, \apjl, 798, L15

\bibitem[{{Kreckel} {et~al}\mbox{.}(2011){Kreckel}, {Platen},
  {Arag{\'o}n-Calvo}, {van Gorkom}, {van de Weygaert}, {van der Hulst},
  {Kova{\v{c}}}, {Yip}, \& {Peebles}}]{kreckel2011}
{Kreckel} K. {et~al.}, 2011, \aj, 141, 4

\bibitem[{{Krumholz} \& {Gnedin}(2011)}]{KrumholzGnedin2011}
{Krumholz} M.~R., {Gnedin} N.~Y., 2011, \apj, 729, 36

\bibitem[{{Kuutma}, {Tamm} \& {Tempel}(2017){Kuutma}, {Tamm}, \&
  {Tempel}}]{kuutma2017}
{Kuutma} T., {Tamm} A., {Tempel} E., 2017, \aap, 600, L6

\bibitem[{{Laigle} {et~al}\mbox{.}(2016){Laigle}, {McCracken}, {Ilbert},
  {Hsieh}, {Davidzon}, {Capak}, {Hasinger}, {Silverman}, {Pichon}, {Coupon},
  {Aussel}, {Le Borgne}, {Caputi}, {Cassata}, {Chang}, {Civano}, {Dunlop},
  {Fynbo}, {Kartaltepe}, {Koekemoer}, {Le F{\`e}vre}, {Le Floc'h}, {Leauthaud},
  {Lilly}, {Lin}, {Marchesi}, {Milvang-Jensen}, {Salvato}, {Sanders},
  {Scoville}, {Smolcic}, {Stockmann}, {Taniguchi}, {Tasca}, {Toft}, {Vaccari},
  \& {Zabl}}]{Laigle2015}
{Laigle} C. {et~al.}, 2016, \apjs, 224, 24

\bibitem[{{Laigle} {et~al}\mbox{.}(2018){Laigle}, {Pichon}, {Arnouts},
  {McCracken}, {Dubois}, {Devriendt}, {Slyz}, {Le Borgne}, {Benoit-L{\'e}vy},
  {Hwang}, {Ilbert}, {Kraljic}, {Malavasi}, {Park}, \& {Vibert}}]{laigle2018}
{Laigle} C. {et~al.}, 2018, \mnras, 474, 5437

\bibitem[{{Laureijs} {et~al}\mbox{.}(2011){Laureijs}, {Amiaux}, {Arduini},
  {Augu{\`e}res}, {Brinchmann}, {Cole}, {Cropper}, {Dabin}, {Duvet}, {Ealet},
  {Garilli}, {Gondoin}, {Guzzo}, {Hoar}, {Hoekstra}, {Holmes}, {Kitching},
  {Maciaszek}, {Mellier}, {Pasian}, {Percival}, {Rhodes}, {Saavedra Criado},
  {Sauvage}, {Scaramella}, {Valenziano}, {Warren}, {Bender}, {Castander},
  {Cimatti}, {Le F{\`e}vre}, {Kurki-Suonio}, {Levi}, {Lilje}, {Meylan},
  {Nichol}, {Pedersen}, {Popa}, {Rebolo Lopez}, {Rix}, {Rottgering},
  {Zeilinger}, {Grupp}, {Hudelot}, {Massey}, {Meneghetti}, {Miller}, {Paltani},
  {Paulin-Henriksson}, {Pires}, {Saxton}, {Schrabback}, {Seidel}, {Walsh},
  {Aghanim}, {Amendola}, {Bartlett}, {Baccigalupi}, {Beaulieu}, {Benabed},
  {Cuby}, {Elbaz}, {Fosalba}, {Gavazzi}, {Helmi}, {Hook}, {Irwin}, {Kneib},
  {Kunz}, {Mannucci}, {Moscardini}, {Tao}, {Teyssier}, {Weller}, {Zamorani},
  {Zapatero Osorio}, {Boulade}, {Foumond}, {Di Giorgio}, {Guttridge}, {James},
  {Kemp}, {Martignac}, {Spencer}, {Walton}, {Bl{\"u}mchen}, {Bonoli},
  {Bortoletto}, {Cerna}, {Corcione}, {Fabron}, {Jahnke}, {Ligori}, {Madrid},
  {Martin}, {Morgante}, {Pamplona}, {Prieto}, {Riva}, {Toledo}, {Trifoglio},
  {Zerbi}, {Abdalla}, {Douspis}, {Grenet}, {Borgani}, {Bouwens}, {Courbin},
  {Delouis}, {Dubath}, {Fontana}, {Frailis}, {Grazian}, {Koppenh{\"o}fer},
  {Mansutti}, {Melchior}, {Mignoli}, {Mohr}, {Neissner}, {Noddle}, {Poncet},
  {Scodeggio}, {Serrano}, {Shane}, {Starck}, {Surace}, {Taylor},
  {Verdoes-Kleijn}, {Vuerli}, {Williams}, {Zacchei}, {Altieri}, {Escudero
  Sanz}, {Kohley}, {Oosterbroek}, {Astier}, {Bacon}, {Bardelli}, {Baugh},
  {Bellagamba}, {Benoist}, {Bianchi}, {Biviano}, {Branchini}, {Carbone},
  {Cardone}, {Clements}, {Colombi}, {Conselice}, {Cresci}, {Deacon}, {Dunlop},
  {Fedeli}, {Fontanot}, {Franzetti}, {Giocoli}, {Garcia-Bellido}, {Gow},
  {Heavens}, {Hewett}, {Heymans}, {Holland}, {Huang}, {Ilbert}, {Joachimi},
  {Jennins}, {Kerins}, {Kiessling}, {Kirk}, {Kotak}, {Krause}, {Lahav}, {van
  Leeuwen}, {Lesgourgues}, {Lombardi}, {Magliocchetti}, {Maguire}, {Majerotto},
  {Maoli}, {Marulli}, {Maurogordato}, {McCracken}, {McLure}, {Melchiorri},
  {Merson}, {Moresco}, {Nonino}, {Norberg}, {Peacock}, {Pello}, {Penny},
  {Pettorino}, {Di Porto}, {Pozzetti}, {Quercellini}, {Radovich}, {Rassat},
  {Roche}, {Ronayette}, {Rossetti}, {Sartoris}, {Schneider}, {Semboloni},
  {Serjeant}, {Simpson}, {Skordis}, {Smadja}, {Smartt}, {Spano}, {Spiro},
  {Sullivan}, {Tilquin}, {Trotta}, {Verde}, {Wang}, {Williger}, {Zhao},
  {Zoubian}, \& {Zucca}}]{euclid1}
{Laureijs} R. {et~al.}, 2011, arXiv e-prints, arXiv:1110.3193

\bibitem[{{Li}, {Narayanan} \& {Dav{\'e}}(2019){Li}, {Narayanan}, \&
  {Dav{\'e}}}]{Li2019}
{Li} Q., {Narayanan} D., {Dav{\'e}} R., 2019, \mnras, 490, 1425

\bibitem[{{Lucero} {et~al}\mbox{.}(2015){Lucero}, {Carignan}, {Elson},
  {Randriamampandry}, {Jarrett}, {Oosterloo}, \& {Heald}}]{Lucero2015}
{Lucero} D.~M., {Carignan} C., {Elson} E.~C., {Randriamampandry} T.~H.,
  {Jarrett} T.~H., {Oosterloo} T.~A., {Heald} G.~H., 2015, \mnras, 450, 3935

\bibitem[{{Maddox} {et~al}\mbox{.}(2015){Maddox}, {Hess}, {Obreschkow},
  {Jarvis}, \& {Blyth}}]{maddox2015}
{Maddox} N., {Hess} K.~M., {Obreschkow} D., {Jarvis} M.~J., {Blyth} S.~L.,
  2015, \mnras, 447, 1610

\bibitem[{{Maiolino} \& {Mannucci}(2019)}]{MZR_review}
{Maiolino} R., {Mannucci} F., 2019, \aapr, 27, 3

\bibitem[{{Malavasi} {et~al}\mbox{.}(2017){Malavasi}, {Arnouts}, {Vibert}, {de
  la Torre}, {Moutard}, {Pichon}, {Davidzon}, {Kraljic}, {Bolzonella}, {Guzzo},
  {Garilli}, {Scodeggio}, {Granett}, {Abbas}, {Adami}, {Bottini}, {Cappi},
  {Cucciati}, {Franzetti}, {Fritz}, {Iovino}, {Krywult}, {Le Brun}, {Le
  F{\`e}vre}, {Maccagni}, {Ma{\l}ek}, {Marulli}, {Polletta}, {Pollo}, {Tasca},
  {Tojeiro}, {Vergani}, {Zanichelli}, {Bel}, {Branchini}, {Coupon}, {De Lucia},
  {Dubois}, {Hawken}, {Ilbert}, {Laigle}, {Moscardini}, {Sousbie}, {Treyer}, \&
  {Zamorani}}]{malavasi2017}
{Malavasi} N. {et~al.}, 2017, \mnras, 465, 3817

\bibitem[{{Malavasi} {et~al}\mbox{.}(2022){Malavasi}, {Langer}, {Aghanim},
  {Gal{\'a}rraga-Espinosa}, \& {Gouin}}]{malavasi2021}
{Malavasi} N., {Langer} M., {Aghanim} N., {Gal{\'a}rraga-Espinosa} D., {Gouin}
  C., 2022, \aap, 658, A113

\bibitem[{{Mannucci} {et~al}\mbox{.}(2010){Mannucci}, {Cresci}, {Maiolino},
  {Marconi}, \& {Gnerucci}}]{mannucci_et_al_2010}
{Mannucci} F., {Cresci} G., {Maiolino} R., {Marconi} A., {Gnerucci} A., 2010,
  \mnras, 408, 2115

\bibitem[{{Marinacci} {et~al}\mbox{.}(2018){Marinacci}, {Vogelsberger},
  {Pakmor}, {Torrey}, {Springel}, {Hernquist}, {Nelson}, {Weinberger},
  {Pillepich}, {Naiman}, \& {Genel}}]{Marinacci2018}
{Marinacci} F. {et~al.}, 2018, \mnras, 480, 5113

\bibitem[{{Moster} {et~al}\mbox{.}(2010){Moster}, {Somerville}, {Maulbetsch},
  {van den Bosch}, {Macci{\`o}}, {Naab}, \& {Oser}}]{Moster2010}
{Moster} B.~P., {Somerville} R.~S., {Maulbetsch} C., {van den Bosch} F.~C.,
  {Macci{\`o}} A.~V., {Naab} T., {Oser} L., 2010, \apj, 710, 903

\bibitem[{{Moutard} {et~al}\mbox{.}(2016){Moutard}, {Arnouts}, {Ilbert},
  {Coupon}, {Hudelot}, {Vibert}, {Comte}, {Conseil}, {Davidzon}, {Guzzo},
  {Llebaria}, {Martin}, {McCracken}, {Milliard}, {Morrison}, {Schiminovich},
  {Treyer}, \& {Van Werbaeke}}]{moutard2016a}
{Moutard} T. {et~al.}, 2016, \aap, 590, A102

\bibitem[{{Muratov} {et~al}\mbox{.}(2015){Muratov}, {Kere{\v{s}}},
  {Faucher-Gigu{\`e}re}, {Hopkins}, {Quataert}, \& {Murray}}]{Muratov2015}
{Muratov} A.~L., {Kere{\v{s}}} D., {Faucher-Gigu{\`e}re} C.-A., {Hopkins}
  P.~F., {Quataert} E., {Murray} N., 2015, \mnras, 454, 2691

\bibitem[{{Naiman} {et~al}\mbox{.}(2018){Naiman}, {Pillepich}, {Springel},
  {Ramirez-Ruiz}, {Torrey}, {Vogelsberger}, {Pakmor}, {Nelson}, {Marinacci},
  {Hernquist}, {Weinberger}, \& {Genel}}]{Naiman2018}
{Naiman} J.~P. {et~al.}, 2018, \mnras, 477, 1206

\bibitem[{{Nelson} {et~al}\mbox{.}(2018){Nelson}, {Pillepich}, {Springel},
  {Weinberger}, {Hernquist}, {Pakmor}, {Genel}, {Torrey}, {Vogelsberger},
  {Kauffmann}, {Marinacci}, \& {Naiman}}]{Nelson2018}
{Nelson} D. {et~al.}, 2018, \mnras, 475, 624

\bibitem[{{Nelson} {et~al}\mbox{.}(2019){Nelson}, {Springel}, {Pillepich},
  {Rodriguez-Gomez}, {Torrey}, {Genel}, {Vogelsberger}, {Pakmor}, {Marinacci},
  {Weinberger}, {Kelley}, {Lovell}, {Diemer}, \& {Hernquist}}]{Nelson2019}
{Nelson} D. {et~al.}, 2019, Computational Astrophysics and Cosmology, 6, 2

\bibitem[{{Noeske} {et~al}\mbox{.}(2007){Noeske}, {Weiner}, {Faber},
  {Papovich}, {Koo}, {Somerville}, {Bundy}, {Conselice}, {Newman},
  {Schiminovich}, {Le Floc'h}, {Coil}, {Rieke}, {Lotz}, {Primack}, {Barmby},
  {Cooper}, {Davis}, {Ellis}, {Fazio}, {Guhathakurta}, {Huang}, {Kassin},
  {Martin}, {Phillips}, {Rich}, {Small}, {Willmer}, \& {Wilson}}]{SFR}
{Noeske} K.~G. {et~al.}, 2007, \apjl, 660, L43

\bibitem[{{Nomoto} {et~al}\mbox{.}(2006){Nomoto}, {Tominaga}, {Umeda},
  {Kobayashi}, \& {Maeda}}]{Nomoto2006}
{Nomoto} K., {Tominaga} N., {Umeda} H., {Kobayashi} C., {Maeda} K., 2006,
  \nphysa, 777, 424

\bibitem[{{Obreschkow} \& {Rawlings}(2009)}]{Obreschkow2009}
{Obreschkow} D., {Rawlings} S., 2009, \mnras, 394, 1857

\bibitem[{{Oppenheimer} \& {Dav{\'e}}(2006)}]{OppenheimerDave2006}
{Oppenheimer} B.~D., {Dav{\'e}} R., 2006, \mnras, 373, 1265

\bibitem[{{Pillepich} {et~al}\mbox{.}(2019){Pillepich}, {Nelson}, {Springel},
  {Pakmor}, {Torrey}, {Weinberger}, {Vogelsberger}, {Marinacci}, {Genel}, {van
  der Wel}, \& {Hernquist}}]{Pillepich2019}
{Pillepich} A. {et~al.}, 2019, \mnras, 490, 3196

\bibitem[{{Pillepich} {et~al}\mbox{.}(2018){Pillepich}, {Springel}, {Nelson},
  {Genel}, {Naiman}, {Pakmor}, {Hernquist}, {Torrey}, {Vogelsberger},
  {Weinberger}, \& {Marinacci}}]{TNG}
{Pillepich} A. {et~al.}, 2018, \mnras, 473, 4077

\bibitem[{{Planck Collaboration} {et~al}\mbox{.}(2016){Planck Collaboration},
  {Ade}, {Aghanim}, {Arnaud}, {Ashdown}, {Aumont}, {Baccigalupi}, {Banday},
  {Barreiro}, {Bartlett}, {Bartolo}, {Battaner}, {Battye}, {Benabed},
  {Beno{\^\i}t}, {Benoit-L{\'e}vy}, {Bernard}, {Bersanelli}, {Bielewicz},
  {Bock}, {Bonaldi}, {Bonavera}, {Bond}, {Borrill}, {Bouchet}, {Boulanger},
  {Bucher}, {Burigana}, {Butler}, {Calabrese}, {Cardoso}, {Catalano},
  {Challinor}, {Chamballu}, {Chary}, {Chiang}, {Chluba}, {Christensen},
  {Church}, {Clements}, {Colombi}, {Colombo}, {Combet}, {Coulais}, {Crill},
  {Curto}, {Cuttaia}, {Danese}, {Davies}, {Davis}, {de Bernardis}, {de Rosa},
  {de Zotti}, {Delabrouille}, {D{\'e}sert}, {Di Valentino}, {Dickinson},
  {Diego}, {Dolag}, {Dole}, {Donzelli}, {Dor{\'e}}, {Douspis}, {Ducout},
  {Dunkley}, {Dupac}, {Efstathiou}, {Elsner}, {En{\ss}lin}, {Eriksen},
  {Farhang}, {Fergusson}, {Finelli}, {Forni}, {Frailis}, {Fraisse},
  {Franceschi}, {Frejsel}, {Galeotta}, {Galli}, {Ganga}, {Gauthier}, {Gerbino},
  {Ghosh}, {Giard}, {Giraud-H{\'e}raud}, {Giusarma}, {Gjerl{\o}w},
  {Gonz{\'a}lez-Nuevo}, {G{\'o}rski}, {Gratton}, {Gregorio}, {Gruppuso},
  {Gudmundsson}, {Hamann}, {Hansen}, {Hanson}, {Harrison}, {Helou},
  {Henrot-Versill{\'e}}, {Hern{\'a}ndez-Monteagudo}, {Herranz}, {Hildebrandt},
  {Hivon}, {Hobson}, {Holmes}, {Hornstrup}, {Hovest}, {Huang}, {Huffenberger},
  {Hurier}, {Jaffe}, {Jaffe}, {Jones}, {Juvela}, {Keih{\"a}nen}, {Keskitalo},
  {Kisner}, {Kneissl}, {Knoche}, {Knox}, {Kunz}, {Kurki-Suonio}, {Lagache},
  {L{\"a}hteenm{\"a}ki}, {Lamarre}, {Lasenby}, {Lattanzi}, {Lawrence}, {Leahy},
  {Leonardi}, {Lesgourgues}, {Levrier}, {Lewis}, {Liguori}, {Lilje},
  {Linden-V{\o}rnle}, {L{\'o}pez-Caniego}, {Lubin}, {Mac{\'\i}as-P{\'e}rez},
  {Maggio}, {Maino}, {Mandolesi}, {Mangilli}, {Marchini}, {Maris}, {Martin},
  {Martinelli}, {Mart{\'\i}nez-Gonz{\'a}lez}, {Masi}, {Matarrese}, {McGehee},
  {Meinhold}, {Melchiorri}, {Melin}, {Mendes}, {Mennella}, {Migliaccio},
  {Millea}, {Mitra}, {Miville-Desch{\^e}nes}, {Moneti}, {Montier}, {Morgante},
  {Mortlock}, {Moss}, {Munshi}, {Murphy}, {Naselsky}, {Nati}, {Natoli},
  {Netterfield}, {N{\o}rgaard-Nielsen}, {Noviello}, {Novikov}, {Novikov},
  {Oxborrow}, {Paci}, {Pagano}, {Pajot}, {Paladini}, {Paoletti}, {Partridge},
  {Pasian}, {Patanchon}, {Pearson}, {Perdereau}, {Perotto}, {Perrotta},
  {Pettorino}, {Piacentini}, {Piat}, {Pierpaoli}, {Pietrobon}, {Plaszczynski},
  {Pointecouteau}, {Polenta}, {Popa}, {Pratt}, {Pr{\'e}zeau}, {Prunet},
  {Puget}, {Rachen}, {Reach}, {Rebolo}, {Reinecke}, {Remazeilles}, {Renault},
  {Renzi}, {Ristorcelli}, {Rocha}, {Rosset}, {Rossetti}, {Roudier},
  {Rouill{\'e} d'Orfeuil}, {Rowan-Robinson}, {Rubi{\~n}o-Mart{\'\i}n},
  {Rusholme}, {Said}, {Salvatelli}, {Salvati}, {Sandri}, {Santos},
  {Savelainen}, {Savini}, {Scott}, {Seiffert}, {Serra}, {Shellard}, {Spencer},
  {Spinelli}, {Stolyarov}, {Stompor}, {Sudiwala}, {Sunyaev}, {Sutton},
  {Suur-Uski}, {Sygnet}, {Tauber}, {Terenzi}, {Toffolatti}, {Tomasi},
  {Tristram}, {Trombetti}, {Tucci}, {Tuovinen}, {T{\"u}rler}, {Umana},
  {Valenziano}, {Valiviita}, {Van Tent}, {Vielva}, {Villa}, {Wade}, {Wandelt},
  {Wehus}, {White}, {White}, {Wilkinson}, {Yvon}, {Zacchei}, \&
  {Zonca}}]{Planck2016}
{Planck Collaboration} {et~al.}, 2016, \aap, 594, A13

\bibitem[{{Poggianti} {et~al}\mbox{.}(2017){Poggianti}, {Jaff{\'e}}, {Moretti},
  {Gullieuszik}, {Radovich}, {Tonnesen}, {Fritz}, {Bettoni}, {Vulcani},
  {Fasano}, {Bellhouse}, {Hau}, \& {Omizzolo}}]{Poggianti2017}
{Poggianti} B.~M. {et~al.}, 2017, \nat, 548, 304

\bibitem[{{Postman} \& {Geller}(1984)}]{PostmanGeller1984}
{Postman} M., {Geller} M.~J., 1984, \apj, 281, 95

\bibitem[{{Poudel} {et~al}\mbox{.}(2017){Poudel}, {Hein{\"a}m{\"a}ki},
  {Tempel}, {Einasto}, {Lietzen}, \& {Nurmi}}]{pouder2017}
{Poudel} A., {Hein{\"a}m{\"a}ki} P., {Tempel} E., {Einasto} M., {Lietzen} H.,
  {Nurmi} P., 2017, \aap, 597, A86

\bibitem[{{Refregier}(2009)}]{euclid2}
{Refregier} A., 2009, Experimental Astronomy, 23, 17

\bibitem[{{Ricciardelli} {et~al}\mbox{.}(2014){Ricciardelli}, {Cava}, {Varela},
  \& {Quilis}}]{ricciardelli2014}
{Ricciardelli} E., {Cava} A., {Varela} J., {Quilis} V., 2014, \mnras, 445, 4045

\bibitem[{{Rojas} {et~al}\mbox{.}(2004){Rojas}, {Vogeley}, {Hoyle}, \&
  {Brinkmann}}]{rojas2004}
{Rojas} R.~R., {Vogeley} M.~S., {Hoyle} F., {Brinkmann} J., 2004, \apj, 617, 50

\bibitem[{{Salerno} {et~al}\mbox{.}(2020){Salerno}, {Mart{\'\i}nez}, {Muriel},
  {Coenda}, {Vulcani}, {Poggianti}, {Moretti}, {Gullieuszik}, {Fritz}, \&
  {Bettoni}}]{salerno2020}
{Salerno} J.~M. {et~al.}, 2020, \mnras, 493, 4950

\bibitem[{{Sales} {et~al}\mbox{.}(2012){Sales}, {Navarro}, {Theuns}, {Schaye},
  {White}, {Frenk}, {Crain}, \& {Dalla Vecchia}}]{sales2012}
{Sales} L.~V., {Navarro} J.~F., {Theuns} T., {Schaye} J., {White} S. D.~M.,
  {Frenk} C.~S., {Crain} R.~A., {Dalla Vecchia} C., 2012, \mnras, 423, 1544

\bibitem[{{Schaap} \& {van de Weygaert}(2000)}]{Schaap2000}
{Schaap} W.~E., {van de Weygaert} R., 2000, \aap, 363, L29

\bibitem[{{Schawinski} {et~al}\mbox{.}(2014){Schawinski}, {Urry}, {Simmons},
  {Fortson}, {Kaviraj}, {Keel}, {Lintott}, {Masters}, {Nichol}, {Sarzi},
  {Skibba}, {Treister}, {Willett}, {Wong}, \& {Yi}}]{Schawinski2014}
{Schawinski} K. {et~al.}, 2014, \mnras, 440, 889

\bibitem[{{Schaye} {et~al}\mbox{.}(2015){Schaye}, {Crain}, {Bower}, {Furlong},
  {Schaller}, {Theuns}, {Dalla Vecchia}, {Frenk}, {McCarthy}, {Helly},
  {Jenkins}, {Rosas-Guevara}, {White}, {Baes}, {Booth}, {Camps}, {Navarro},
  {Qu}, {Rahmati}, {Sawala}, {Thomas}, \& {Trayford}}]{eagle}
{Schaye} J. {et~al.}, 2015, \mnras, 446, 521

\bibitem[{{Schmidt}(1959)}]{Schmidt1959}
{Schmidt} M., 1959, \apj, 129, 243

\bibitem[{{Scodeggio} {et~al}\mbox{.}(2018){Scodeggio}, {Guzzo}, {Garilli},
  {Granett}, {Bolzonella}, {de la Torre}, {Abbas}, {Adami}, {Arnouts},
  {Bottini}, {Cappi}, {Coupon}, {Cucciati}, {Davidzon}, {Franzetti}, {Fritz},
  {Iovino}, {Krywult}, {Le Brun}, {Le F{\`e}vre}, {Maccagni}, {Ma{\l}ek},
  {Marchetti}, {Marulli}, {Polletta}, {Pollo}, {Tasca}, {Tojeiro}, {Vergani},
  {Zanichelli}, {Bel}, {Branchini}, {De Lucia}, {Ilbert}, {McCracken},
  {Moutard}, {Peacock}, {Zamorani}, {Burden}, {Fumana}, {Jullo}, {Marinoni},
  {Mellier}, {Moscardini}, \& {Percival}}]{scodeggio2018}
{Scodeggio} M. {et~al.}, 2018, \aap, 609, A84

\bibitem[{{Smith} {et~al}\mbox{.}(2017){Smith}, {Bryan}, {Glover}, {Goldbaum},
  {Turk}, {Regan}, {Wise}, {Schive}, {Abel}, {Emerick}, {O'Shea}, {Anninos},
  {Hummels}, \& {Khochfar}}]{Smith2017}
{Smith} B.~D. {et~al.}, 2017, \mnras, 466, 2217

\bibitem[{{Sorini} {et~al}\mbox{.}(2022){Sorini}, {Dav{\'e}}, {Cui}, \&
  {Appleby}}]{Sorini2022}
{Sorini} D., {Dav{\'e}} R., {Cui} W., {Appleby} S., 2022, \mnras, 516, 883

\bibitem[{{Sousbie}(2011)}]{Sousbie2011}
{Sousbie} T., 2011, \mnras, 414, 350

\bibitem[{{Sousbie}, {Pichon} \& {Kawahara}(2011){Sousbie}, {Pichon}, \&
  {Kawahara}}]{SousbiePK2011}
{Sousbie} T., {Pichon} C., {Kawahara} H., 2011, \mnras, 414, 384

\bibitem[{{Springel}(2005)}]{Springel2005}
{Springel} V., 2005, \mnras, 364, 1105

\bibitem[{{Springel} {et~al}\mbox{.}(2018){Springel}, {Pakmor}, {Pillepich},
  {Weinberger}, {Nelson}, {Hernquist}, {Vogelsberger}, {Genel}, {Torrey},
  {Marinacci}, \& {Naiman}}]{Springel2018}
{Springel} V. {et~al.}, 2018, \mnras, 475, 676

\bibitem[{{Takada} {et~al}\mbox{.}(2014){Takada}, {Ellis}, {Chiba}, {Greene},
  {Aihara}, {Arimoto}, {Bundy}, {Cohen}, {Dor{\'e}}, {Graves}, {Gunn},
  {Heckman}, {Hirata}, {Ho}, {Kneib}, {Le F{\`e}vre}, {Lin}, {More},
  {Murayama}, {Nagao}, {Ouchi}, {Seiffert}, {Silverman}, {Sodr{\'e}},
  {Spergel}, {Strauss}, {Sugai}, {Suto}, {Takami}, \& {Wyse}}]{PFS2014}
{Takada} M. {et~al.}, 2014, \pasj, 66, R1

\bibitem[{{Tegmark} {et~al}\mbox{.}(2004){Tegmark}, {Blanton}, {Strauss},
  {Hoyle}, {Schlegel}, {Scoccimarro}, {Vogeley}, {Weinberg}, {Zehavi},
  {Berlind}, {Budavari}, {Connolly}, {Eisenstein}, {Finkbeiner}, {Frieman},
  {Gunn}, {Hamilton}, {Hui}, {Jain}, {Johnston}, {Kent}, {Lin}, {Nakajima},
  {Nichol}, {Ostriker}, {Pope}, {Scranton}, {Seljak}, {Sheth}, {Stebbins},
  {Szalay}, {Szapudi}, {Verde}, {Xu}, {Annis}, {Bahcall}, {Brinkmann},
  {Burles}, {Castander}, {Csabai}, {Loveday}, {Doi}, {Fukugita}, {Gott},
  {Hennessy}, {Hogg}, {Ivezi{\'c}}, {Knapp}, {Lamb}, {Lee}, {Lupton}, {McKay},
  {Kunszt}, {Munn}, {O'Connell}, {Peoples}, {Pier}, {Richmond}, {Rockosi},
  {Schneider}, {Stoughton}, {Tucker}, {Vanden Berk}, {Yanny}, {York}, \& {SDSS
  Collaboration}}]{Tegmark2004}
{Tegmark} M. {et~al.}, 2004, \apj, 606, 702

\bibitem[{{Trager} {et~al}\mbox{.}(2000){Trager}, {Faber}, {Worthey}, \&
  {Gonz{\'a}lez}}]{MZR}
{Trager} S.~C., {Faber} S.~M., {Worthey} G., {Gonz{\'a}lez} J.~J., 2000, \aj,
  120, 165

\bibitem[{{Tremonti} {et~al}\mbox{.}(2004){Tremonti}, {Heckman}, {Kauffmann},
  {Brinchmann}, {Charlot}, {White}, {Seibert}, {Peng}, {Schlegel}, {Uomoto},
  {Fukugita}, \& {Brinkmann}}]{Tremonti2004}
{Tremonti} C.~A. {et~al.}, 2004, \apj, 613, 898

\bibitem[{{Tudorache} {et~al}\mbox{.}(2022){Tudorache}, {Jarvis}, {Heywood},
  {Ponomareva}, {Maddox}, {Frank}, {Adams}, {Bowler}, {Whittam}, {Baes}, {Pan},
  {Rajohnson}, {Sinigaglia}, \& {Spekkens}}]{madalina}
{Tudorache} M.~N. {et~al.}, 2022, \mnras, 513, 2168

\bibitem[{{Urban} {et~al}\mbox{.}(2017){Urban}, {Werner}, {Allen},
  {Simionescu}, \& {Mantz}}]{highZ2}
{Urban} O., {Werner} N., {Allen} S.~W., {Simionescu} A., {Mantz} A., 2017,
  \mnras, 470, 4583

\bibitem[{{Vulcani} {et~al}\mbox{.}(2019){Vulcani}, {Poggianti}, {Moretti},
  {Gullieuszik}, {Fritz}, {Franchetto}, {Fasano}, {Bettoni}, \&
  {Jaff{\'e}}}]{vulcani2019}
{Vulcani} B. {et~al.}, 2019, \mnras, 487, 2278

\bibitem[{{Wegner} {et~al}\mbox{.}(2019){Wegner}, {Salzer}, {Taylor}, \&
  {Hirschauer}}]{wegner2019}
{Wegner} G.~A., {Salzer} J.~J., {Taylor} J.~M., {Hirschauer} A.~S., 2019, \apj,
  883, 29

\bibitem[{{Weinberger} {et~al}\mbox{.}(2017){Weinberger}, {Springel},
  {Hernquist}, {Pillepich}, {Marinacci}, {Pakmor}, {Nelson}, {Genel},
  {Vogelsberger}, {Naiman}, \& {Torrey}}]{Weinberger2017}
{Weinberger} R. {et~al.}, 2017, \mnras, 465, 3291

\bibitem[{{Wetzel}, {Tinker} \& {Conroy}(2012){Wetzel}, {Tinker}, \&
  {Conroy}}]{Wetzel2012}
{Wetzel} A.~R., {Tinker} J.~L., {Conroy} C., 2012, \mnras, 424, 232

\bibitem[{{Wetzel} {et~al}\mbox{.}(2013){Wetzel}, {Tinker}, {Conroy}, \& {van
  den Bosch}}]{wetzel2013}
{Wetzel} A.~R., {Tinker} J.~L., {Conroy} C., {van den Bosch} F.~C., 2013,
  \mnras, 432, 336

\bibitem[{{Williams} {et~al}\mbox{.}(2009){Williams}, {Quadri}, {Franx}, {van
  Dokkum}, \& {Labb{\'e}}}]{williams}
{Williams} R.~J., {Quadri} R.~F., {Franx} M., {van Dokkum} P., {Labb{\'e}} I.,
  2009, \apj, 691, 1879

\bibitem[{{Winkel} {et~al}\mbox{.}(2021){Winkel}, {Pasquali}, {Kraljic},
  {Smith}, {Gallazzi}, \& {Jackson}}]{winkel_2021}
{Winkel} N., {Pasquali} A., {Kraljic} K., {Smith} R., {Gallazzi} A., {Jackson}
  T.~M., 2021, \mnras, 505, 4920

\bibitem[{{Zel'dovich}(1970{\natexlab{a}})}]{zeldovich2}
{Zel'dovich} Y.~B., 1970{\natexlab{a}}, Astrophysics, 6, 164

\bibitem[{{Zel'dovich}(1970{\natexlab{b}})}]{zeldovich1}
{Zel'dovich} Y.~B., 1970{\natexlab{b}}, \aap, 5, 84

\bibitem[{{Zheng} {et~al}\mbox{.}(2022){Zheng}, {Liao}, {Hu}, {Gao}, {Grand},
  {Gu}, \& {Guo}}]{zheng2022}
{Zheng} H., {Liao} S., {Hu} J., {Gao} L., {Grand} R. J.~J., {Gu} Q., {Guo} Q.,
  2022, \mnras, 514, 2488

\bibitem[{{Zolotov} {et~al}\mbox{.}(2015){Zolotov}, {Dekel}, {Mandelker},
  {Tweed}, {Inoue}, {DeGraf}, {Ceverino}, {Primack}, {Barro}, \&
  {Faber}}]{Zolotov2015}
{Zolotov} A. {et~al.}, 2015, \mnras, 450, 2327

\end{thebibliography}

\bsp	
\label{lastpage}
\end{document}